\documentclass[a4paper,12pt]{article}
\usepackage{graphicx}
\usepackage{dcolumn}
\usepackage{bm}
\usepackage{amssymb}
\usepackage{amsmath}
\usepackage{hyperref}
\usepackage{authblk}
\usepackage{cite}
\usepackage{float}
\usepackage{caption}
\usepackage{tikz}

\begin{document} 

\title{L\'{e}vy-like flights and fractal geometry of finite point sets}
\author[1]{Konstantinos Chalas\footnote{email: kchalas@sissa.it}}
\author[2]{F.~K.~Diakonos\footnote{email: fdiakono@phys.uoa.gr}}
\author[3]{A.~S.~Kapoyannis\footnote{email: akapog@phys.uoa.gr}}
\affil[1]{SISSA and INFN Sezione di Trieste, via Bonomea 265, 34136 Trieste, Italy}
\affil[2,3]{Nuclear and Particle Physics Section, Faculty of Physics, 
University of Athens, GR-15784 Greece}

\maketitle

\begin{abstract}

We study L\'{e}vy-like and truncated L\'{e}vy-like flights with step probability distribution of the form $r^{-1+\nu}$ for 
negative, positive, and zero $\nu$, focusing on the appearance of fractal geometry characteristics in the generated point sets. Forming ensembles of such point sets with fixed multiplicity, we develop simulation techniques leading to the desired value of correlation dimension in a vast continuous interval of scales. In particular, we demonstrate the possibility to produce ensembles of data sets with a low number of points with the needed properties. Furthermore, we show that the positive 
$\nu$ distributions, apart from a region near the upper scale limit, show fractal behaviour that extends to infinitesimally low scales. As an example, we apply our findings to producing simulations
relevant to the search for critical fluctuations, related to QCD critical endpoint, in heavy-ion collision experiments.


\end{abstract}

\section{Introduction}\label{sec:intro}

L\'{e}vy flights \cite{Levy1937, Montroll1983} provide a powerful method to generate random point sets with a prescribed fractal dimension in the range \(0 < d_F < 3\) \cite{Alemany1994}. The fractal structure of the resulting point set originates from the heavy-tailed step-length distribution of the form 
\begin{equation}\label{eq:st_pr}
P(r) \sim r^{-1+\nu} = r^{-1 \pm \mu},\;
\mu =|\nu|\;,
\end{equation}
which gives rise to scale-invariant spatial correlations and reflects the long-range influence of rare, large (formally unbounded) jumps. In fact, L\'{e}vy flights are referred exclusively to the case $-2 \le \nu <0$ when $P(r)$ is stable \cite{Levy1937}. In this case, $P(r)$ has to be regularized for $r \to 0$ (usually with a small cutoff) to ensure normalizability. L\'{e}vy flights have found widespread application across a broad range of systems with scale-invariant features, such as anomalous diffusion and transport in complex and disordered systems \cite{Metzler2000}, optimal search strategies and foraging patterns of animals in sparse environments \cite{Viswanathan1999, Viswanathan2011}, optimization and computational algorithms with enhanced global search efficiency in complex landscapes \cite{Yang2010,Chawla2018,Li2022}, to name a few. However, in practice, scale invariance and the associated fractal geometry usually occurs between specific scales, which, in general, may differ by only a few decades \cite{Malcai1997}. A natural way to take this restriction, in point sets generated through random walks, into account, is to use the so called truncated L\'{e}vy flights \cite{Mantegna1994} where an upper cutoff in the size of step-length is introduced. Truncated L\'{e}vy flights have been used in systems where the cutoff scale plays a significant role such as in econophysics \cite{Mantegna1994,Xiong2010,Constantinides2013}, human mobility \cite{Schlink2011}, anomalous transport in finite disordered systems \cite{Sokolov2004,Negrete2009}, etc. The presence of the upper cutoff breaks scale invariance inducing multifractality in the geometric characteristics of point sets generated by truncated L\'{e}vy flights \cite{Nakao2000,Vinogradov2010}.

Introducing an upper bound on the step-length $\ell$ naturally extends the range of permissible exponents in the power-law distribution $r^{-1+\nu}$, allowing step-length distributions with $\nu \geq 0$. In this regime, for $\nu>0$, there is no need for a lower cutoff in $r$, unlike in truncated L\'{e}vy flights. In this work, we adopt the term \emph{L\'{e}vy-like} to refer to all flights whose step-length distribution follows $P(r) \sim r^{-1+\nu}$ with $\nu$ taking arbitrary negative, zero, or positive values, assuming only the necessary cutoffs are present. Correspondingly, we use the term \emph{truncated L\'{e}vy-like flights} when both upper and lower cutoffs on the step-length size are imposed. Although these processes represent a natural generalization of L\'{e}vy and truncated L\'{e}vy flights, a unified framework for analyzing their scaling behavior and the fractal geometry of the resulting point sets is still lacking. Developing such a framework is one of the main objectives of the present work. In particular we are interested to evolve a systematic way for controlling and maximizing the range of the scaling region which in many cases goes beyond the adjustment of upper and lower cut-off parameters. As a tool towards this goal we will use the correlation integral, $C(R)$ \cite{Grassberger1983}, to seek for the fractal correlation dimension, $d_F$, of the produced set of points. For the calculation of $C$ we use the ring technique developed in \cite{Kapoyannis2022}, which allows for fast calculations. As will become evident in the subsequent analysis, a notable outcome of this correlation-dimension approach is that, in many cases, the scaling region is not directly associated with the heavy tail of the underlying step-length distribution.

Furthermore, as will be shown in the following sections, the properties of interest depend strongly on the total number of steps in the walk\footnote{Throughout this manuscript, the terms “walk” and “flight” are used interchangeably.} we perform. This in turn rises some interesting and fundamental questions. Is it possible to have the same properties with a data set of fewer points than the steps of the walk? Or is it possible to combine point sets
without altering their properties? These questions arise naturally because frequently we have to form 
sets with given number of points and they will be addressed in Section \ref{sec:alter_Nm}. There we show how we can extract a fewer number of points drawn from a higher
number set without altering their distribution. We, also, discuss the conditions under which we can join different data sets. In Section \ref{sec:step} we record the mathematical formulas of the step probability distributions we use throughout our paper, present the two types of flights we use and explain how we calculate the correlation dimension. In Section \ref{sec:dif_mul} we show how the $C$ distributions for negative and positive $\nu$ change under the variation of the 
number of steps. In Section \ref{sec:dif_nu} we show the form of the $C$ distributions for different 
values of negative and positive $\nu$. In both Sections \ref{sec:dif_mul} and \ref{sec:dif_nu} the
steep probabilities maintain only the necessary lower or upper limit. In Section \ref{sec:ad_lim}
we discuss the effect of the imposition of an additional limit on the form of the $C$ distributions
of negative and positive $\nu$. Also, we discuss the effect of the application of a working window
which excludes data outside of it. In Section \ref{sec:nu=0} we investigate the distributions for
zero $\nu$ which require both limits to be present. We find interesting applications for low multiplicity data sets. 
In Section \ref{sec:lim_pm} we link the properties of $C$ distributions with negative and positive $\nu$ 
when both upper and lower limits are present.
In Section \ref{sec:d_F=1/3} we apply the developed techniques to produce numerous applications for a specific fractal dimension. In the Conclusions we summarize our results, while in the Appendix we prove some formulas needed in the applications of Section \ref{sec:d_F=1/3}.

\section{Altering $N_m$ in a simulation}\label{sec:alter_Nm}

In walks, i.e., when each step starts from the point reached by the previous one, the
correlation integral, $C$, generally depends on the number of steps taken. In practice,
however, one may wish to construct a set containing a prescribed number of points.
We therefore examine how the number of points in a set drawn from a walk can be adjusted.

Let us first consider a set of points with multiplicity $n$. We will show that a subset
of $k<n$ points, selected at random from this set, forms a new set that follows the same
$C$ distribution. Our proof holds in the most general setting, where the $n$-step walk may
be heterogeneous, i.e., the $i$-th step may be generated by a different step probability
distribution than the $(i-1)$-th step. It also applies to sets in which each point is
generated independently of the previous one.

We assign to each point an index equal to the number of steps required to reach it.
The $n$ points thus form $n\left(n-1\right)/2$ pairs. Consider one such pair, formed
by the points $i$ and $j$. The contribution to the correlation integral from this
specific pair—namely, the correlation integral of the two-point set defined by the
$i$-th and $j$-th steps—is denoted by $\tilde C_{i,j}$. The correlation distribution
of the full set of $n$ points is then given by:
\begin{equation}\label{eq:Cn,general}
C_n = \sum\limits_{i \ne j} p_{i,j}^n \tilde C_{i,j}\;,
\end{equation}
where $p_{i,j}^n$ is the probability of finding the pair $i,j$ within the pairs that can be formed by $n$ points. All the pairs have equal probability and all the possible pairs are $n\left(n-1\right)/2$, so:
\[
p_{i,j}^n = \frac{1}{\frac{n\left(n-1\right)}{2}} = \frac{2}{n\left(n-1\right)}\;.
\]
The most general $C$ distribution is then:
\begin{equation}\label{eq:Cn}
C_n = \frac{2}{n\left(n-1\right)}\sum\limits_{i \ne j} \tilde C_{i,j}
\end{equation}
We pick $k$ points randomly from the set of $n$. Then their distribution is:
\begin{equation}
C_k = \sum\limits_{i \ne j} p_{i,j}^k \tilde C_{i,j}\;,
\end{equation}
where $p_{i,j}^k$ is the probability of finding the pair $i,j$ within the pairs which can be formed by 
$k$ points, picked randomly from the set of $n$. We first find all the possible pairs that can be formed from all possible sets of $k$ points. We can form 
$\left( {\begin{array}{*{20}{c}} n\\ k \end{array}} \right)$ 
different sets of $k$ points. For each set the 
pairs that can be formed are $\frac{k\left( k-1 \right)}{2}$. So the possible pairs are:
\[
\left( {\begin{array}{*{20}{c}} n\\k \end{array}} \right)\frac{k\left(k-1\right)}{2}\;.
\]
Now the sets of $k$ points that include the specific $i$ and $j$ points are 
\[
\left( {\begin{array}{*{20}{c}}{n - 2}\\{k - 2}\end{array}} \right)\;.
\]
Each of these sets form 1 pair between $i$ and $j$ points. So:
 

\[
p_{i,j}^k = 
\dfrac{{\left( {\begin{array}{*{20}{c}} {n - 2}\\{k - 2}\end{array}} \right) \cdot 1}}{{\left( {\begin{array}{*{20}{c}}n\\k\end{array}} \right)\dfrac{{k\left(k-1\right)}}{2}}} =
\dfrac{1}{\dfrac{n\left( n - 1 \right)}{2}} = 
p_{i,j}^n
\]

Thus, this proves that the correlation function of the $k$ points drawn randomly from the set
of $n$ is the same as the correlation function of the $n$ points
\begin{equation}
C_k = C_n\;.
\end{equation}
The validity of last equation is shown in Fig.~\ref{fig:pick}, where we compare the $C$ distribution
of a complete walk of $10^3$ steps with the $C$ distributions of sets of lower number of points
drawn randomly from the complete walk. However, low number sets may need more events to decrease
fluctuations.
 
There is a useful way to write the correlation integral for a set of points 
produced in a walk where now all the steps have the same probability distribution. 
Specifically, going back to eq.~(\ref{eq:Cn}), we see that the pairs formed by two points which
have the same number of steps $M$ in between should follow the same distribution, ${\tilde C}_M$.
This follows from the fact that
they are produced in the same way, irrespectively of the number of steps taken to arrive at the
the first of them. Also, in a walk of $n$ steps, there are totally $n-M$ such pairs.
So eq.~(\ref{eq:Cn}) can be rewritten in a way that we can group the correlation integrals of all
the pairs in the correlation integrals of pairs which are $M$ steps away:
\begin{equation}\label{eq:Cn,M}
C_n 
=\frac{2}{n\left( {n - 1} \right)}\sum\limits_{M = 1}^{n - 1} {\left( {n - M} \right){\tilde C}_M}
\end{equation}

In another procedure, a set of points are produced in an non-successive procedure with the  
same probability distribution. Then, all the pairs between every two points have the same 
distribution, $\tilde C$, thus:

\begin{figure}[H]
\centering
\vspace{-0.0cm}
\includegraphics[scale=0.5,trim=2.in 0.3in 2.in 0.2in,angle=0]{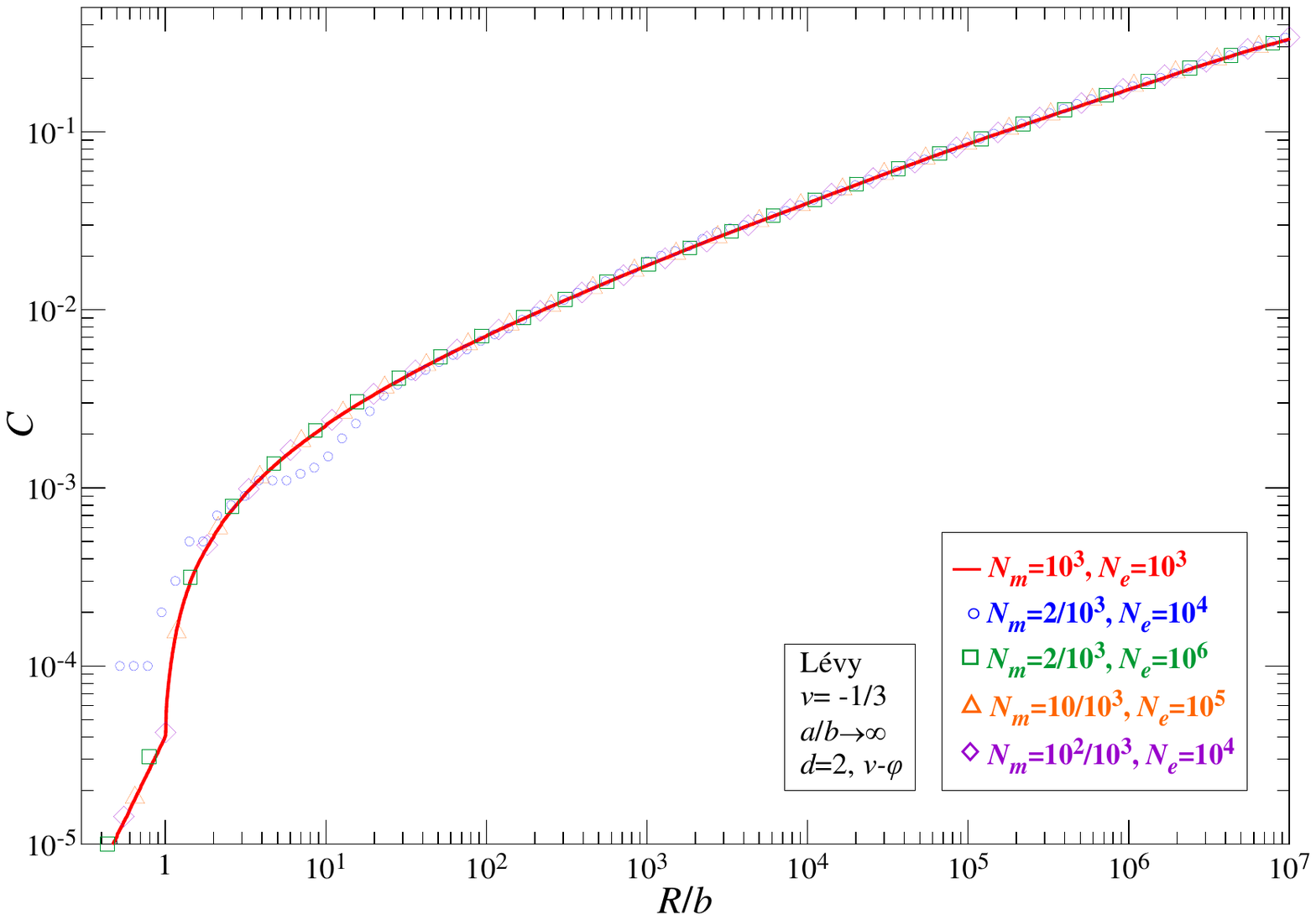}
\caption{\label{fig:pick} {\small 
The $C(R)$ distribution of an arbitrary point set (a L\'{e}vy walk with 
step probability exponent $\nu=-1/3$ for the $\nu-\phi$ case, embedding space dimension $d=$2,
lower step limit $b$ and no upper step limit). With continuous line we depict the $C(R)$ of
all the points of the set which is $N_m=10^3$. The various symbols correspond to the 
$C(R)$ curves for sets of different number of points drawn randomly from the full
set consisting of $N_m$ steps. In the caption $N_m=j/k$ denotes (here and in the rest of the paper)
$j$ points taken from a full set of $k$ points. The increased fluctuations for $N_m=2$ and
number of events $N_e=10^4$ decrease when events increase to $N_e=10^6$.}}
\end{figure}

\noindent
\begin{equation}\label{eq:Cn,non-cons}
C_n = \frac{2}{n\left( {n - 1} \right)}\sum\limits_{i \ne j} {\tilde C}_{i,j} 
=\frac{2}{n\left( {n - 1} \right)}\sum\limits_{i \ne j} {\tilde C}
= \frac{2}{n\left( {n - 1} \right)}\frac{n\left( {n - 1} \right)}{2}\tilde C
\Rightarrow C_n = \tilde C
\end{equation}
So the distribution is independent from the number of points, $n$, contained in the set. 
From this observation it follows that at a set of points produced through a non-consequent 
procedure with the same step probability, we can join the sets forming larger sets without 
altering the correlation integral distribution. But we have to note that the first point in each
of the sets we will be joining has to be produced in the same way as the rest of the points
contained in each set.

\section{Producing the walk and evaluating the fractal dimension}\label{sec:step}

\subsection{Step Probabilities}

In this subsection we will present the step probability densities $p(r)$, the corresponding 
cumulative distributions, $F(r)$ and the inverse 
cumulative functions $\rho(r)$. Choosing $\rho$ as a uniformly distributed random number in the interval [0,1), then
$r$ will follow the density $p(r)$.

{\it $\nu < 0$}

When $\nu=-\mu < 0$, the step probability (\ref{eq:st_pr}) has a non-integrable
singularity at $r=0$. In order to have a probability distribution function (PDF),
a lower limit $b$ has to be set in the step
that can be taken. On the contrary, there is no need to impose an upper limit, $a$, in the step,
which can be left to $a \rightarrow \infty $. The relevant equations become then:
\begin{equation}\label{eq:nn_pdf}
p(r) = \mu b^{\mu}r^{-1-\mu}\;,
\end{equation}
\begin{equation}\label{eq:nn_cu}
F(r) = \left[ 1 - \left( \frac{b}{r} \right)^\mu \right]\;,
\end{equation}
\begin{equation}\label{eq:nn_incu}
r = b \left( {1-\rho} \right)^{-1/\mu}\;.
\end{equation}
However, the additional limit $a$ can be set, if needed, so we have the relations:
\begin{equation}\label{eq:nn_pdfa}
p(r) = \mu b^{\mu} \left[1- \left(\frac{b}{a}\right)^{\mu}\right]^{-1} r^{-1-\mu}\;,
\end{equation}
\begin{equation}\label{eq:nn_cua}
F(r) = \left[1- \left(\frac{b}{a}\right)^{\mu}\right]^{-1}
 \left[ 1 - \left( \frac{b}{r} \right)^\mu \right]\;,
\end{equation}
\begin{equation}\label{eq:nn_incua}
r = b \left\{ {1-\left[1- \left(\frac{b}{a}\right)^{\mu}\right] \rho} \right\}^{-1/\mu}\;.
\end{equation}

\noindent
{\it $\nu > 0$}

When $\nu=\mu > 0$, the step probability (\ref{eq:st_pr}) has a non-integrable
singularity at $r\rightarrow \infty$. In order to have a probability distribution function,
an upper limit $a$ has to be set in the potential step. On the contrary, there is no need 
to impose a lower limit, $b$, in the step,
which can be set to $b=0$, leading to: 
\begin{equation}\label{eq:pn_pdf}
p(r) = \mu a^{-\mu}r^{-1+\mu}\;,
\end{equation}
\begin{equation}\label{eq:pn_cu}
F(r) = \left( \frac{r}{a} \right)^{\mu}  \;,
\end{equation}
\begin{equation}\label{eq:pn_incu}
r = a \rho ^{1/\mu} \;.
\end{equation}
Again, another non-zero lower limit $b$ can be set, with the following formulas:
\begin{equation}\label{eq:pn_pdfb}
p(r) = \mu a^{-\mu} \left[1- \left(\frac{b}{a}\right)^{\mu}\right]^{-1} r^{-1+\mu}\;,
\end{equation}
\begin{equation}\label{eq:pn_cub}
F(r) = \left[1- \left(\frac{b}{a}\right)^{\mu}\right]^{-1}
 \left[ \left( \frac{r}{a} \right)^\mu - \left( \frac{b}{a} \right)^\mu \right]\;,
\end{equation}
\begin{equation}\label{eq:pn_incub}
r = a \left\{ {\left[1- \left(\frac{b}{a}\right)^{\mu}\right]\rho +\left( \frac{b}{a} \right)^\mu } \right\}^{1/\mu}\;.
\end{equation}

\noindent
{\it $\nu = 0$}

When $\nu=0$, the step probability (\ref{eq:st_pr}) has two non-integrable
singularities at $r=0$ and at $r\rightarrow \infty$. In order to have a probability distribution function,
both an upper limit, $a$ and a lower limit, $b$, have to be set in the step.
In each case we have to calculate a suitable normalisation factor to form a proper PDF $p(r)$. So, then, we have:
\begin{equation}\label{eq:zn_pdf}
p(r) = \left[ \ln \left( \frac{a}{b} \right) \right]^{-1} r^{-1} \;,
\end{equation}
\begin{equation}\label{eq:zn_cu}
F(r) = \left[ \ln \left( \frac{a}{b} \right) \right]^{-1} 
\left[ \ln \left( \frac{r}{b} \right) \right] \;,
\end{equation}
\begin{equation}\label{eq:zn_incu}
r = b \left( \frac{a}{b} \right)^\rho \;.
\end{equation}

All distributions for any value of $\nu$ remain invariant under a rescaling with a constant $\lambda$, which rescales 
$b\rightarrow \lambda b$, $a\rightarrow \lambda a$ and $R\rightarrow \lambda R$. 
So we can produce in our simulation sets of 
points for arbitrary values of $b$, $a$ and $\nu$ which follow a distribution $C(R)$. If we then 
multiply all the points with an arbitrary scaling constant $\lambda$, the resulting $C'$ will be
\begin{equation}\label{eq:rescale}
C'(\lambda R)=C(R)
\end{equation}

\subsection{The walk}

Having at hand different step probabilities we can complete the walk.
We will use two ways of producing
the visiting points. In the first we will take steps with a probability of the step length $r$ 
according to (\ref{eq:st_pr}), beginning from last point visited. 
Then, the direction of the step will be
decided randomly, so that the direction of each step is isotropic in the embedding space.
In $d=1$ space we will randomly move forwards or backwards. 
In $d=2$ space we will randomly choose an angle $\phi$ in the interval $[0,2\pi)$. In $d=3$ space
additionally to the angle $\phi$ we will randomly choose an angle $\theta$ in the interval 
$[0,\pi)$. This way of generation of the walk will be denoted as $\nu-\phi$ and the
displacement vector of the produced $i$-th step, $\Delta\vec{r}_i$, will be
\begin{align*}
d=1:& \;\;\;\Delta\vec{r}_i=\left ( ar \right ) \;, a=\pm 1 \\
d=2:& \;\;\;\Delta\vec{r}_i=\left ( r\cos(\phi),r\sin(\phi)  \right ) \;, \phi \in [0,2\pi) \\
d=3:& \;\;\;\Delta\vec{r}_i=\left ( r\cos(\phi)\sin(\theta),r\sin(\phi)\sin(\theta),r\cos(\theta) \right ) \;, \phi \in [0,2\pi) \;, \theta \in [0,\pi)
\end{align*}
In the second way we will produce independent one-dimensional walks in every one of the $d$
available dimensions of our embedding space in the same manner as in the $\nu-\phi$ case with $d=$1.
The $d$ produced numbers will be the co-ordinates of the visited point.
We will call this way of generation
as $\nu-\nu$ and the vector of the produced $i$-th step, $\Delta\vec{r}_i$, will be
\begin{align*}
d=1:& \;\;\;\Delta\vec{r}_i=\left ( a_1 r_1 \right ) \;, a_1=\pm 1 \\
d=2:& \;\;\;\Delta\vec{r}_i=\left ( a_1 r_1,a_2 r_2  \right ) \;, a_1=\pm 1, a_2=\pm 1 \\
d=3:& \;\;\;\Delta\vec{r}_i=\left ( a_1 r_1,a_2 r_2,a_3 r_3 \right ) \;, a_1=\pm 1, a_2=\pm 1,
a_3=\pm 1
\end{align*}
In both cases, the reached point after the $i$-th step will be produced by adding the vector step to the point
reached by the $i-1$-th step (Markovian procedure)
\begin{equation}\label{eq:walk}
\vec{r}_i=\vec{r}_{i-1}+\Delta\vec{r}_i
\end{equation}

\subsection{Calculating correlation dimension}

For the extraction of $d_F$ we will use a ``local'' slope of $\ln C$ as function of the logarithm of
the scale $\ln R$. This slope is formed around a specific point of $C$ which corresponds to a 
specific $R_i$ and it is linked to it. To calculate it, though, we use $i_m$ points with greater scale
and $i_m$ points with lower scale than $R_i$. Then taking these $2 i_m +1$ points of $\ln C$ we
find the corresponding least squares straight line. The slope of this line is the $d_F$ which we
correspond to $R_i$. Our aim is to smooth out statistical fluctuations, without erasing the 
functional dependence of $C$ on $R$. For this reason the number of $i_m$ has to be chosen
appropriately in each case. A linear part of $\ln C$ in a scale interval of $\ln R$ should have a
constant $d_F$ in the interval. In our method this is
the case in this interval apart from the $i_m$ points
at the edges of the interval. At the edges we find a
$d_F$ which smoothly changes values.
So our calculations of $d_F$ are mostly trustworthy at linear parts of
$C$ in the logarithmic plot.

\section{Distributions for different Multiplicities} \label{sec:dif_mul}

In this section we shall see how the number of steps affects the $C$ distribution in the case of
negative and positive $\nu$.
We begin by investigating the shape of correlation integral, $C$, in a walk of $N_m$ L\'{e}vy
steps, when the exponent $\nu$ in (\ref{eq:st_pr}) is negative. In this case, there is no
need for an upper limit and we shall not set such a limit at the moment. The necessary lower
limit is set at $b$.
Due to scaling invariance, we can produce in our simulations sets of 
points for an arbitrary value of $b$, e.g. $b=1$. Then our plots are valid for all cases of $b$ if we
plot them as function of $R/b$. 

In Fig.~\ref{fig:nNm}(i) we depict walks with varying $N_m$, called multiplicity and sufficient
number of events, $N_e$, for each case so that to suppress statistical uncertainties sufficiently.
The exponent $\nu$ remains fixed, here at the value of -1/3.
We show results for three embedding spaces, $d=$1, 2, 3 and two ways of producing the walks,
$\nu-\phi$ and $\nu-\nu$.

As it is evident from (\ref{eq:nn_pdf}) no step is allowed to be lower than $b$. Thus, when
$N_m=2$ and in the $\nu-\phi$ cases, two points of the set cannot have distance less $b$.
In the $\nu-\nu$ cases, two points of the set cannot have distance less than $b\sqrt{d}$,
since the projection of this distance in each dimension cannot be less than $b$. 
To have a unified description, between the $\nu-\phi$ and $\nu-\nu$ cases, as far the lowest
available scale of $C$ for $N_m=2$ is concerned, we define
\begin{equation}\label{eq:b_ef}
b_{ef}=b, \;{\rm for}\; \nu-\phi\;\;\;\;\;
b_{ef}=b\sqrt{d}, \;{\rm for}\; \nu-\nu
\end{equation}
So the $C$ distribution for $N_m=2$ tends to zero for $R\rightarrow b_{ef}$.
For $N_m>2$, it is possible for
two points to have distance less than $b_{ef}$, due to the random changes in the orientation of
the steps. However, the $C$ function retains a rapid decrease at scales of the order of $b_{ef}$
for $N_m>2$.
At scales less than $b_{ef}$ and for $N_m>2$ the embedding space effect appears and the
relevant slope tends to $d$.
In Fig.~\ref{fig:nNm} we set $b$ so that $b_{ef}$ to be the same for $\nu-\phi$ and $\nu-\nu$ cases.
So if $b_1$ is the lower step limit for $\nu-\phi$ cases then we set this limit to be 
$b_2=b_1/\sqrt{d}$ for $\nu-\nu$ cases.

We see that at low $N_m$ it is not possible to obtain a part of the $C$ function which is almost
linear in the logarithmic scale. Such a part with slope equal to $\left|\nu\right|$ starts to appear for a very large number of steps
and it is increasing with the further increase of $N_m$. This is the case in \cite{Alemany1994}, where 
$N_m=10^4$.
The linear part of $C$ forms at intermediate scales. At scales approaching $b_{ef}$ and 
maximum available values
$C$ departs from linear behaviour.
This linear part of correlation integral in the $\nu-\phi$
case and for all $d$ is approximated by the line:
\begin{equation}\label{eq:nn_ap_nuph}
C\left( R \right) \approx \frac{2}{N_m} \left( \frac{R}{b} \right)^{|\nu|}
\end{equation}
In the $\nu-\nu$ case this line becomes:

\begin{figure}[H]
\centering
\vspace{-0.7cm}
\hspace{-2.5cm}(i)\hspace{2.5cm}
\includegraphics[scale=0.58,trim=2.in 0.3in 2.in 0.2in,angle=0]{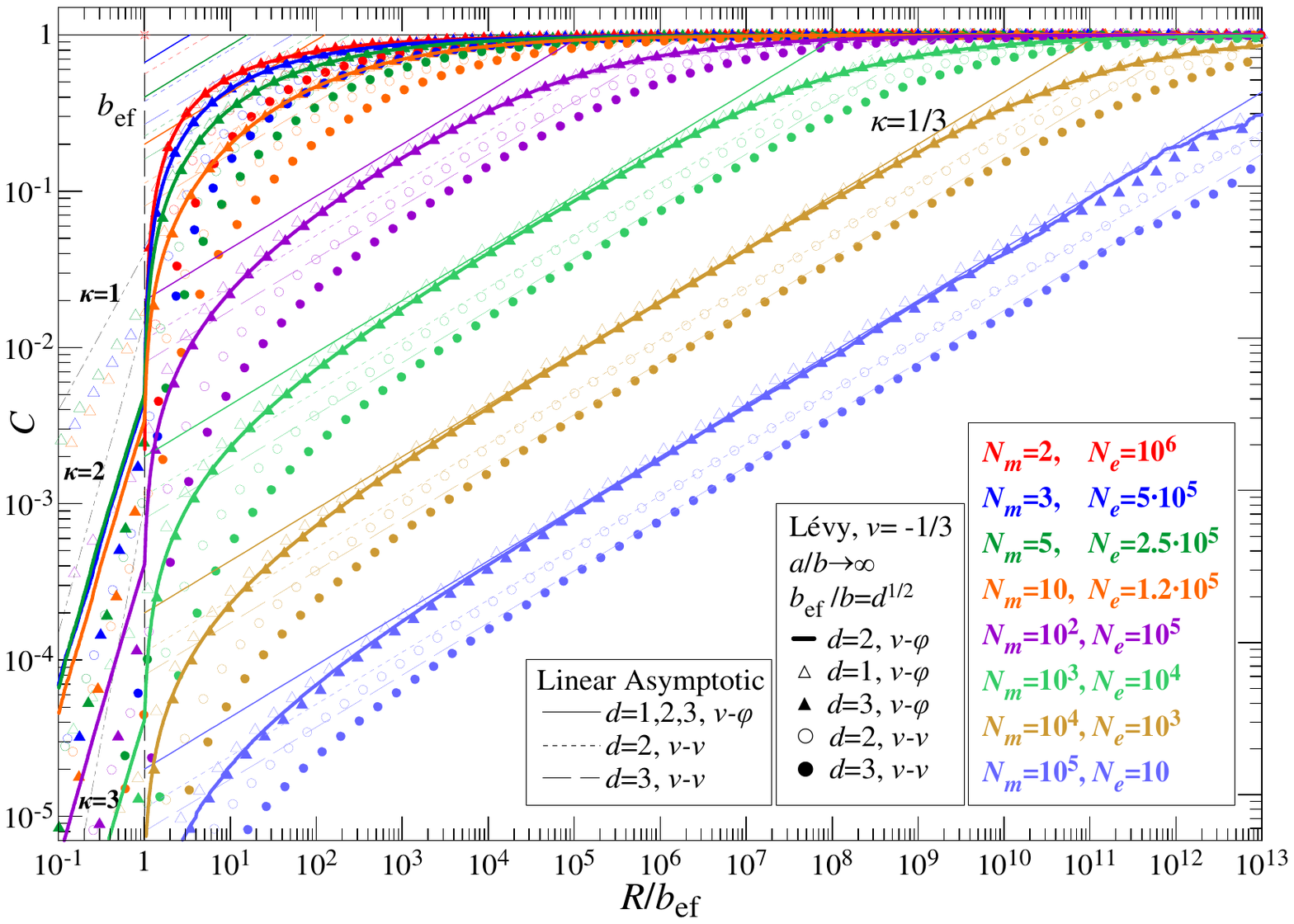}\\
\hspace{-2.5cm}(ii)\hspace{2.5cm}
\includegraphics[scale=0.58,trim=2.in 0.3in 2.in 0.2in,angle=0]{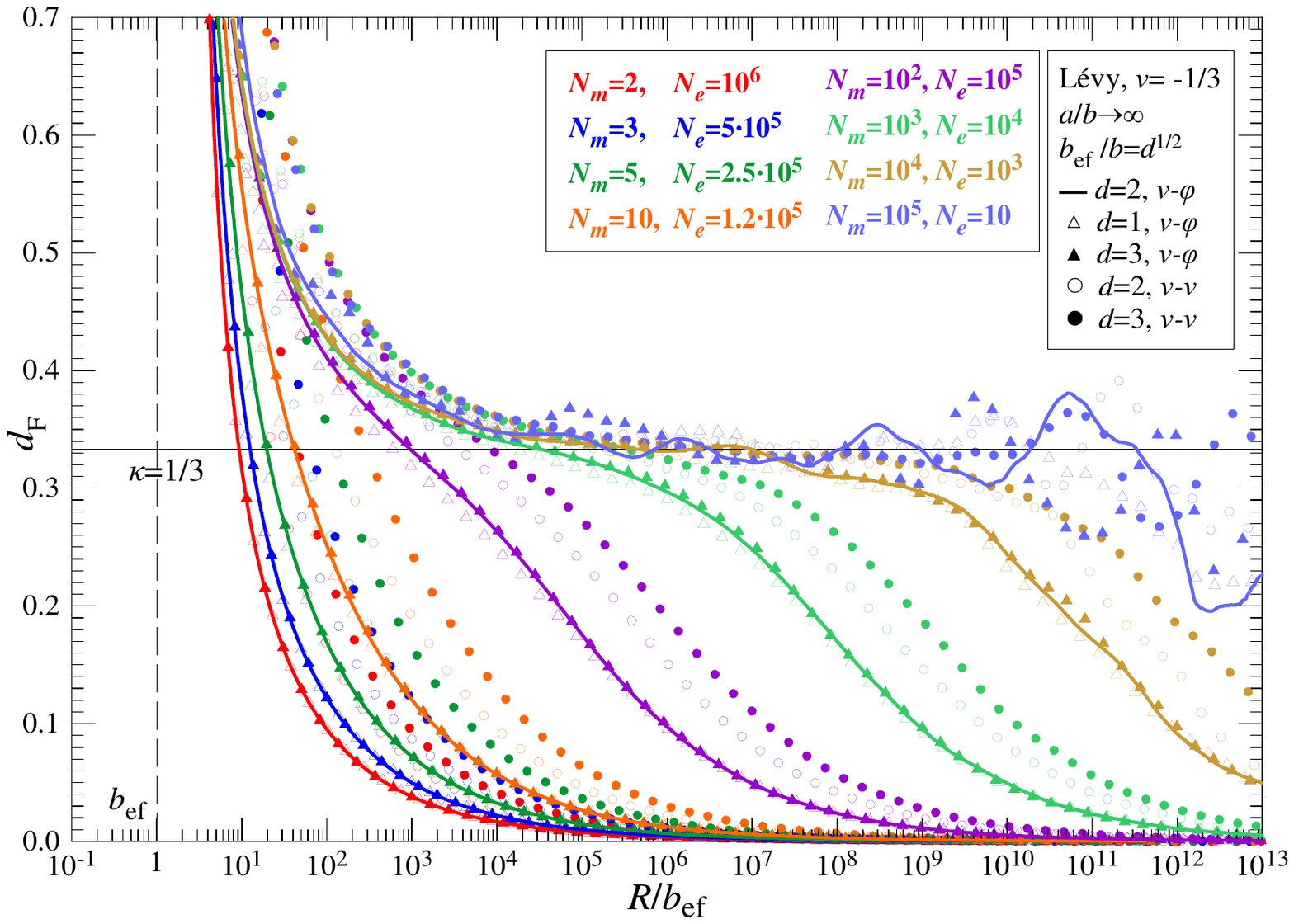}
\caption{\label{fig:nNm} {\small L\'{e}vy walks with varying
multiplicities $N_m$ for fixed step probability exponent $\nu=-1/3<0$. The walks are carried out in
$d=$1,2 and 3 embedding spaces. We show the $\nu-\phi$ and $\nu-\nu$ cases 
(coinciding for $d=1$). No upper limit $a$ is imposed in the steps. 
Lower limit $b_{ef}$ is the same in $\nu-\phi$ and $\nu-\nu$ cases.
With thin lines we present the straight lines which are approached by the distributions
in each case (shrinked to a point for $N_m=2$).
 (i) The correlation integral $C$ of each set as function of the scale $R/b_{ef}$
in log-log plot. (ii) The local slopes $d_F$ of the sets presented in (i).}}
\end{figure}

\begin{figure}[H]
\centering
\vspace{-0.7cm}
\hspace{-2.5cm}(i)\hspace{2.5cm}
\includegraphics[scale=0.58,trim=2.in 0.3in 2.in 0.2in,angle=0]{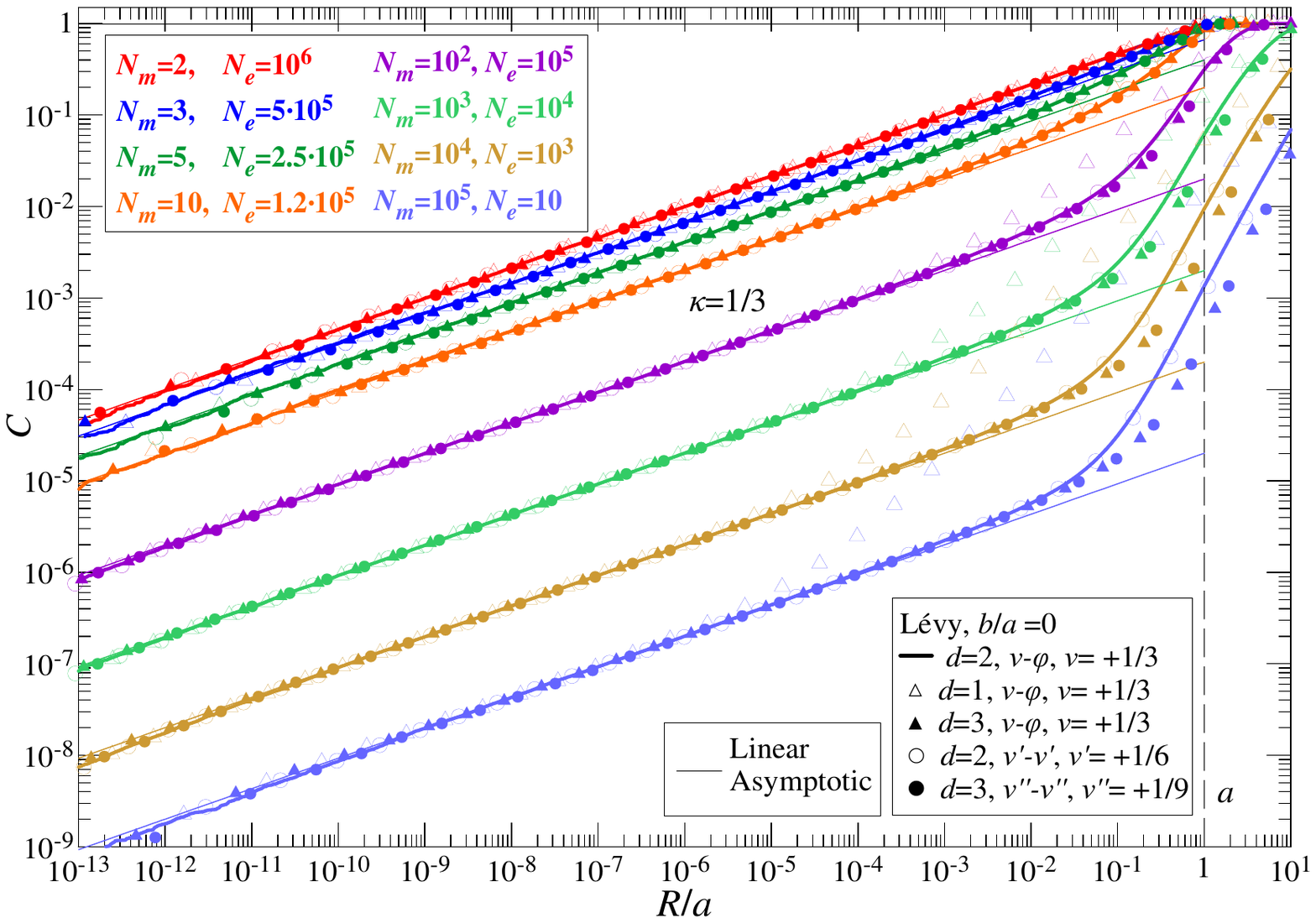}\\
\hspace{-2.5cm}(ii)\hspace{2.5cm}
\includegraphics[scale=0.58,trim=2.in 0.3in 2.in 0.2in,angle=0]{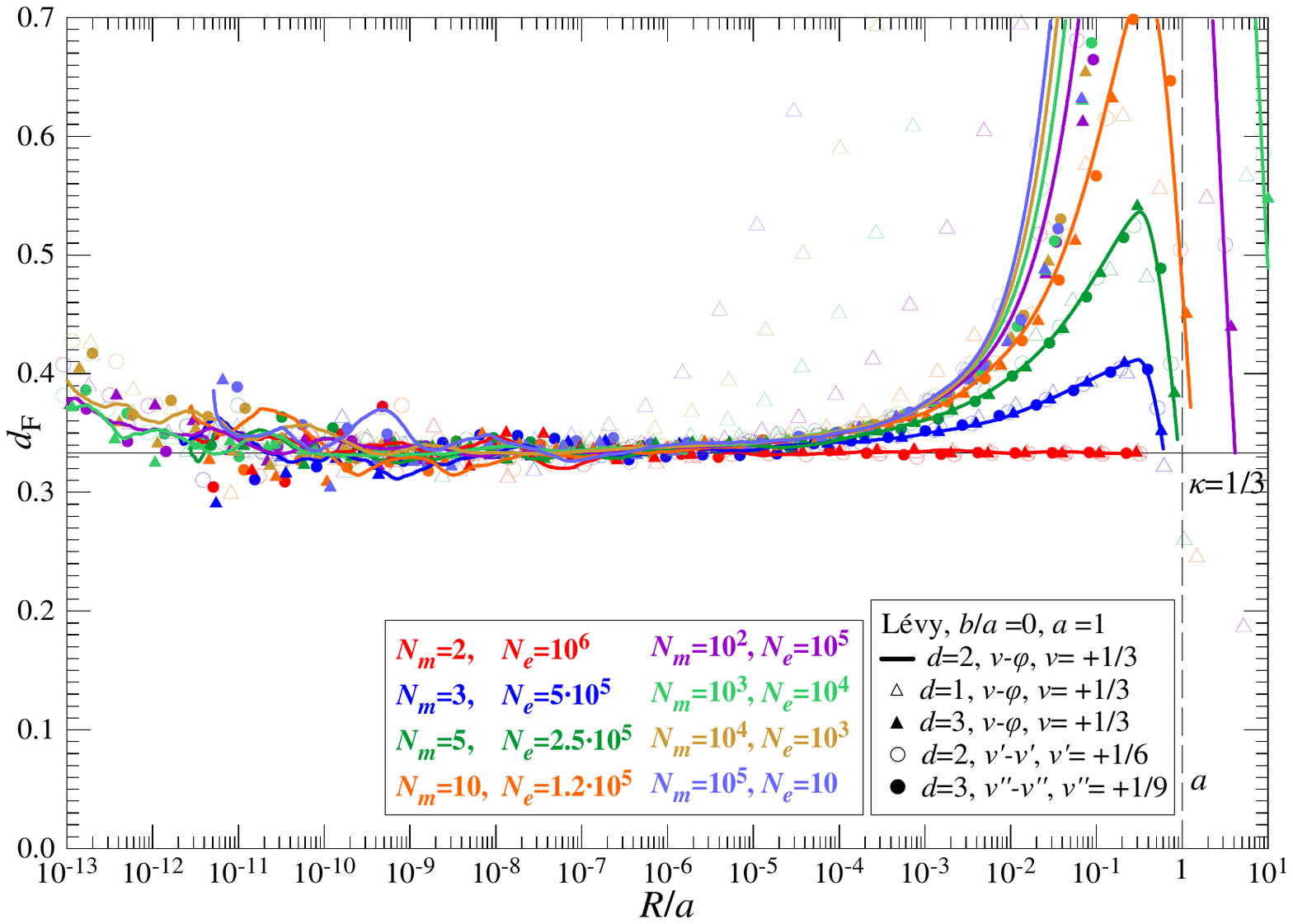}
\caption{\label{fig:pNm} {\small L\'{e}vy walks with varying
multiplicities $N_m$ for fixed step probability exponent $\nu=+1/3>0$. The walks are carried out in
$d=$1,2 and 3 embedding spaces. We show the $\nu-\phi$ and $\nu-\nu$ cases 
(coinciding for $d=1$). No lower limit $b$ is imposed in the steps. 
Upper limit is $a$ in all cases.
With thin lines we present the straight lines which are approached by the distributions
in each case. In the $\nu'-\nu'$ cases, $\nu'=\nu/d$.
 (i) The correlation integral $C$ of each set as function of the scale $R/a$
in log-log plot. (ii) The local slopes $d_F$ of the sets presented in (i).}}
\end{figure}

$\;$\vspace{-0.5cm}
\begin{equation}\label{eq:nn_ap_nunu}
C\left( R \right) \approx \frac{\left(2/d\right)}{N_m} \left( \frac{R}{b} \right)^{|\nu|}
\end{equation}

In Fig.~\ref{fig:nNm}(ii) we depict the corresponding $d_F$. We observe that at large
multiplicities, $N_m>10^4$, for this specific value of $\nu$, $d_F$ starts to attain a value equal to
 $|\nu|$ in some scale interval.
As we can see a very large multiplicity is needed to form a linear part of $C$ when $\nu<0$.
However, if we want to simulate events with low $N_m$, 
then we can use the results of section 
\ref{sec:alter_Nm}, produce a walk with very large number of steps and then form random sets from
it with specific number of $N_m$.
Finally we observe that at scales lower than $b_{ef}$ the slope tends to the embedding space 
dimension $d$.

We next investigate the shape of $C$ in walks of $N_m$ L\'{e}vy
steps, when the exponent $\nu$ in (\ref{eq:st_pr}) is positive. In this case, there is no
need for a lower limit. The upper
limit, necessary for the step probability to be integrable, is set to $a$.
The distributions are presented as function of $R/a$, which suggests that for any $a'$ we can have
the same value of $C$ at a certain scale, $R'$, so that $R'/a'=R/a$. 

In Fig.~\ref{fig:pNm}(i) we depict walks with varying $N_m$ and
exponent $\nu$ fixed at the value of +1/3.
We show results for three embedding spaces dimensions, $d=$1,2,3 and two ways of producing the walks,
$\nu-\phi$ and $\nu-\nu$.
As it is evident from (\ref{eq:pn_pdf}) no step is allowed to be greater than $a$. Thus, when
$N_m=2$ and in the $\nu-\phi$ cases, two points of the set cannot have distance greater than $a$.
Thus, for this case $C(a)=1$.
In the $\nu-\nu$ cases, two points of the set cannot have distance greater than $a\sqrt{d}$,
however the correlation integral has almost acquired the unit value at scale $R=a$.
So all the distributions exhibit an alteration of their behaviour as they approach at scale $a$.
The scale at which $C$ approaches unit increases with $N_m$, since the greater distance in this
case is of the order of $\sim N_m a$.

Below the scale $a$ all the distributions start to form a linear part in the log-log plot,
which extends nearly limitless\footnote{Machine accuracy of the used computer certainly sets limit to
these calculations.} to infinitesimal scales (since we are working in logarithmic scale). The slope of this 
linear part equals $\nu$.
In contrast to the case where $\nu<0$, here the linear part of $C$ forms immediately after $R=a$
when $N_m=2$. This attribute makes the $\nu>0$ case ideal for forming simulations with low 
multiplicities.
We observe that when $d=1$ the $C$ distribution acquires the $\nu$ slope at lower 
scales, compared to cases of $d=2$ and $d=3$.
Also, the $d=3$ distribution acquires the relative slope at higher scales compared to
the $d=2$ case. These effects become more prominent as $N_m$ increases.

The linear part of correlation integral is approximated, for both the $\nu-\nu$ and $\nu-\phi$
cases, by
\begin{equation}\label{eq:pn_ap}
C\left( R \right) \approx \frac{2}{N_m} \left( \frac{R}{a} \right)^{\nu}\;.
\end{equation}

Lastly, we observe that, when $\nu>0$, in order to form in embedding space of dimension $d$ 
a walk with
dimension $\nu$ from
one-dimensional walks ($\nu'-\nu'$ case), the step exponents of these walks should be $\nu'=\nu/d$. 
This is contrasted to the case with $\nu<0$, where the step exponents of the one-dimensional walks,
$\nu'$
have the same absolute value with the fractal dimension, $|\nu|$, of the produced $d$-dimensional walk,
$|\nu'|=|\nu|$.
This fact may be attributed to the following situation.
When $\nu<0$ some steps may receive extremely high values. As a result the part of the
$d$-space where the walk takes place is not covered sufficiently to produce a rise in the 
dimensionality
of the set. On the contrary, when $\nu>0$ the walk occurs within a part of $d$-space bounded
by limits of the order of $a$. This results to local sufficient covering of this part of the space.

\vspace{-0.0 cm}
\section{Distributions for different step exponents $\nu$} \label{sec:dif_nu}

In this section we will depict the $C$ distributions for walks with varying exponents $\nu$
in the step probabilities, for a fixed number of steps.
We show results for embedding
spaces $d=$1, 2 and 3, as well as, for the $\nu-\phi$ and $\nu-\nu$ cases.

\begin{figure}[H]
\centering
\vspace{-0.7cm}
\hspace{-2.5cm}(i)\hspace{2.5cm}
\includegraphics[scale=0.57,trim=2.in 0.3in 2.in 0.2in,angle=0]{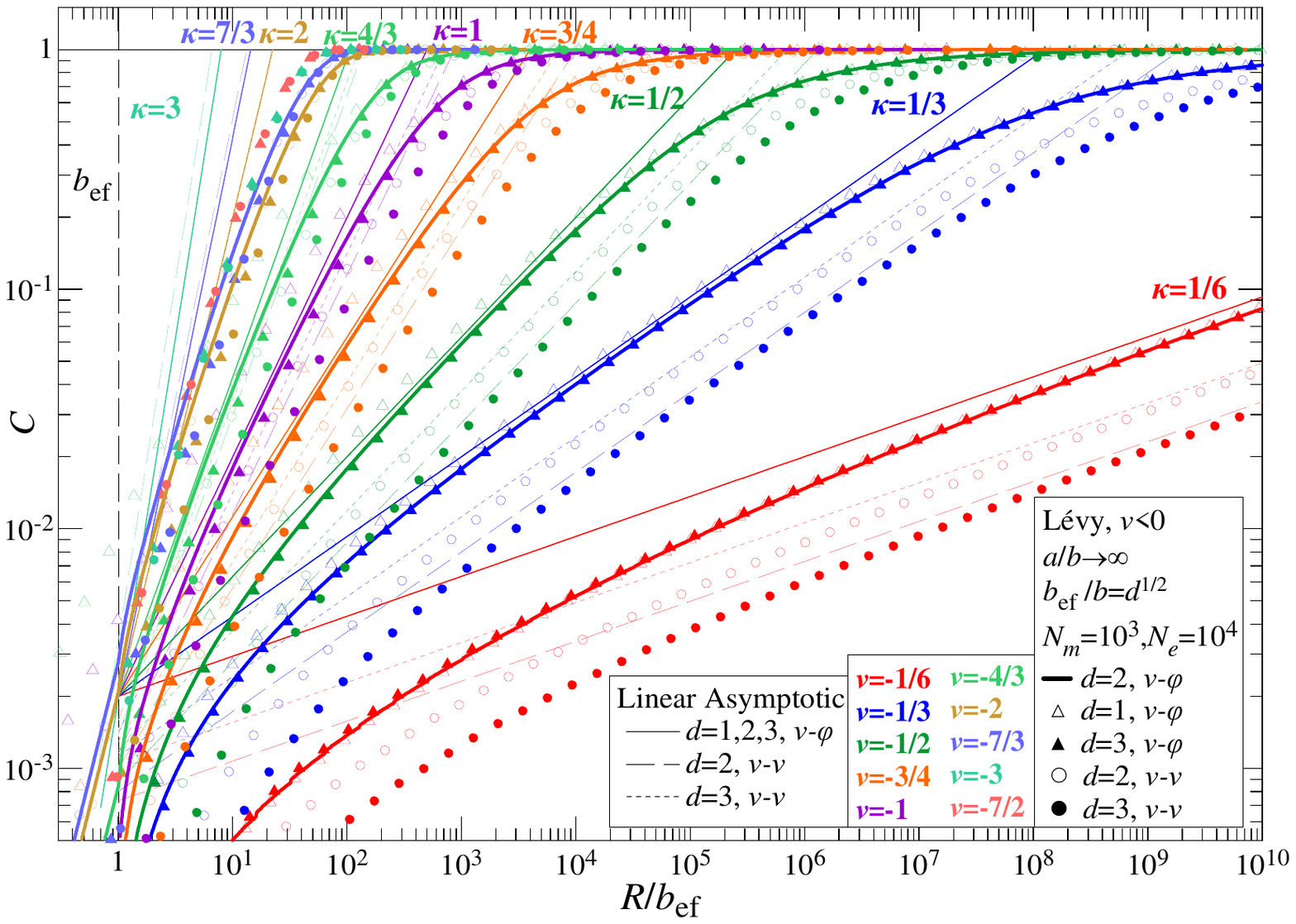}\\
\includegraphics[scale=0.57,trim=0.5in 0.3in 6.cm 0.2in,angle=0]{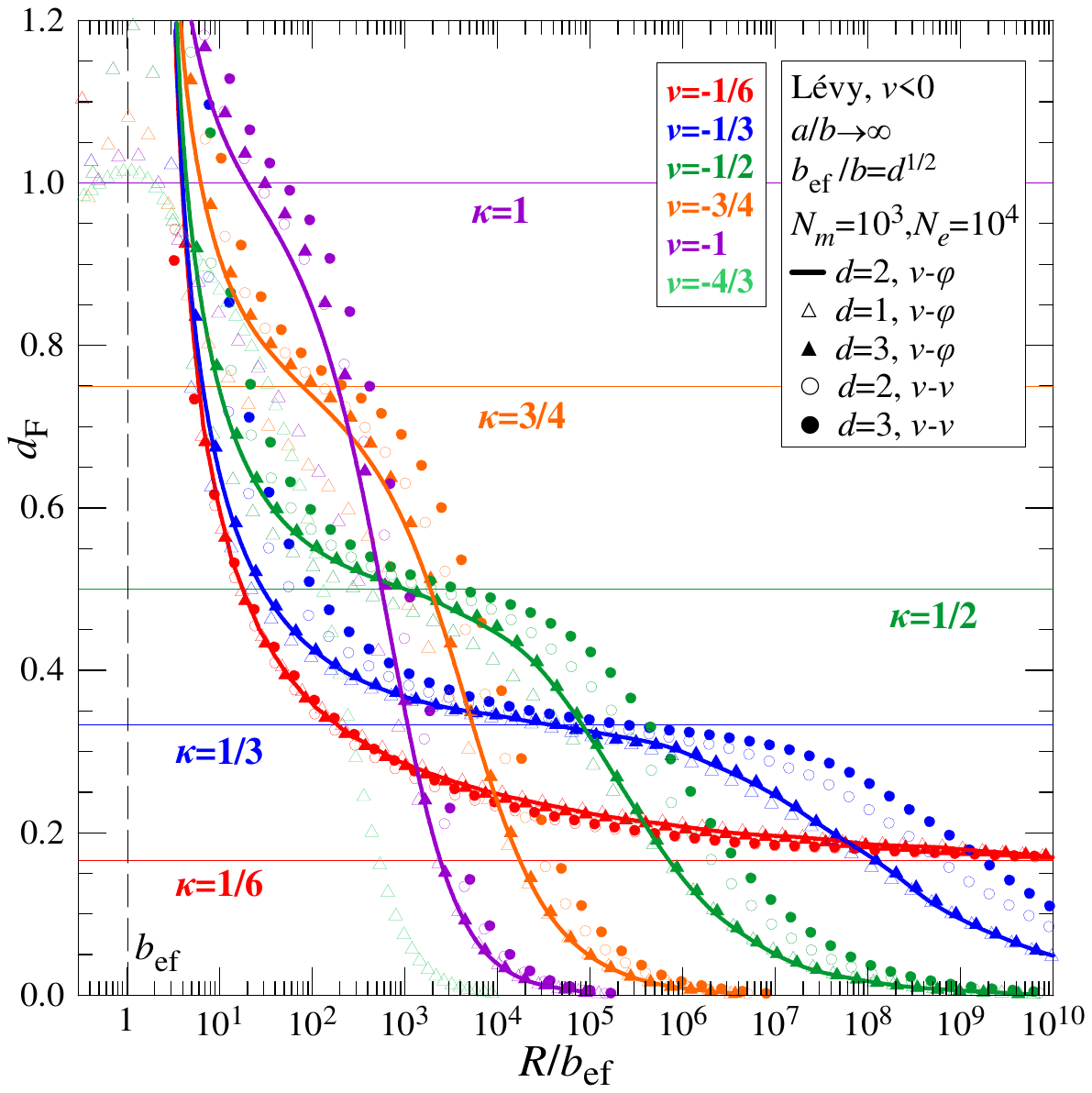}
\includegraphics[scale=0.57,trim=1.7in 0.3in 19.5cm 0.2in,angle=0]{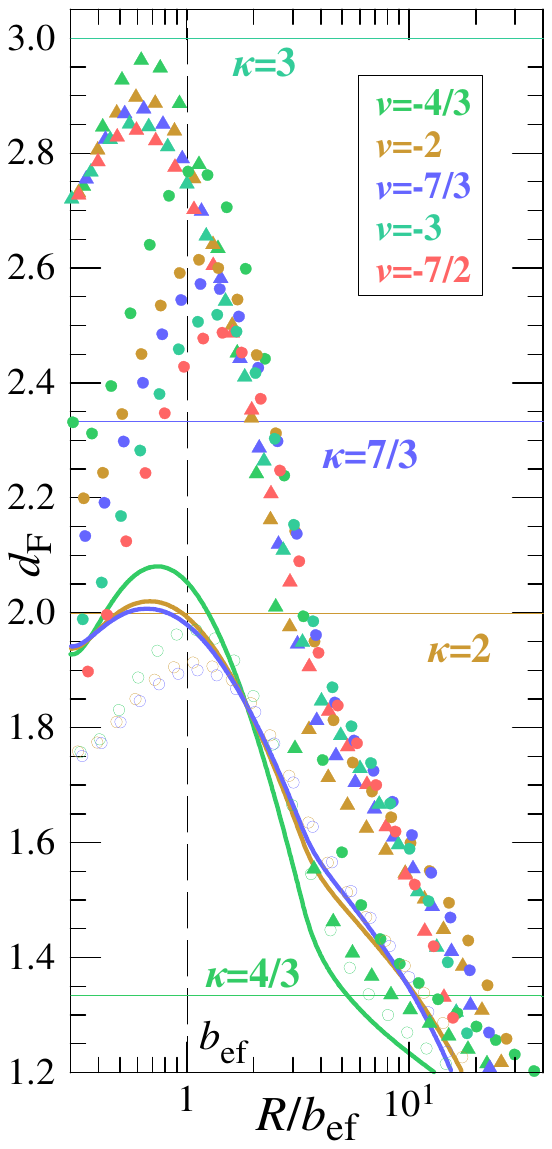}\\
\vspace{-0.4cm}(ii)\hspace{17.cm}\vspace{-0.2cm}
\caption{\label{fig:nnu} {\small L\'{e}vy walks with varying
step probability exponents $\nu<0$ for fixed multiplicity $N_m=10^3$ and number of events
$N_m=10^4$. The walks are carried out in
$d=$1,2 and 3 dimensions of embedding spaces. We show the $\nu-\phi$ and $\nu-\nu$ cases 
(coinciding for $d=1$). No upper limit $a$ is imposed in the steps. 
Lower limit is $b$ in $\nu-\phi$ and $b/\sqrt{d}$ in $\nu-\nu$ cases.
With thin lines we present the straight lines which are approached by the distributions
in each case.
 (i) The correlation integral $C$ of each set as function of the scale $R/b_{ef}$
in log-log plot. (ii) The local slopes $d_F$ of the sets presented in (i). Lower absolute
values of $\nu$ in left part and higher absolute values of $\nu$ in right part of the graph.}}
\end{figure}

\begin{figure}[H]
\centering
\vspace{-0.7cm}
\hspace{-2.5cm}(i)\hspace{2.5cm}
\includegraphics[scale=0.58,trim=2.in 0.3in 2.in 0.2in,angle=0]{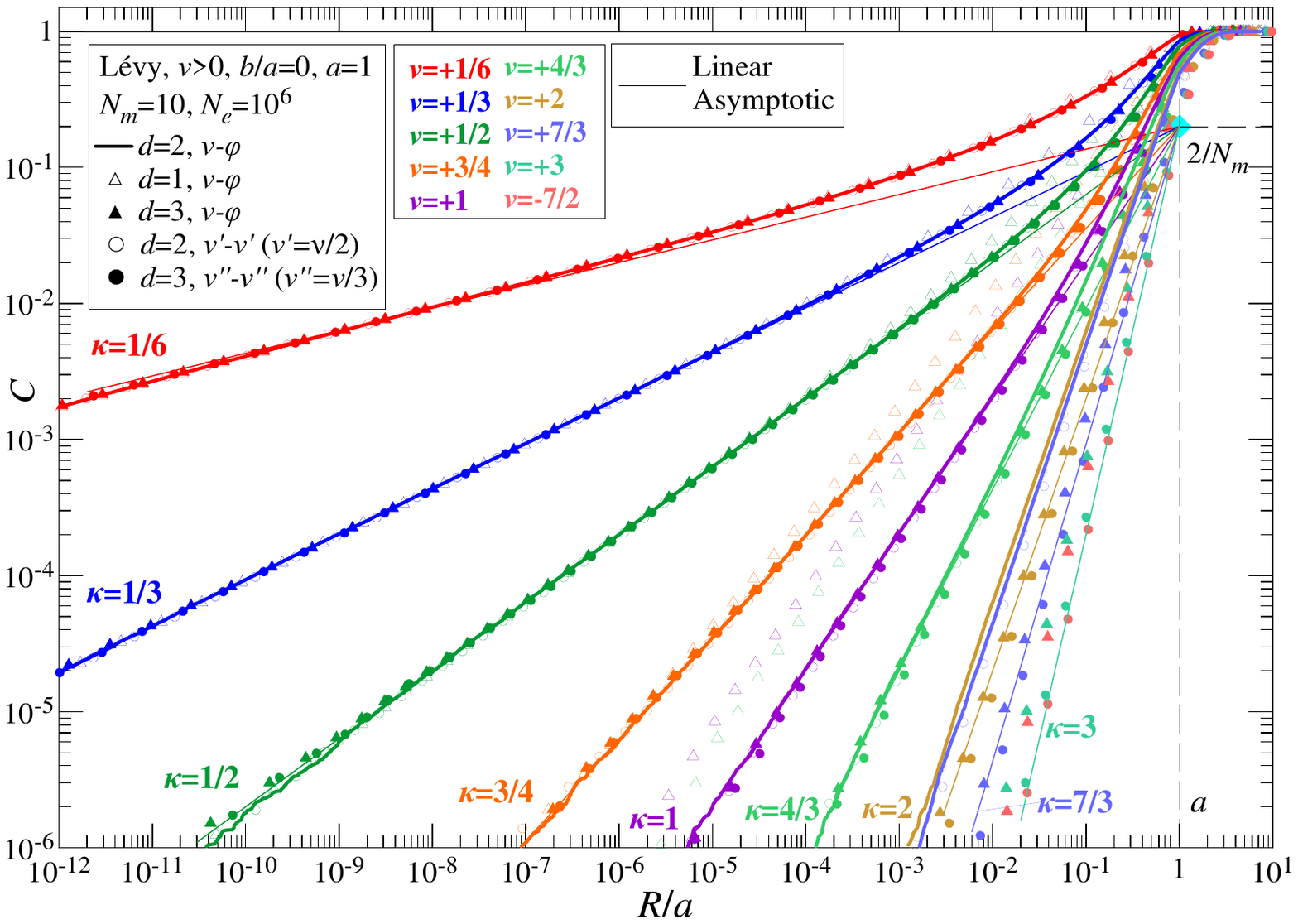}\\
\hspace{-2.5cm}(ii)\hspace{2.5cm}
\includegraphics[scale=0.58,trim=2.in 0.3in 2.in 0.2in,angle=0]{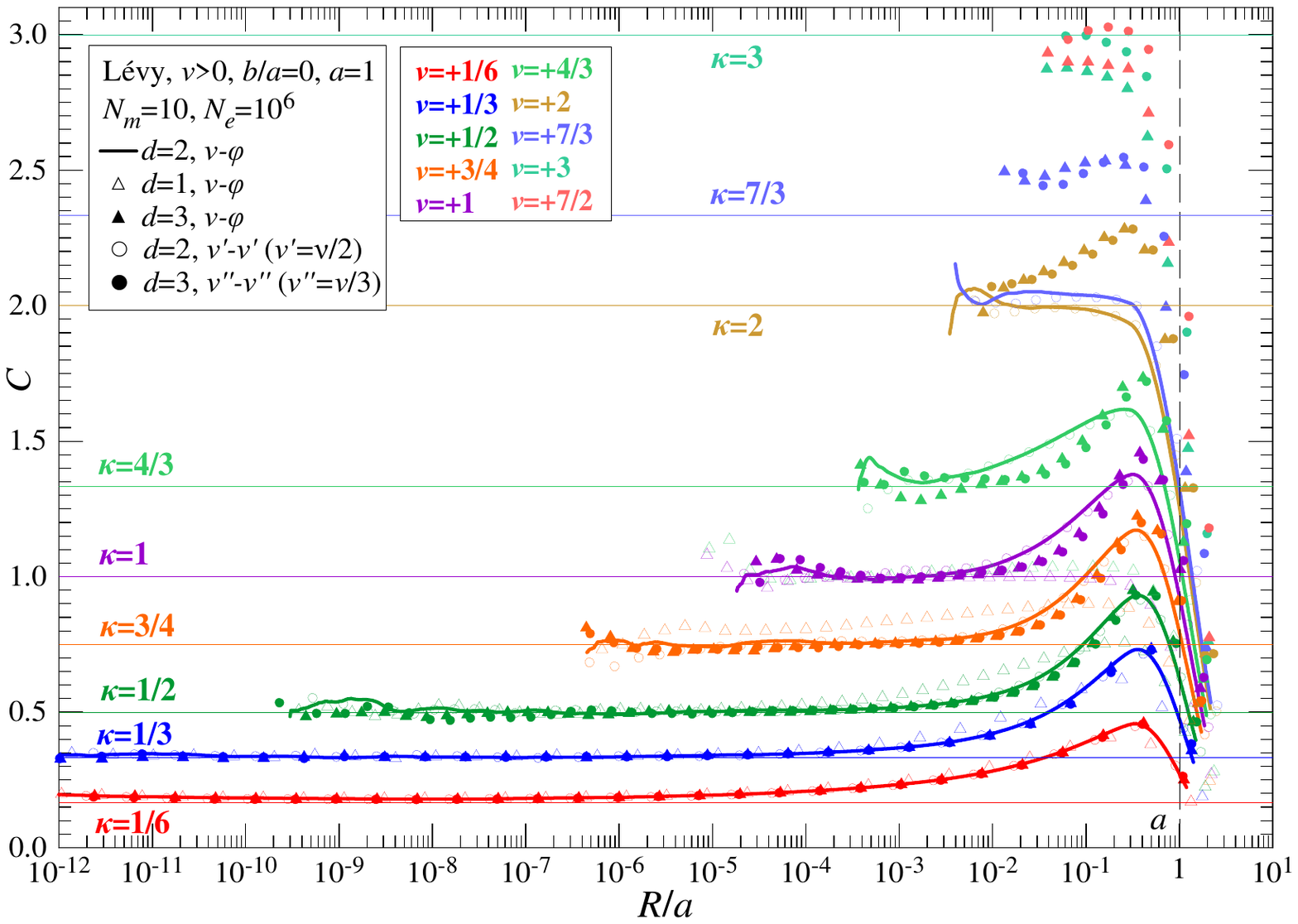}
\caption{\label{fig:pnu} {\small L\'{e}vy walks with varying
step probability exponents $\nu>0$ for fixed multiplicity $N_m=10$ and number of events
$N_m=10^6$. The walks are carried out in
$d=$1,2 and 3 dimensions of embedding spaces. We show the $\nu-\phi$ and $\nu-\nu$ cases 
(coinciding for $d=1$). No lower limit $b$ is imposed in the steps. 
Higher limit is $a$ in $\nu-\phi$ and $\nu-\nu$ cases.
With thin lines we present the straight lines which are approached by the distributions
in each case.
 (i) The correlation integral $C$ of each set as function of the scale $R/a$
in log-log plot. (ii) The local slopes $d_F$ of the sets presented in (i).}}
\end{figure}

We begin with negative $\nu$. In Fig.~\ref{fig:nnu} we depict walks for varying $\nu<0$ and
multiplicity $N_m=10^3$ for sufficient number of events $N_e=10^4$.
Again here, as in the previous section, we have adjusted $b$ so that the $\nu-\phi$ and $\nu-\nu$
cases have the same $b_{ef}$. We proceed with positive $\nu$. In Fig.~\ref{fig:pnu} 
we depict walks for varying $\nu>0$, multiplicity $N_m=10$ and number of events $N_e=10^6$.

Our observations for both signs of $\nu$ are the following:
 a) For negative $\nu$, the $C$ distributions acquire the expected slope $|\nu|$
at medium scales, if $|\nu|<d$. The increase of number of steps, in this case, is expected to 
improve the situation. For positive $\nu$, the $C$ distributions acquire the expected slope $\nu$
at low scales without bound, if $\nu<d$.
 b) As $|\nu|$ approaches $d$, the linear part of $d$ shrinks,
 c) When $|\nu|$ exceeds $d$, then the relevant slope approaches or has as an upper limit $d$, instead of $|\nu|$,
in accordance to the fractal projection theorem (e.g. $\nu=\pm 4/3$ for $d=1$,
$\nu=\pm 7/3$ for $d=2$ and $\nu=\pm 7/2$ for $d=3$).

We end with some remarks on the shape of the $C$ distributions depicted in the current and in the
previous section. 
The negative $\nu$ walks exhibit an abrupt change for scales below $b_{ef}$, related to the 
lower limit. They begin to tend to a slop equal to the dimension of the embedding space. 
For positive $\nu$ the $C$ distributions tend to unit for scales above the upper limit $a$ in a continuous manner.
Exception to both cases are the $N_m=2$ distributions. For $\nu<0$ the $C$ distribution ends
at $b_{ef}$. For $\nu>0$ and for the $\nu-\phi$ cases the $C$ distributions are simply a perfect
straight line (in log-log plot) reaching unit at scale $a$.
Lastly, for negative $\nu$, if we exclude the scales below $b_{ef}$, the $C$ distributions are concave 
and linear. For positive $\nu$, if we exclude the scales near and above $a$, the $C$ distributions are 
convex and linear.

\section{Setting additional limits} \label{sec:ad_lim}

The step probabilities for negative and positive $\nu$ cases require a lower or higher limit, 
respectively. 
Additional limits, though, can be set in each case, if needed.
Thus, a higher limit $a$ can be set for $\nu<0$ (eq.~(\ref{eq:nn_pdfa})) or a lower limit $b$ 
for $\nu>0$ (eq.~(\ref{eq:pn_pdfb})).

\begin{figure}[H]
\centering
\vspace{-0.0cm}
\hspace{0.4cm}\includegraphics[scale=0.3,trim=0.4in 0.3in 0.7in 0.2in,angle=0]{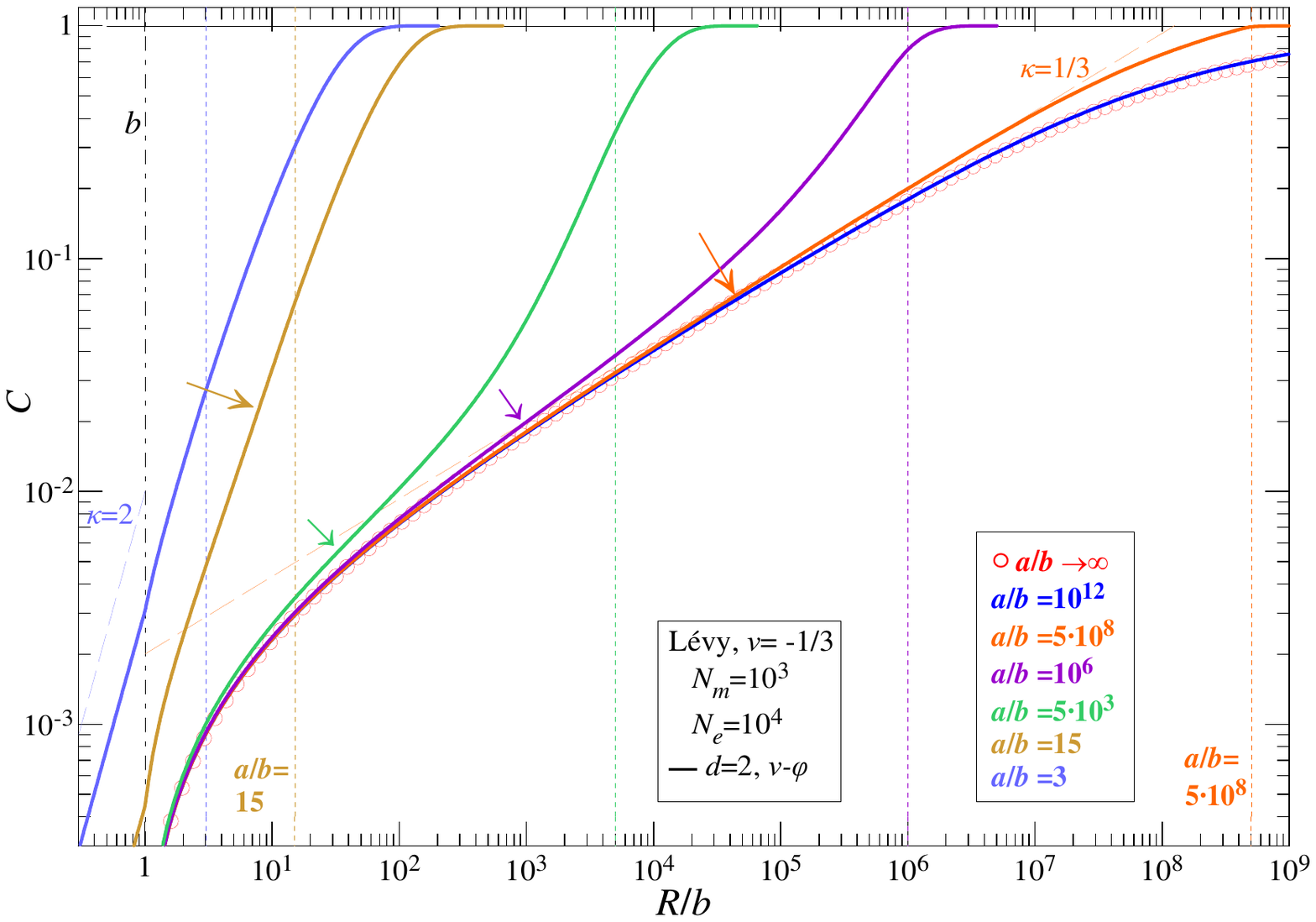}
\hspace{0.8cm}\includegraphics[scale=0.3,trim=0.4in 0.3in 0.7in 0.2in,angle=0]{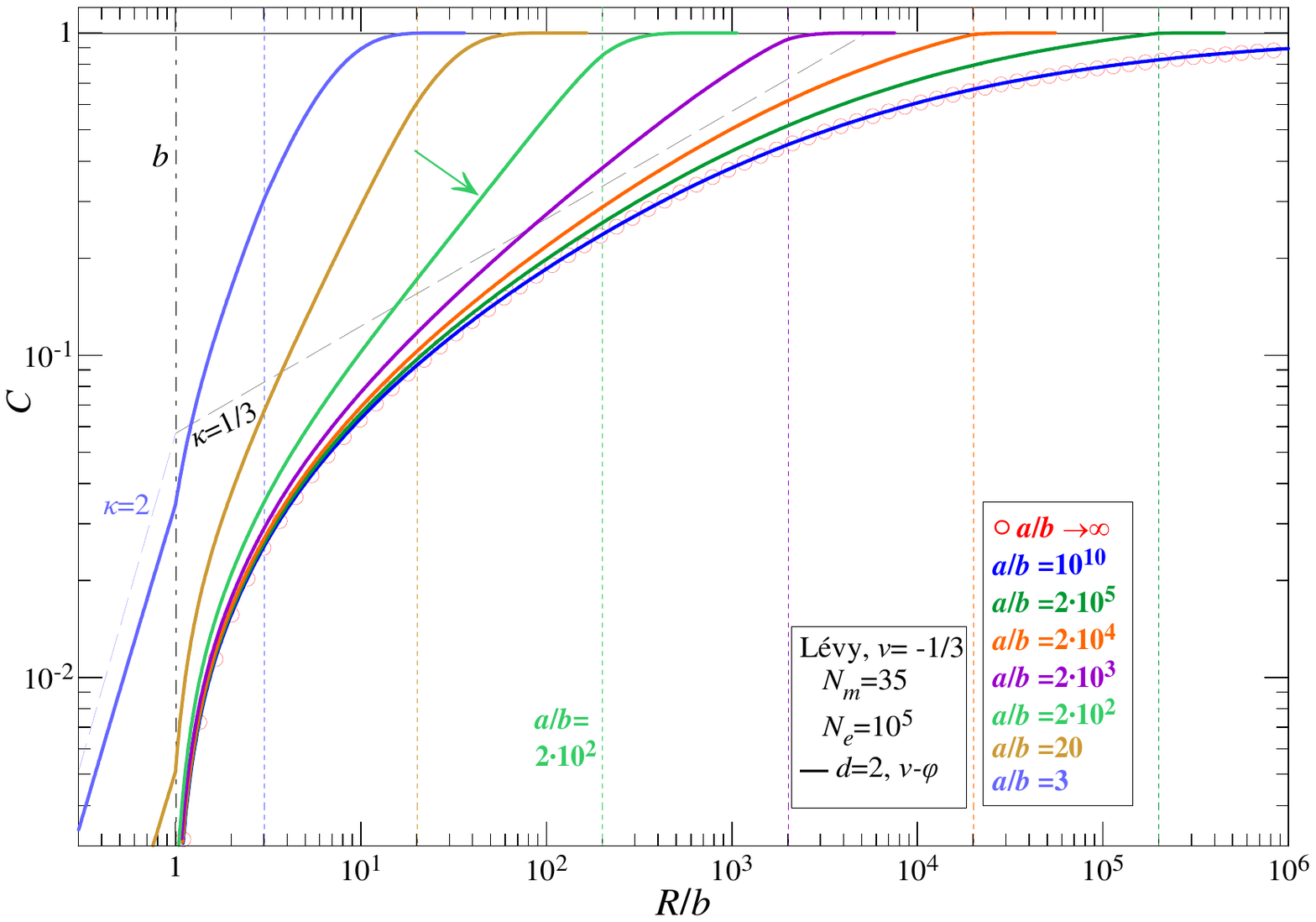}\\
\vspace{-0.4cm}(Ai)\hspace{8.cm}   (Bi) \hspace{7.6cm} $\;$ \\
\hspace{0.4cm}\includegraphics[scale=0.3,trim=0.4in 0.3in 0.7in 0.2in,angle=0]{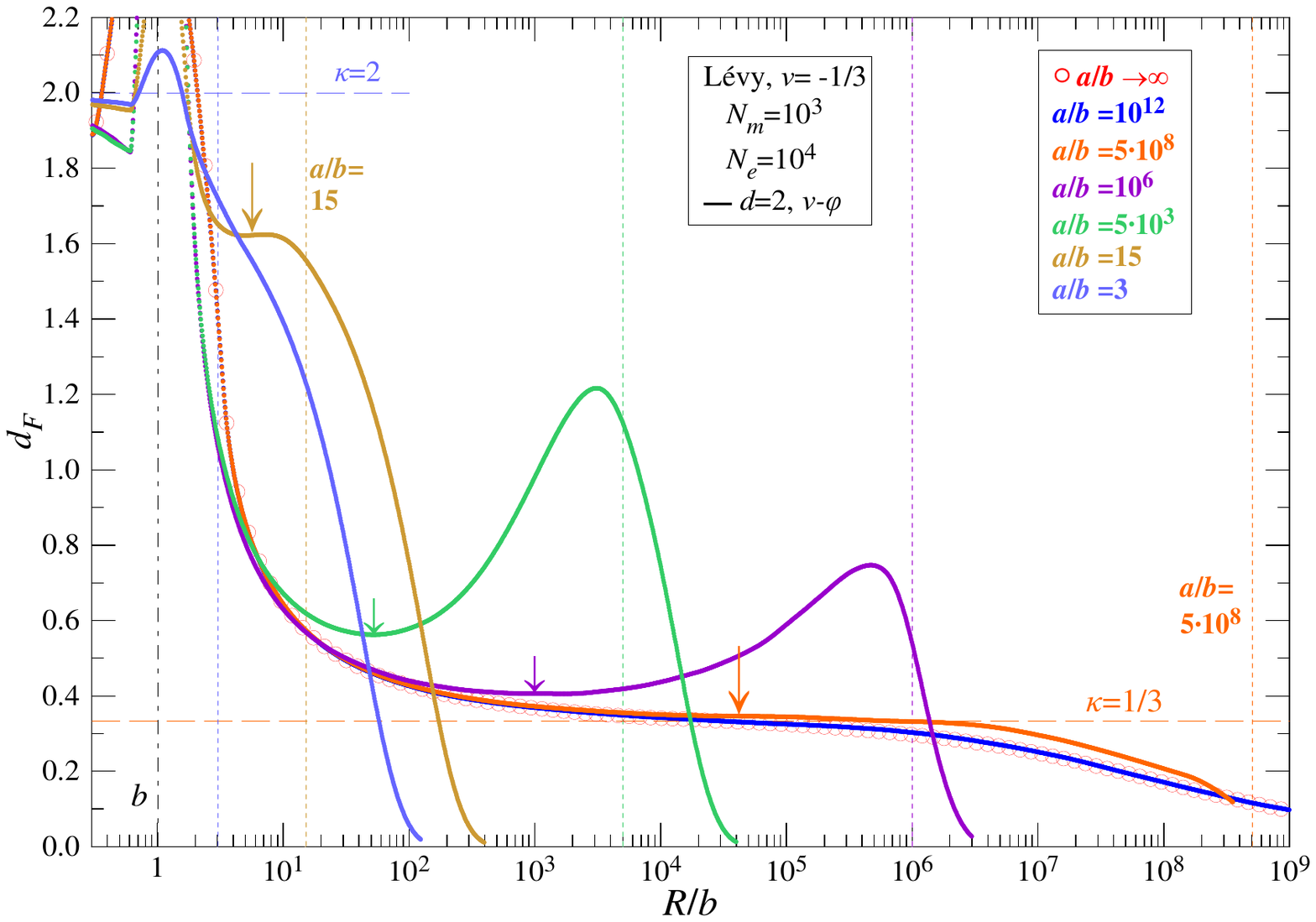}
\hspace{0.8cm}\includegraphics[scale=0.3,trim=0.4in 0.3in 0.7in 0.2in,angle=0]{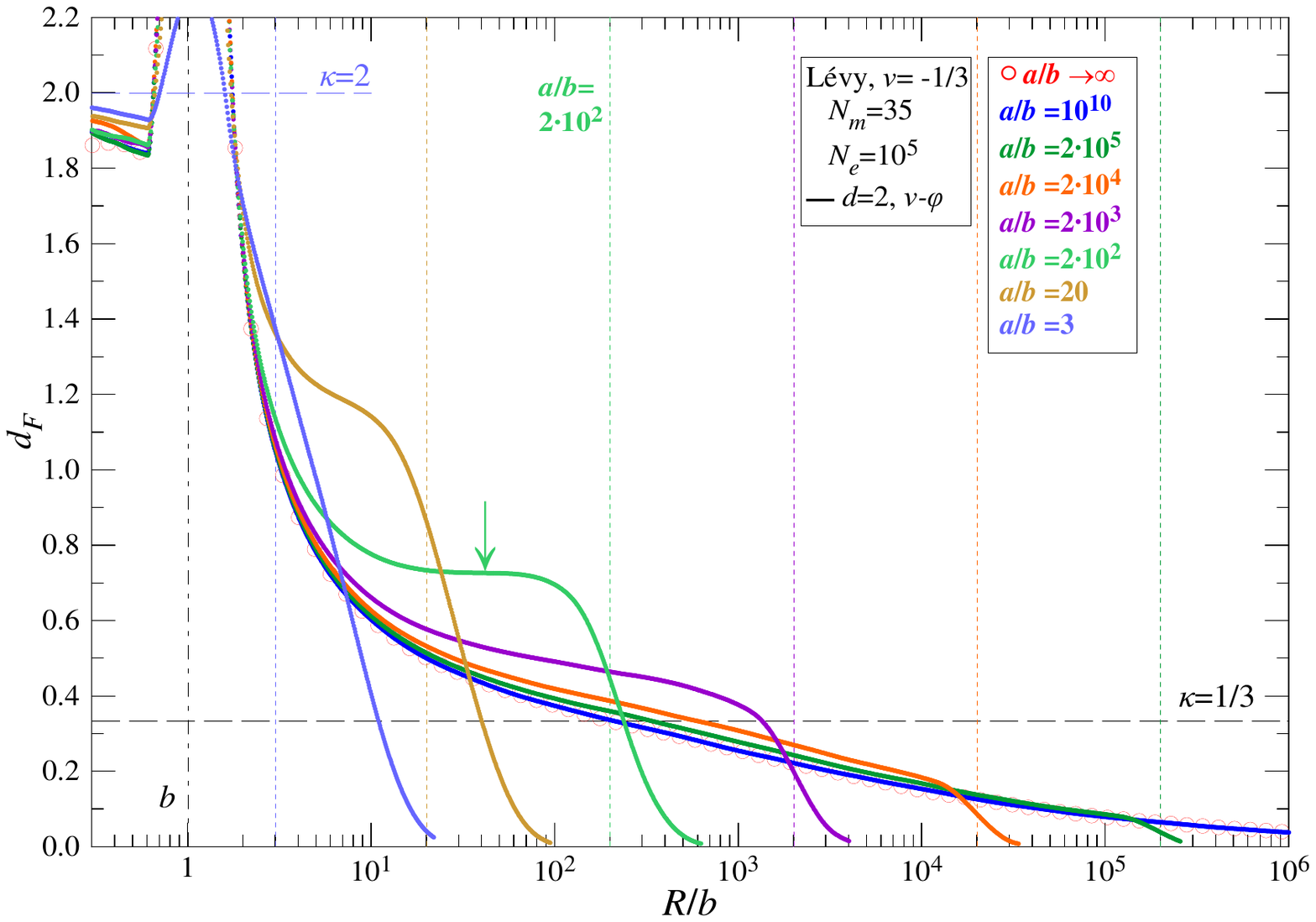}\\
\vspace{-0.4cm}(Aii)\hspace{7.8cm}   (Bii) \hspace{7.4cm} $\;$ \\
\hspace{0.4cm}\includegraphics[scale=0.3,trim=0.4in 0.3in 0.7in 0.2in,angle=0]{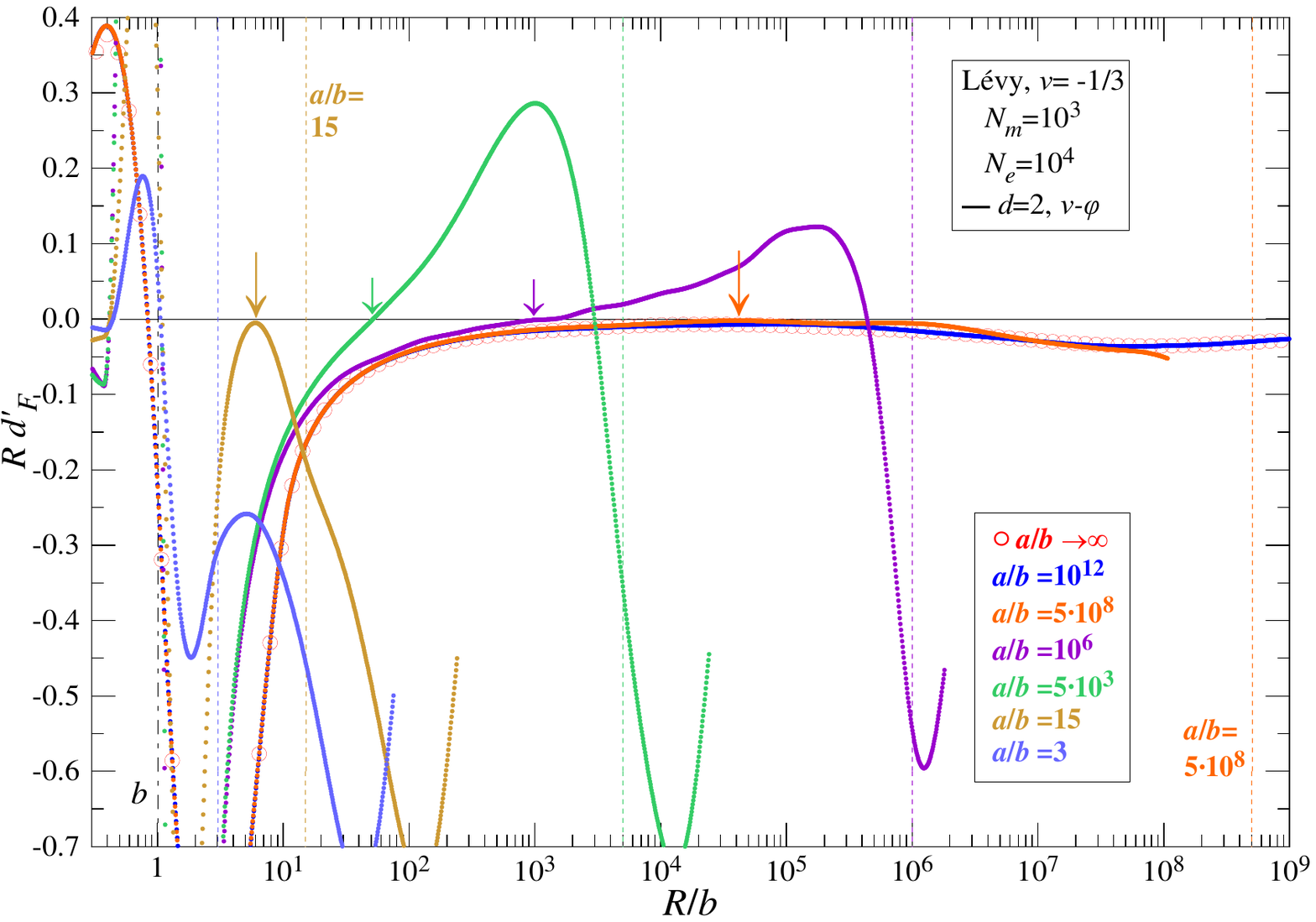}
\hspace{0.8cm}\includegraphics[scale=0.3,trim=0.4in 0.3in 0.7in 0.2in,angle=0]{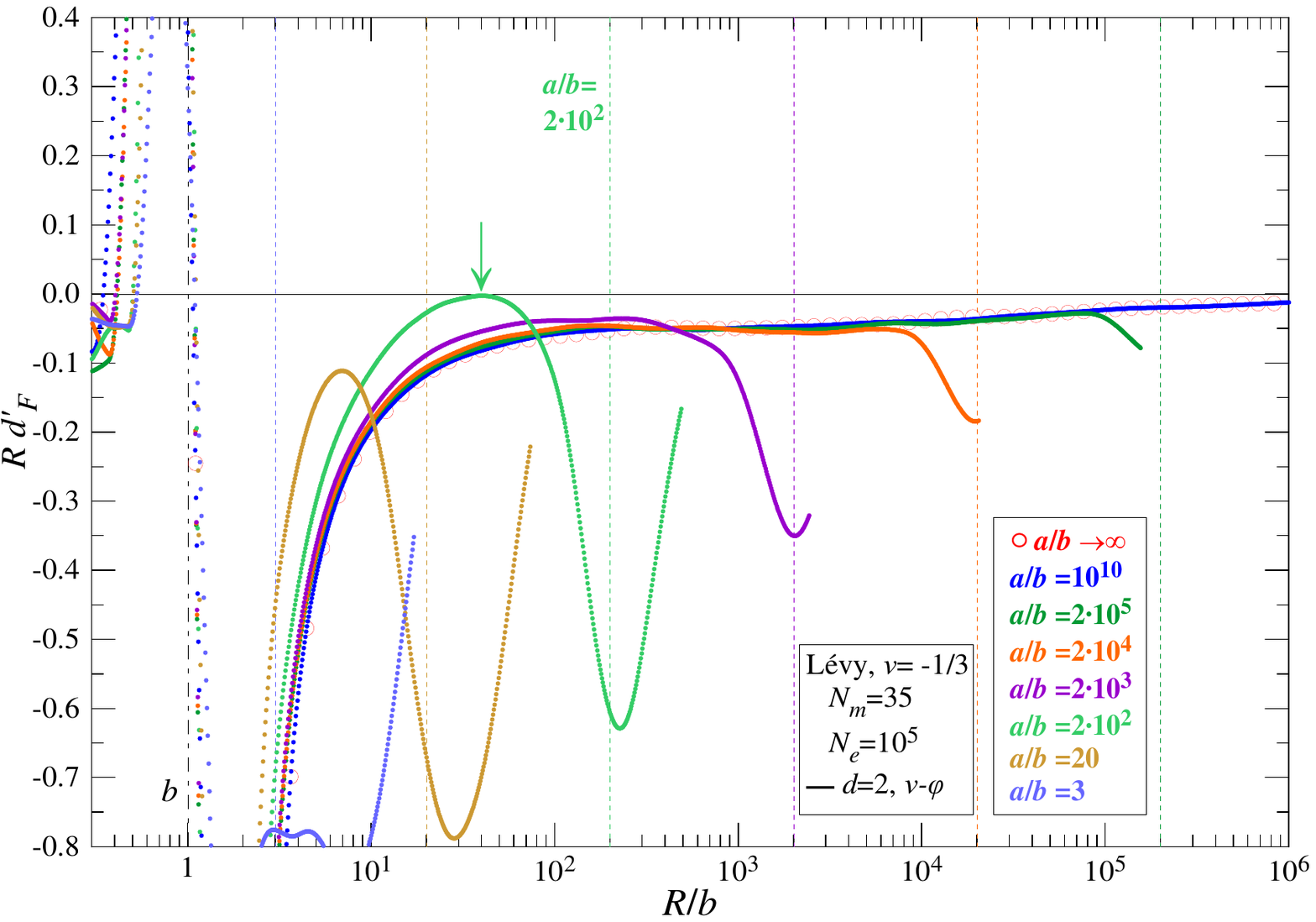}\\
\vspace{-0.4cm}(Aiii)\hspace{7.6cm}   (Biii) \hspace{7.2cm} $\;$ \\
\caption{\label{fig:add_a} {\small L\'{e}vy walks with 
step probability exponent $\nu=-1/3$ and varying upper step limits $a$ for $\nu-\phi$ case.
In (A) multiplicity $N_m=10^3$ and number of events
$N_e=10^4$, while in (B) $N_m=35$ and $N_e=10^5$. (i) The correlation integral $C$ of each set as function of the scale $R/b$
in log-log plot.
(ii) The local slopes $d_F=d(lnC)/d(lnR)$ of the walks presented in (i).
(iii) The derivatives $R\cdot d_F'=$ $R\cdot d(d_F)/dR$ of the slopes presented in (ii).
With long arrows optimal linear parts in log-log plot of $C(R)$, $d_{F1}$ and with short
arrows local minimum slopes, $d_{F2}$, are denoted.}}
\end{figure}

In case of negative $\nu$ the higher limit $a$ does not practically alter the distribution, if
$a/b \gg 1$ (in Fig.~\ref{fig:add_a}(A) the high values of $a/b=10^{12}$ in (i) and $a/b=10^{10}$ 
in (ii) produce almost the same results
with the case where a higher limit is absent). However such a high limit can greatly facilitate
the handling of data.
Additionally, for very high number of steps the 
introduction of 
high $a$ can improve the expected linearity of the $C(R)$ distribution, leading to a value of 
$d_F$ very close to $|\nu|$ (e.g. in Fig.~\ref{fig:add_a}(Ai) the value of $a/b=5\cdot10^8$).
Setting $a$ to lower values gradually distorts the $C$ distribution with respect to the 
distribution where $a$ does not exist. The deviation is greater at high scales and smaller at
low scales. The deviation strengthens as $a$ is lowered for constant $b$.
In general, the slope of $C$, $d_F$, is then changed and the appearance of the $d_F=|\nu|$
which is exhibited for a scale interval (a linear part in the $C(R)$ log-log diagram)
can disappear. The slopes of $C(R)$ show a transient character with different $d_F$ for
different scales.

The local extremums of $d_F$ can be traced by solving the equation
\begin{equation}\label{eq:dF'=0}
d'_F=0 \Rightarrow \frac{d(d_F)}{dR}=0\;.
\end{equation}
We are interested in cases where $d_F$ remains constant for a scale interval, so these
cases are covered by eq.~(\ref{eq:dF'=0}). Additionally the $C$ distribution for absent $a$
has a behaviour where its log-log slope decreases as scales increase from the $b$ region and
then it acquires an almost constant value before it falls off to zero.
We want to explore whether we can find such a ``minimal'' behaviour even if we apply an upper 
limit. This condition is translated to the equation
\begin{equation}\label{eq:dF''=0}
d''_F=0 \Rightarrow \frac{d^2(d_F)}{dR^2}=0\;.
\end{equation}

\noindent
So we are interested in scale intervals where the 1st and 2nd derivative of $d_F$ vanishes.
After all, at a perfect constant function $d_F$, all the derivatives should vanish, so the
simultaneous fulfilment of eqs.~(\ref{eq:dF'=0})-(\ref{eq:dF''=0}) is a minimum requirement.
We call the achieved slope in these cases $d_{F1}$, where the index ``1'' refers to the best linear
log-log slope of $C$ but in a situation where this slope remains always decreasing for scales
greater than $b$.

\begin{figure}[H]
\centering
\vspace{-0.0cm}
\hspace{0.4cm}\includegraphics[scale=0.3,trim=0.4in 0.3in 0.7in 0.2in,angle=0]{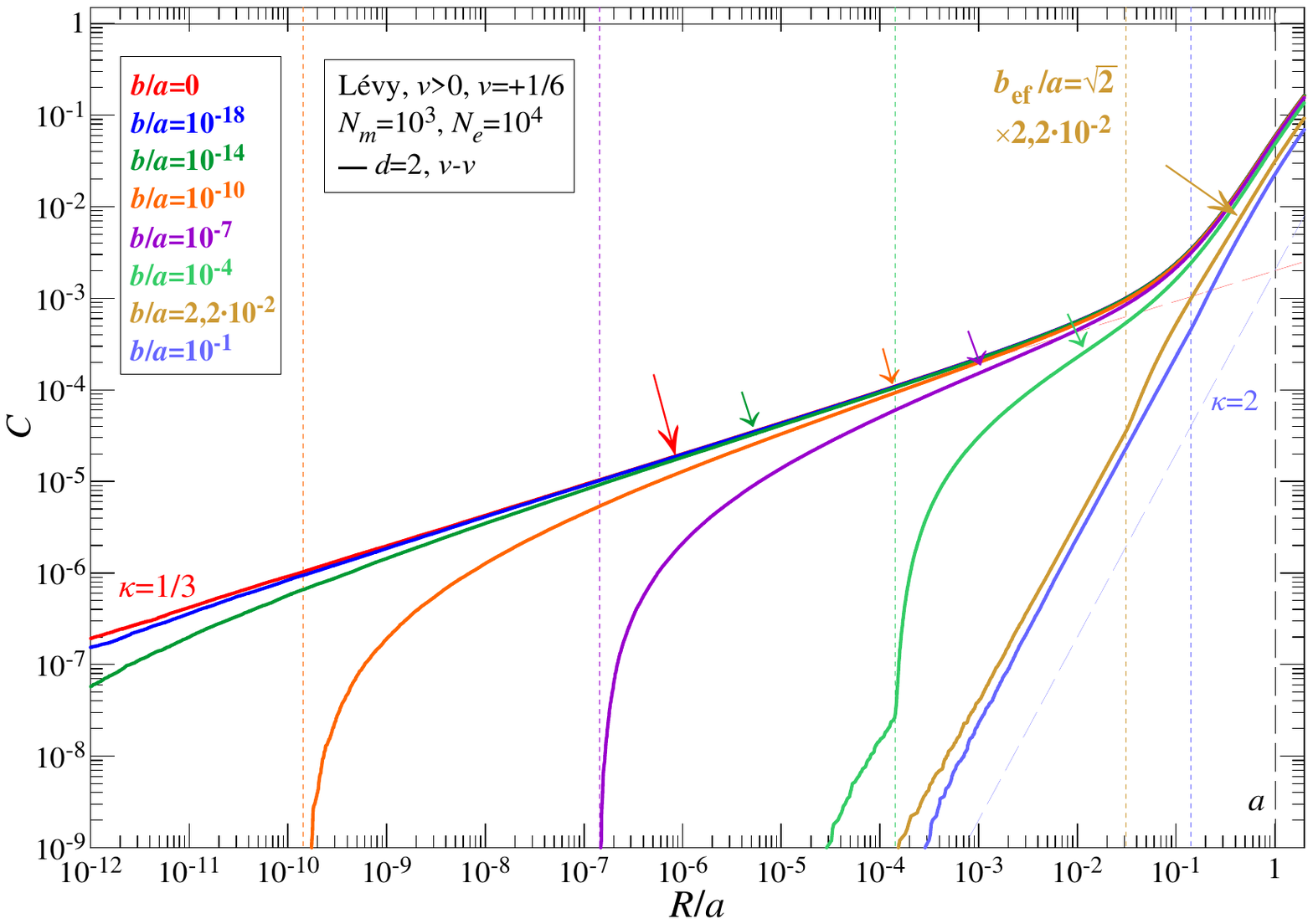}
\hspace{0.8cm}\includegraphics[scale=0.3,trim=0.4in 0.3in 0.7in 0.2in,angle=0]{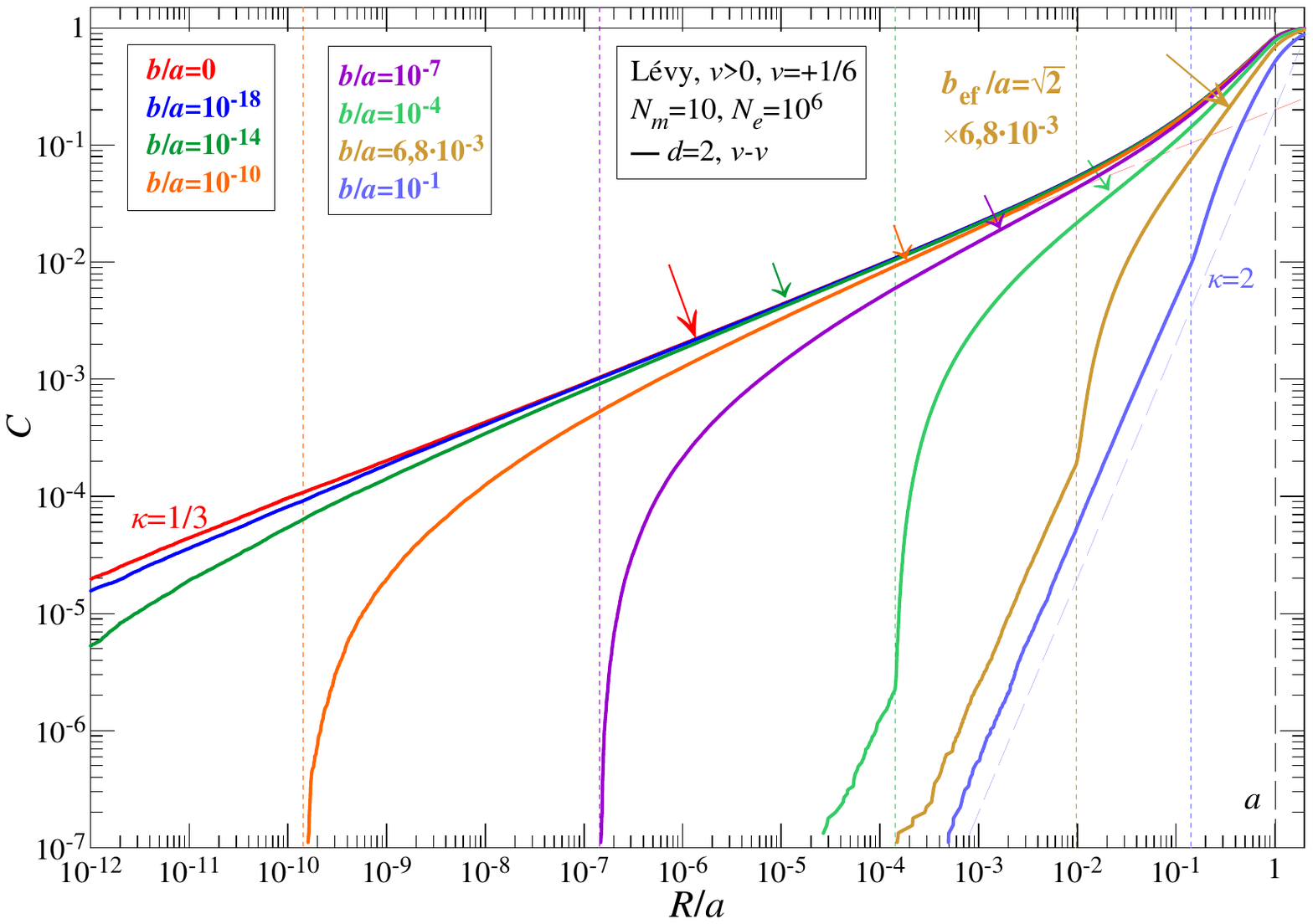}\\
\vspace{-0.3cm}(Ai)\hspace{7.8cm}   (Bi) \hspace{7.9cm} $\;$ \\
\hspace{0.4cm}\includegraphics[scale=0.3,trim=0.4in 0.3in 0.7in 0.2in,angle=0]{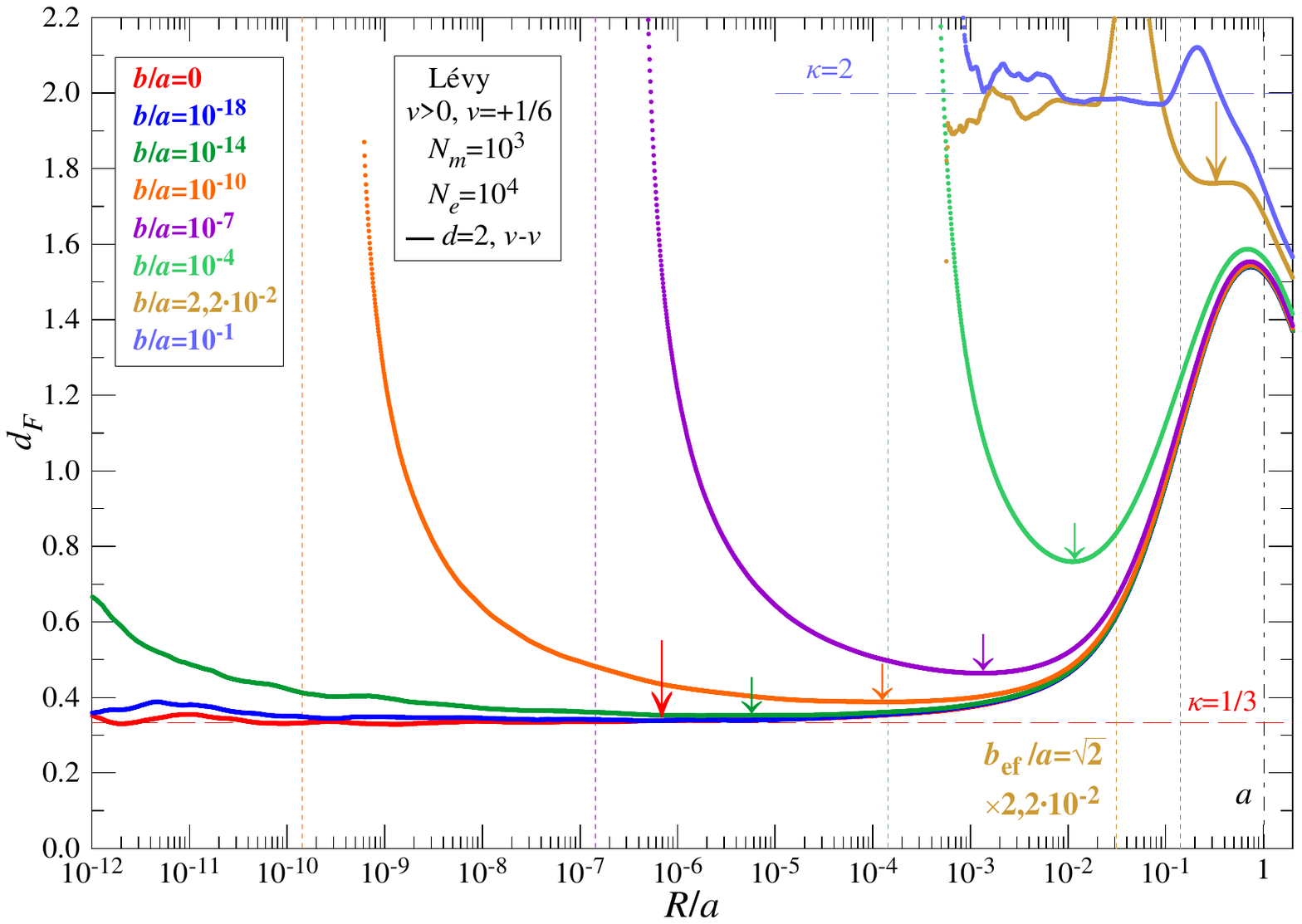}
\hspace{0.8cm}\includegraphics[scale=0.3,trim=0.4in 0.3in 0.7in 0.2in,angle=0]{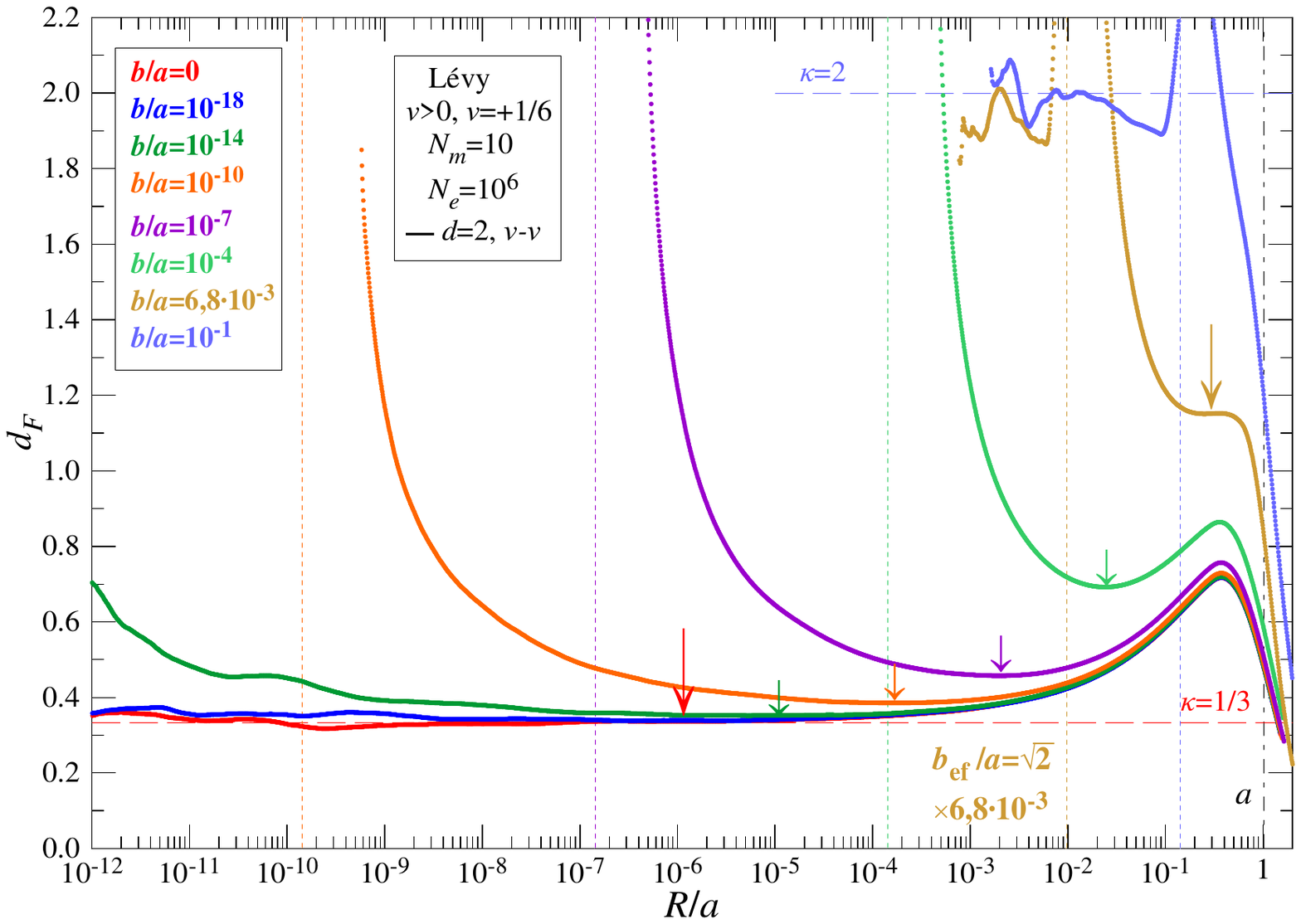}\\
\vspace{-0.3cm}(Aii)\hspace{7.7cm}   (Bii) \hspace{7.8cm} $\;$ \\
\hspace{0.4cm}\includegraphics[scale=0.3,trim=0.4in 0.3in 0.7in 0.2in,angle=0]{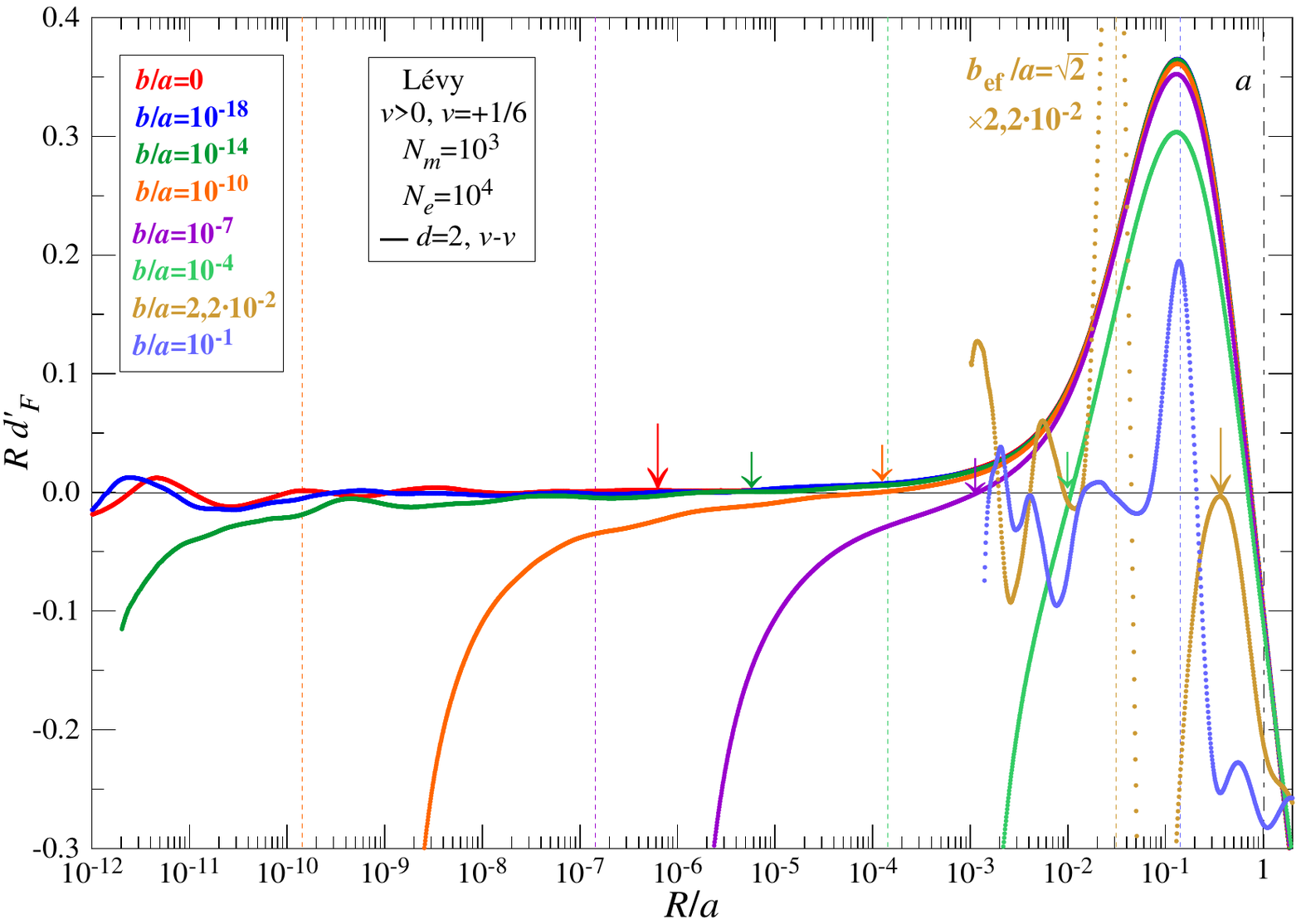}
\hspace{0.8cm}\includegraphics[scale=0.3,trim=0.4in 0.3in 0.7in 0.2in,angle=0]{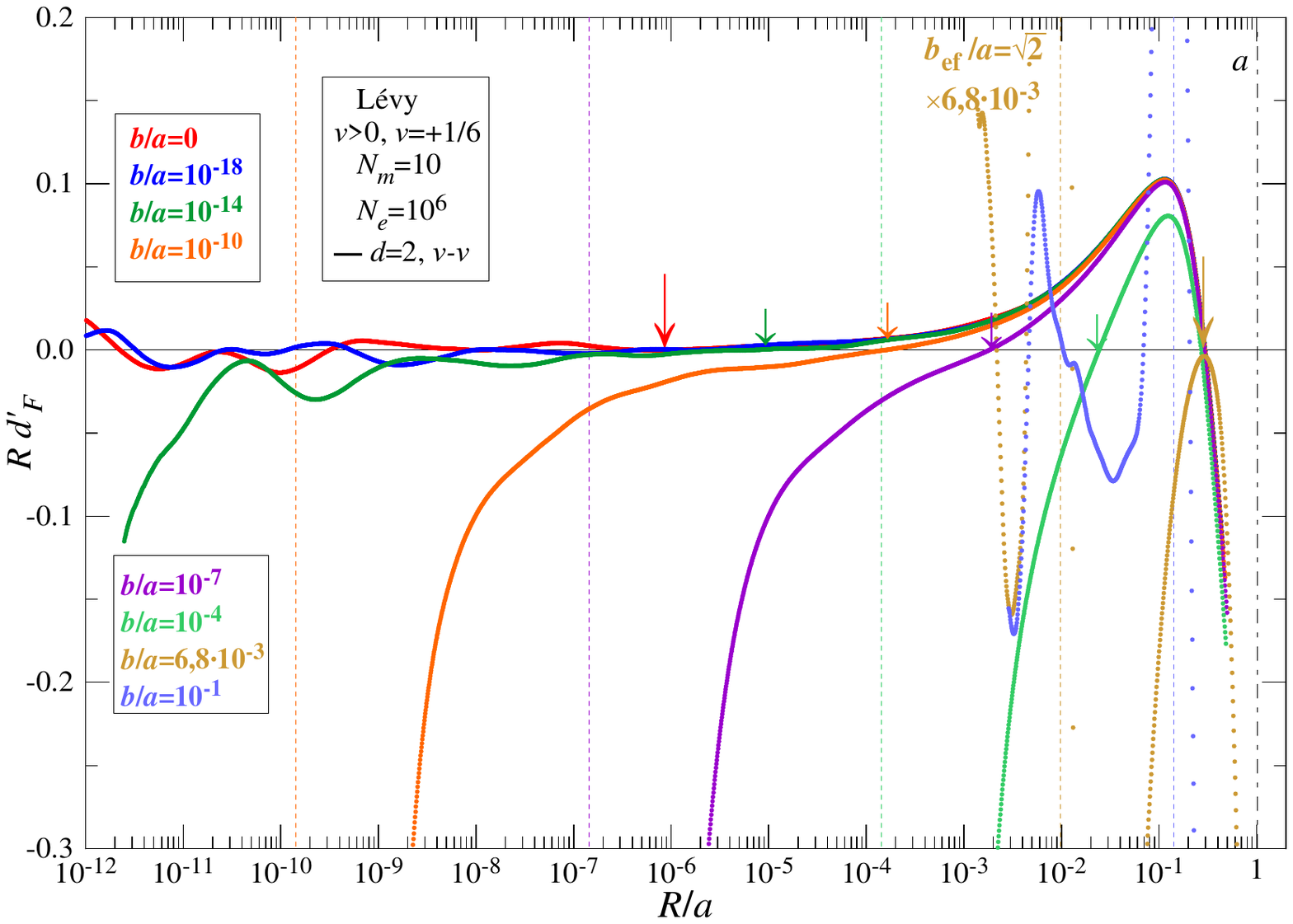}\\
\vspace{-0.3cm}(Aiii)\hspace{7.6cm}   (Biii) \hspace{7.6cm} $\;$ \\
\caption{\label{fig:add_b} {\small L\'{e}vy walks with 
step probability exponent $\nu=+1/3$ and varying lower step limits $b$ for $\nu-\nu$ case.
In (A) multiplicity $N_m=10^3$ and number of events
$N_e=10^4$, while in (B) $N_m=10$ and $N_e=10^6$. (i) The correlation integral $C$ of each set as function of the scale $R/a$
in log-log plot.
(ii) The local slopes $d_F=d(lnC)/d(lnR)$ of the walks presented in (i).
(iii) The derivatives $R\cdot d_F'=$ $R\cdot d(d_F)/dR$ of the slopes presented in (ii).
With long arrows optimal linear parts in log-log plot of $C(R)$, $d_{F1}$ and with short
arrows local minimum slopes, $d_{F2}$, are denoted.}}
\end{figure}

We observe that the $C$ distributions of negative $\nu$, even without an explicit
higher limit $b$ have two intrinsic limits. The one is the necessary lower limit $b$.
The other is set by the higher scale available in the data set, where $C$ reaches the unit value.
This is pushed to higher scale values as the number of steps, $N_m$, increases but always remains
finite for finite number of steps.
The introduction of a new upper limit $a$ can ``interact'' with these intrinsic limits and produce,
under certain conditions, new optimal linear parts of the $C(R)$ distribution. 
For high values of steps
there are two scale intervals where such linear parts can appear.
The setting of $a$ automatically changes the intrinsic high limit scale, transferring this role to $a$.
A linear part of $C$ can be formed at scales close to $a$ for specific choice of $a$ for certain $N_m$.
If $a$ moves close to $b$ the behaviour of $C$ at scales close to $b$ changes and another linear part 
of $C$ can be formed at scales close to $b$ for specific choice of $a$ for certain $N_m$.
In Fig.~\ref{fig:add_a}(iii) we plot $Rd'_F$, which is easy to depict for all scales.
The interesting linear parts can be traced at 
scales where eqs.~(\ref{eq:dF'=0})-(\ref{eq:dF''=0}) are satisfied simultaneously.
The extremal points of $d'_F$ satisfy $d''_F=0$. But $(Rd'_F)'=d'_F+Rd''_F$. If $d'_F=0$, then
$(Rd'_F)'=Rd''_F$. So the zero points of $Rd'_F$ which are, also, local extremal points, satisfy both
eqs.~(\ref{eq:dF'=0})-(\ref{eq:dF''=0}). 

As mentioned, for very high $N_m$ the linear part corresponds to $d_F \simeq |\nu|$.
As $N_m$ decreases this linear part is formed at lower scales and
the produced $d_F$ is higher than $|\nu|$. Still there are two optimal linear parts which can be 
formed
for two specific choices of $a$ for the given $N_m$. These can be traced by applying 
eqs.~(\ref{eq:dF'=0})-(\ref{eq:dF''=0}), as can be seen in Fig.~\ref{fig:add_a}(Aiii).
As $N_m$ decreases further the two linear parts move closer.
At a certain $N_m$ the two linear parts merge into a single one. In Fig.~\ref{fig:add_a}(B) which
refers to $d=2$ and the $\nu-\phi$ case this occurs for $N_m \simeq$35. 
So for number of steps greater than a minimum number we can find two best linear parts in the
$C$ distribution with slopes $d_{F1}$ (these are depicted with long arrows in Fig.~\ref{fig:add_a}).
For a specific number of steps, $N_m$, these two slopes $d_{F1}$ can be formed for two certain 
values of higher limits $a$.
For $a$ between these two values, the $C$ distribution exhibits a different behaviour, where its
slope, for scales greater than $b$ decreases and then increases, before it finally decreases to 
zero. So at a certain scale interval we can find a minimum local slope, for a spesific choice of $a$,
where eq.~(\ref{eq:dF'=0}) holds, but not eq.~(\ref{eq:dF''=0}). We will denote this slope as 
$d_{F2}$ and it
is the next to best we can find after $d_{F1}$, since the slope will not change so rapidly as
scales change. These slopes are denoted with short arrows in Fig.~\ref{fig:add_a}. Obviously the
slopes $d_{F1,2}$ are of higher values than $|\nu|$.

For lower values of steps the negative $\nu$ distributions cannot exhibit optimal linear parts
and the local slopes $d_F$ exhibit transient behaviour.
Also, for scales below $b_{ef}$ the $C$ distribution acquires a scaling behaviour where the fractal
 dimension is close to the dimension of the embedding space, $d$. This behaviour is present
for any $N_m>2$.

In case of positive $\nu$ the lower limit $b$ does not practically alter the distribution, if
$a/b \ll 1$ (in Fig.~\ref{fig:add_b}(A) the low values of $b/a=10^{-18}$ produce almost the same 
results with the case where a lower limit is absent).
When we introduce finite $b$, the $C$ distribution is distorted and the distortion increases as $b$ 
approaches $a$.
Also the $C$ distribution diminishes abruptly as $b_{ef}$ is approached
from the higher scales. Below $b_{ef}$ the set acquires the dimension $d$ of the embedding space
and this part of $C$ remains linear in the log-log diagram.

For the case of $C$ distributions with positive $\nu$ and no lower step limit, we observe that their
explicit limits are the necessary upper limit $a$ and the zero scale (the distribution can
infinitely continue with the same slope till zero in the log-log plot, limited though in practice by  
machine accuracy). If we search for linear parts in the way we did for negative $\nu$,
we will always find one for absent $b$ at lower scales with $d_F$ close to $\nu$. This part
will always exist for all $N_m$.
When we introduce a finite $b$, in order to trace optimal linear parts
of $C(R)$, in the same sense of negative $\nu$, we may again apply 
eqs.~(\ref{eq:dF'=0})-(\ref{eq:dF''=0}). Such a linear part can be found for $b$ approaching 
$a$ for every $N_m$. For example such linear part can be seen in Fig.~\ref{fig:add_b}, in (A) for 
$N_m=10^3$ and $b/a=2.2 \cdot 10^2$ and in (B) for $N_m=10$ and $b/a=6.8 \cdot 10^3$.
So we can have two slopes of the type $d_{F1}$ (long arrows in Fig.~\ref{fig:add_b}).
Applying only eq.~(\ref{eq:dF'=0}) we can also find slopes of the type $d_{F2}$ 
(short arrows in Fig.~\ref{fig:add_b}).

\begin{figure}[H]
\centering
\vspace{-0.0cm}
\hspace{-0.5cm}(i)\includegraphics[scale=0.3,trim=0.4in 0.3in 0.7in 0.2in,angle=0]{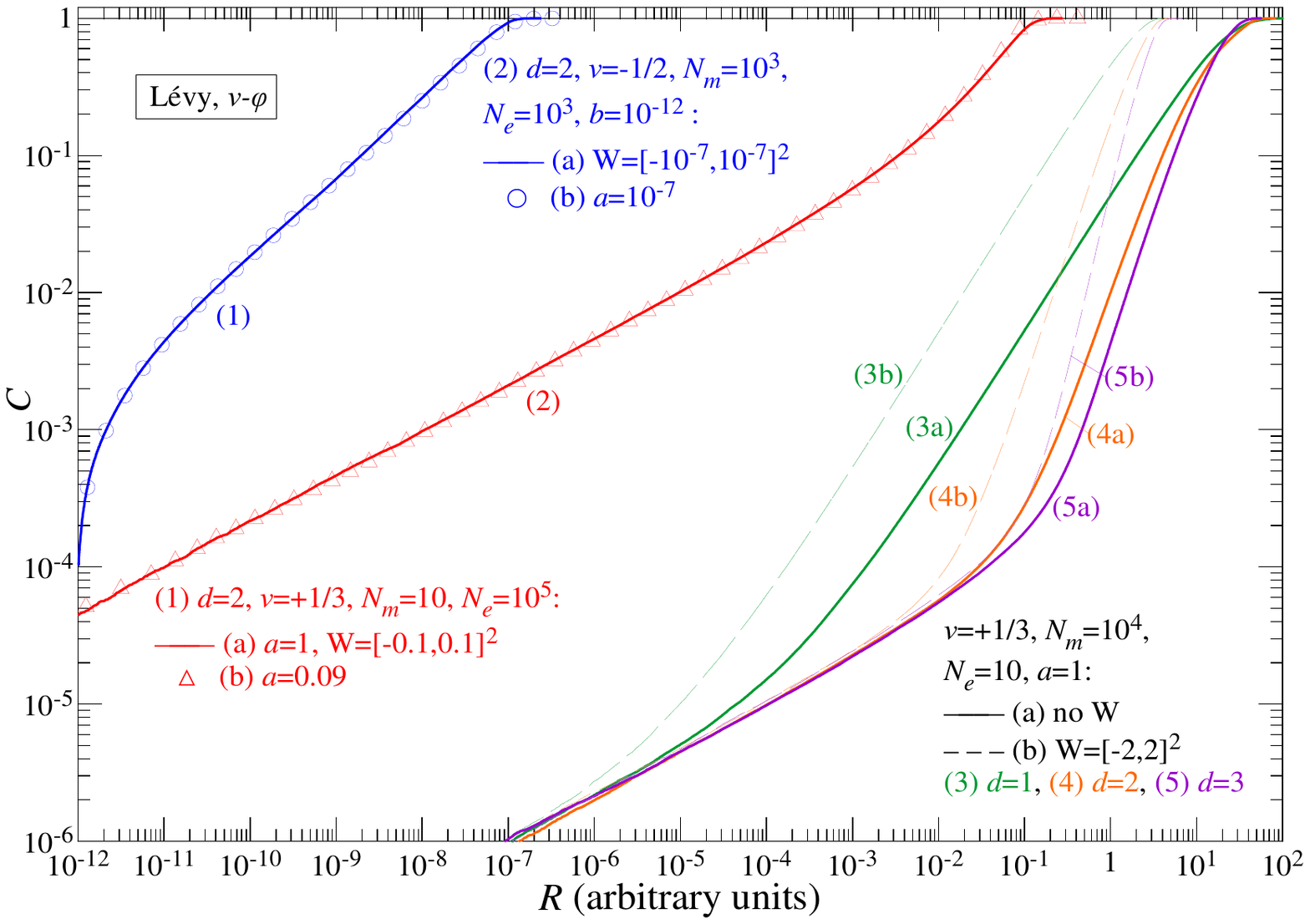}
(ii)\includegraphics[scale=0.3,trim=0.4in 0.3in 0.7in 0.2in,angle=0]{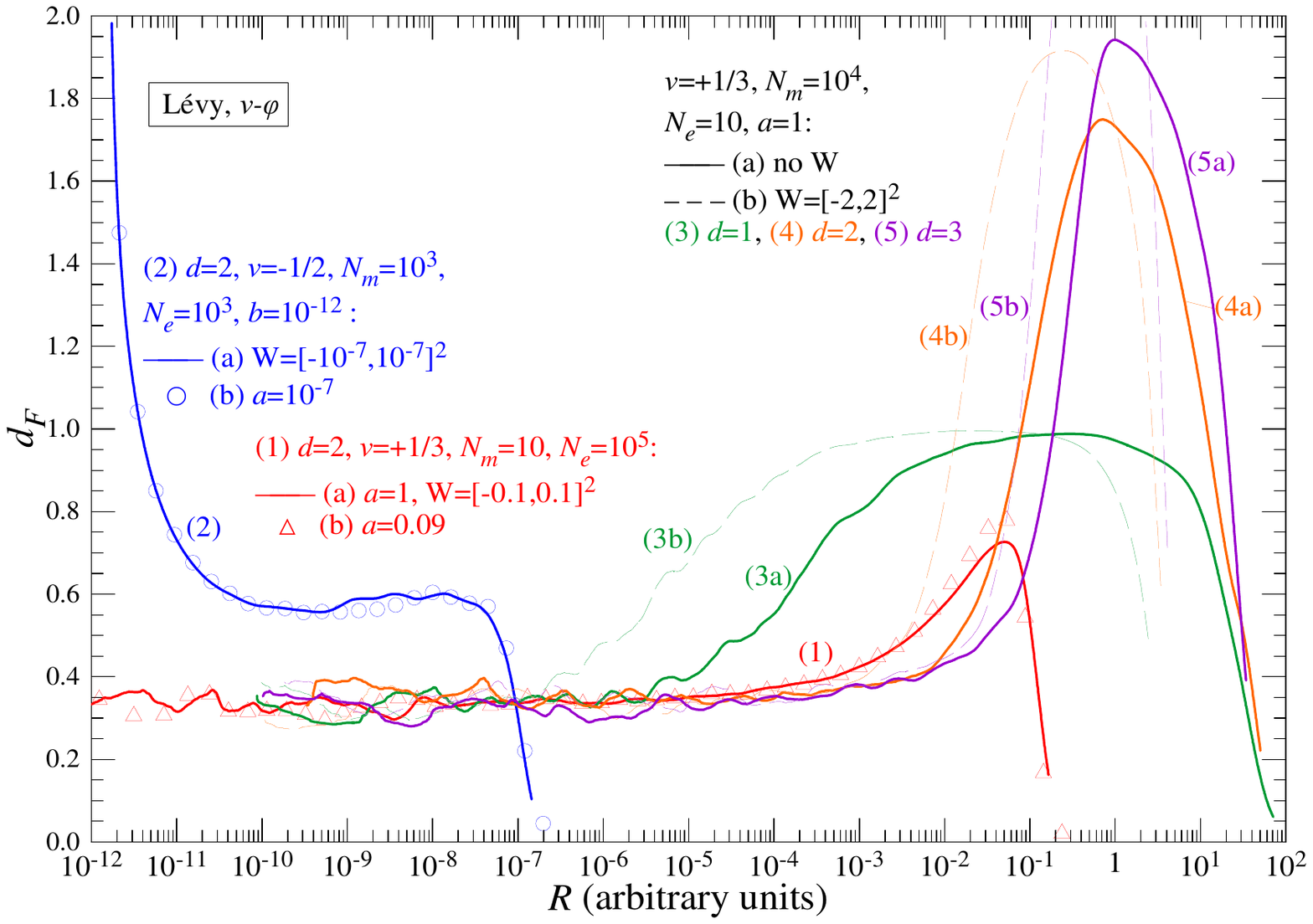}\\
\caption{\label{fig:window} {\small Effect of working window (W) on $C$ distributions of L\'{e}vy walks.
Curves (1) ($\nu>0$) and curves (2) ($\nu<0$) show that the working window has almost the same
effect as a suitable upper limit $a$.
Curves (3)-(5) show that the working window for $\nu>0$ can limit the scale interval
necessary to reach the slope $\nu$, an effect that becomes more significant as $d$ increases.
In (i) the $C$ function and in (ii) the corresponding local log-log slopes. The graphs remain
unchanged in the rescaling $a'=\lambda a$, $b'=\lambda b$ and $R'=\lambda R$.}}
\end{figure}

The additional limits $a$ or $b$ can help to place boundaries on the interval of scales
where we want to conduct our analysis. Another technique which can limit the space where
our data exist is to place a working window. In this practice, while producing our L\'{e}vy 
steps when a
point falls outside this window is rejected and another step is produced until the visited
point falls within the window. 
This technique is in general equivalent to setting an upper limit $a$ in the step distribution.
For low scales the application of a working window is not expected to change the results, since
most pairs with low distances remain undisturbed by the window.
These can be observed in Fig.~\ref{fig:window}, where all the data are produced with the
first step taken from the origin and the working windows are also centred at the origin.
Line (1a) corresponds to a positive $\nu$ case
with low multiplicity and an upper limit $a$=1, where we have imposed a window with width lower
than $a$. Then we produce a data set with the same characteristics of line (1a), but with a lower value
of $a$ and without a window. The produced results (curve (1b)) are practically the same. We, also,
repeat the calculations for a negative $\nu$ case and find almost the same results either
with an upper limit $a$ or a window with a width close to $a$ (lines (2)).

However, the working window can, under certain circumstances, change the behaviour 
at scales close to the boundaries and, also, can limit the boundary phenomena.
Such a case is depicted in Fig.~\ref{fig:window}(I)-(II), where $\nu$ is positive and the
multiplicity $N_m$ is high. We see that, as the embedding space dimension, $d$, increases,
the scale interval needed to reach the slope equal to $\nu$ decreases.

\section{Distributions for zero $\nu$} 
\label{sec:nu=0}

\begin{figure}[H]
\centering
\vspace{-0.5cm}
\hspace{-0.5cm}(i)\includegraphics[scale=0.57,trim=0.4in 0.3in 0.7in 0.2in,angle=0]
{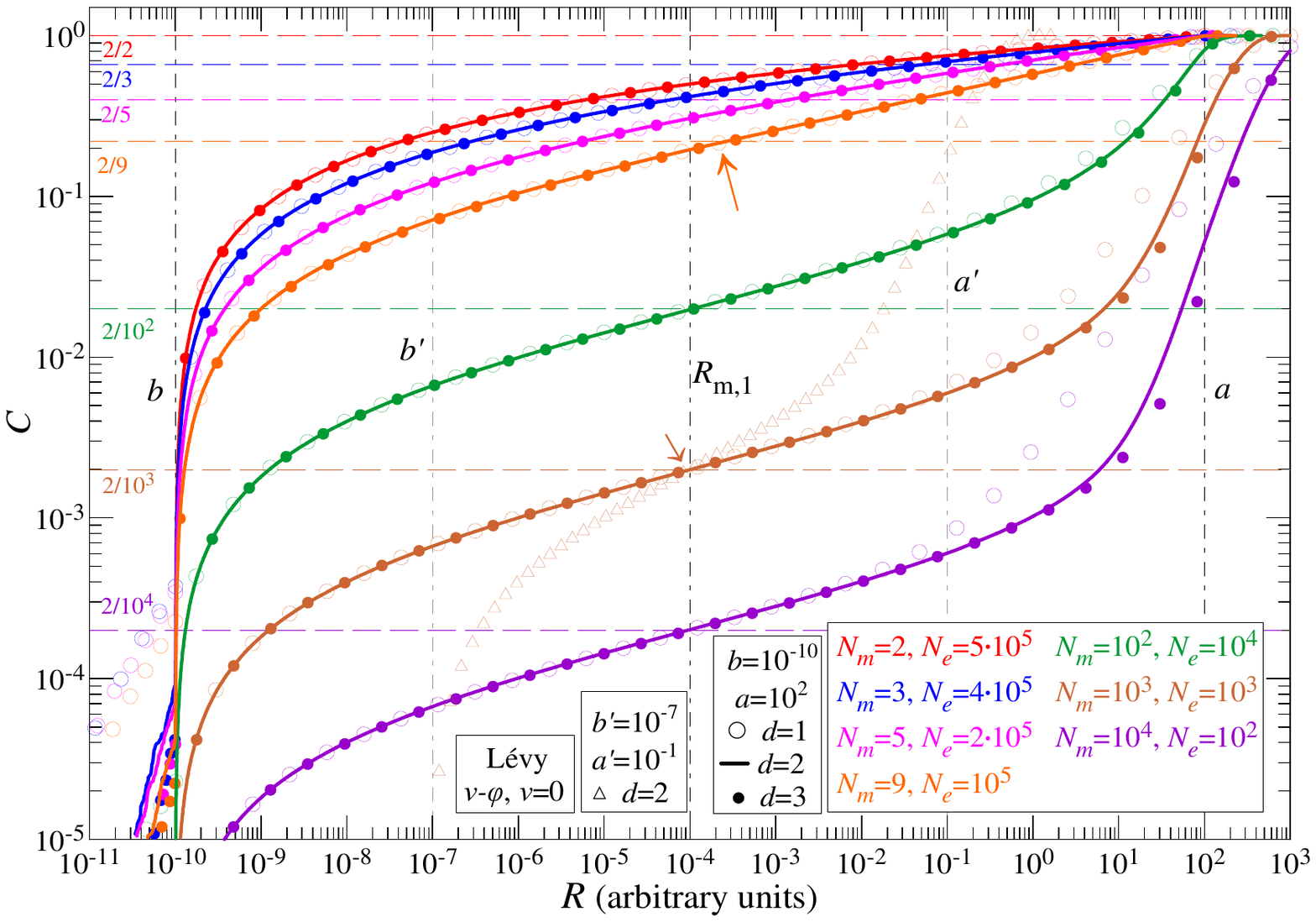}\\
\hspace{-0.7cm}(ii)\includegraphics[scale=0.57,trim=0.4in 0.3in 0.7in 0.2in,angle=0]{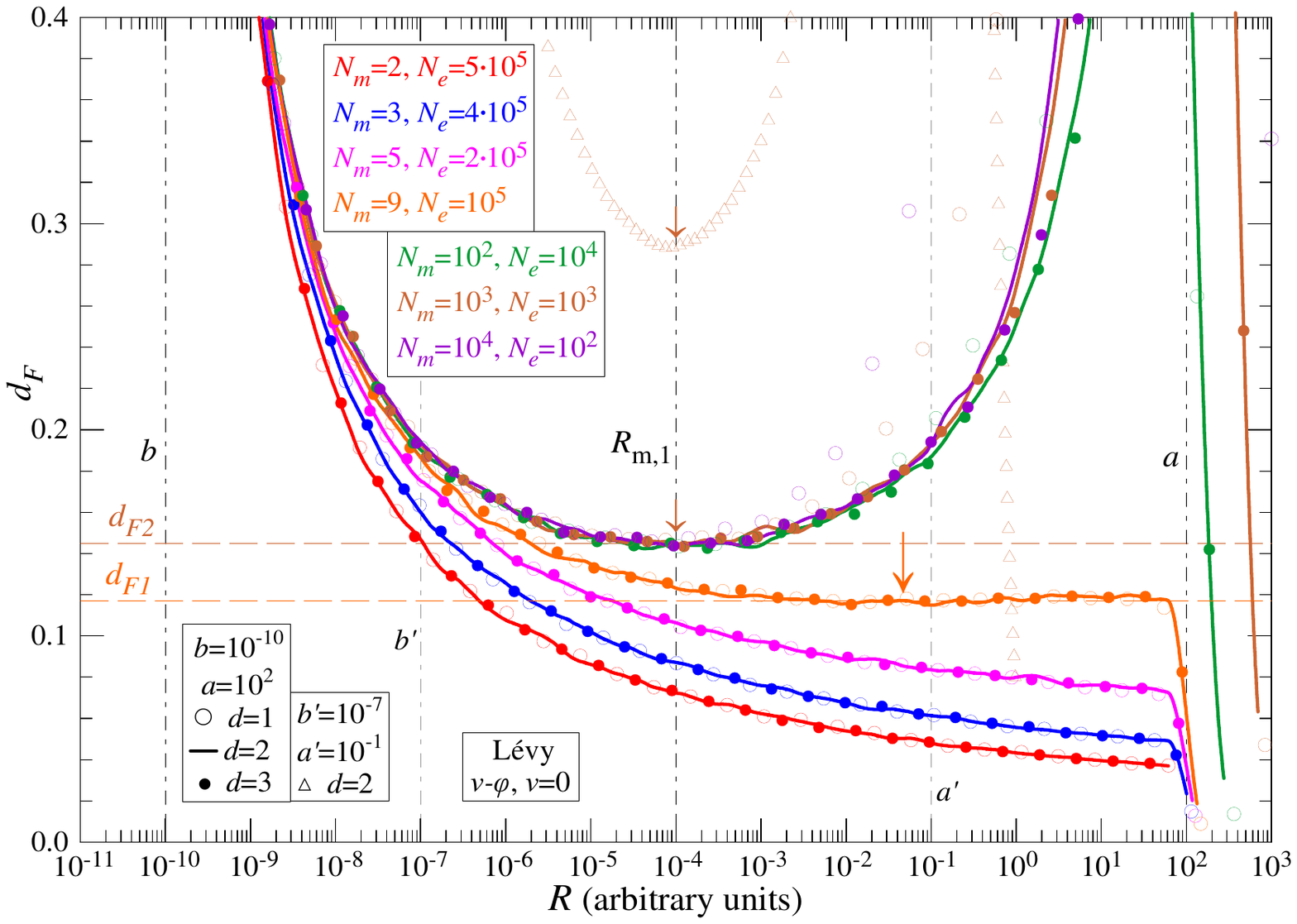}
\caption{\label{fig:nu=0_Nm_nu-phi} {\small 
L\'{e}vy walks with 
step probability exponent $\nu=0$ for the $\nu-\phi$ case, $d=$1,2,3 dimensions, fixed step 
limits $a$ and $b$ and with 
varying number of steps. 
Another case for $d=$2 and different step limits $a'$ and $b'$ is depicted.
At high number of steps the slope at $R_{m,1}$ stabilises at $d_{F2}$ 
(short arrows).
At a certain low number of steps a slope of type $d_{F1}$ is formed (long arrows). 
The $C$ distributions are depicted in (i), while the local log-log slopes, $d_F$, are depicted in (ii).
The graphs remain unchanged in the rescaling $a''=\lambda a$, $b''=\lambda b$ and $R''=\lambda R$.
}}
\end{figure}

The case of L\'{e}vy walks with $\nu=$0 necessities the existence of both limits, $a$ and $b$,
according to eq.~(\ref{eq:zn_pdf}). We begin to study this case by investigating how the $C$
distribution changes for fixed upper and lower step limits while we change the number of steps,
$N_m$. The results are shown on Fig.~\ref{fig:nu=0_Nm_nu-phi} for the $\nu-\phi$ case and 
different dimensions
$d$ of embedding space. The interesting result we extract is that as $N_m$ is increasing the $C$
distribution begins to acquire almost the same slopes for intermediate scales which no longer
depend on $N_m$ (the $C$ curve moves downwards as $N_m$ increases but it stays parallel to the
curves of lower $N_m$). This is evident more clearly in Fig.~\ref{fig:nu=0_Nm_nu-phi}(II) where
the $d_F$ curves start to converge to the same curve at intermediate scales as $N_m$ is increased  
sufficiently. This is true for all $d=$1,2,3 which we used in our analysis.
Also, the pattern shows that between the limits $a$ and $b$ there is a minimum slope that can be 
achieved and this slope is higher than zero. This minimum slope is of the type $d_{F2}$ which we
showed in the previous section and occurs, for all $d$, at about the scale
\begin{equation}\label{eq:R_m1}
R_{m,1} \cong \sqrt {b \cdot a} = b \left( \frac{a}{b} \right)^{1/2}
\end{equation}

\begin{figure}[H]
\centering
\vspace{-0.5cm}
\hspace{-0.5cm}(i)\includegraphics[scale=0.57,trim=0.4in 0.3in 0.7in 0.2in,angle=0]{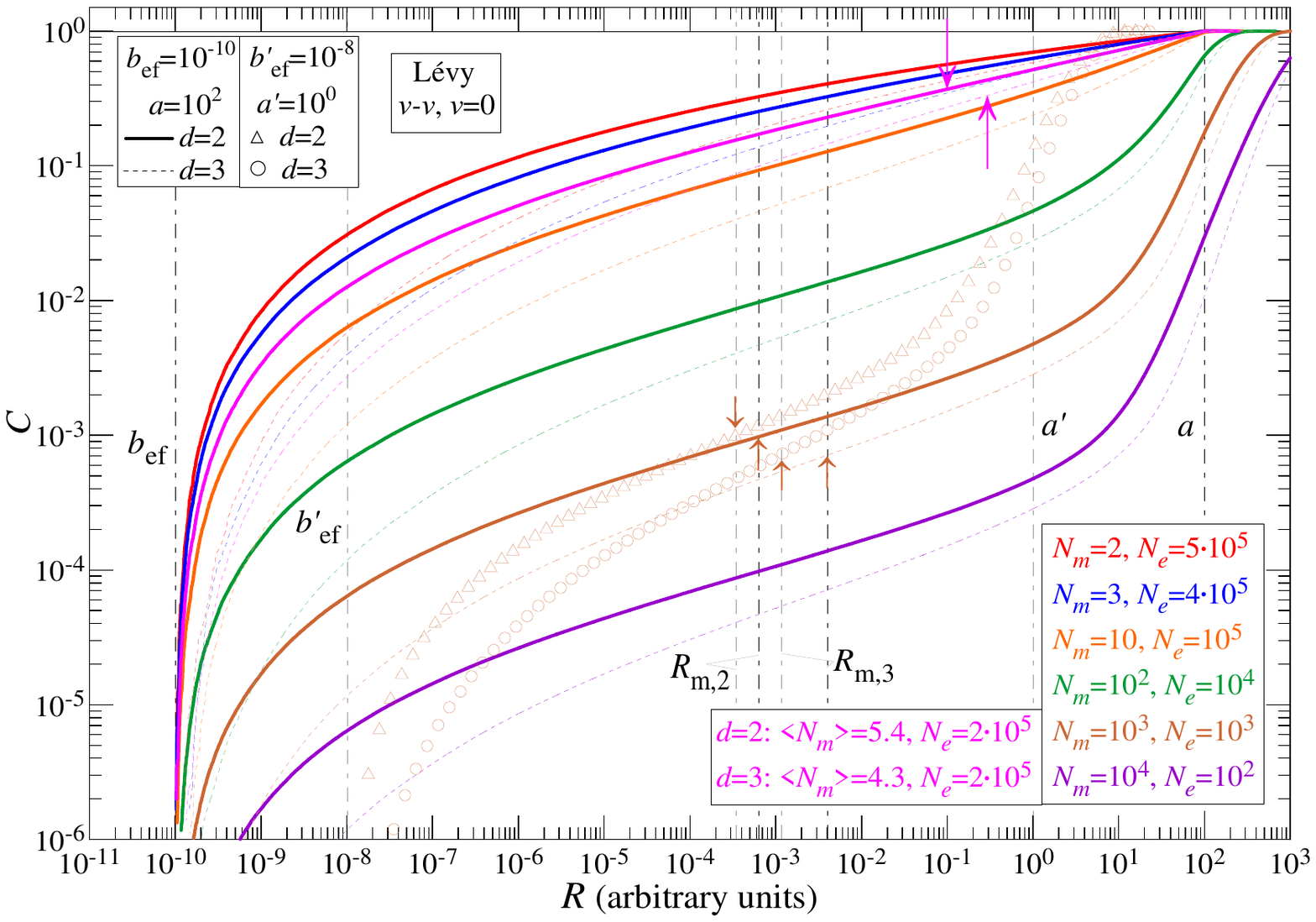}\\
\hspace{-0.7cm}(ii)\includegraphics[scale=0.57,trim=0.4in 0.3in 0.7in 0.2in,angle=0]{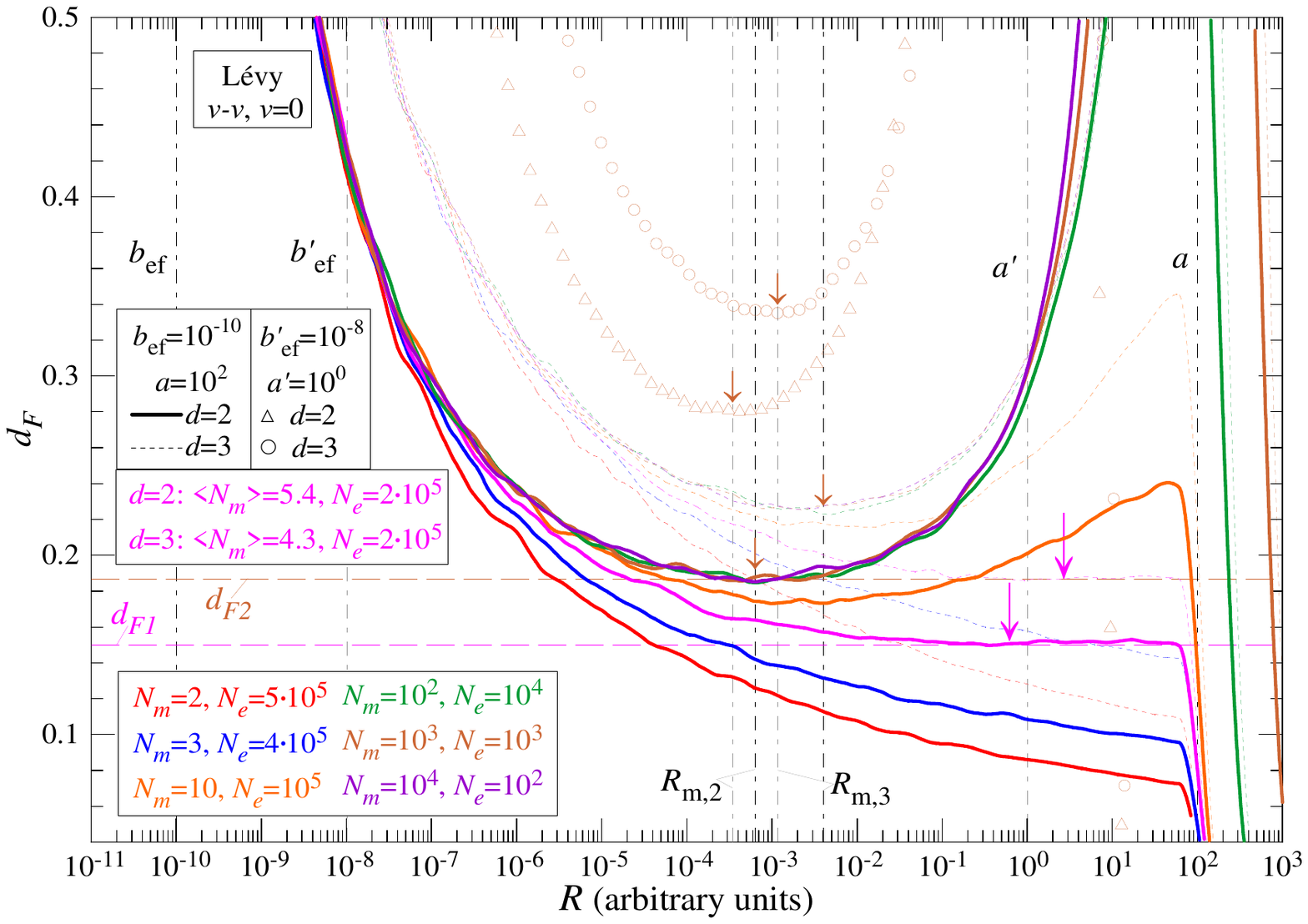}
\caption{\label{fig:nu=0_Nm_nu-nu} {\small 
L\'{e}vy walks with 
step probability exponent $\nu=0$ for the $\nu-\nu$ case, $d=$2,3 dimensions, fixed step 
limits $a$ and $b$ and with 
varying number of steps. 
Two other cases for $d=$2 and 3 and different step limits $a'$ and $b'$ are depicted.
At high number of steps the slopes at $R_{m,d}$ ($d$=2,3) stabilise at $d_{F2}$ 
(short arrows).
At a certain low number of steps a slope of type $d_{F1}$ is formed (long arrows). 
The $C$ distributions are depicted in (i), while the local log-log slopes, $d_F$, are depicted in (ii).
The graphs remain unchanged in the rescaling $a''=\lambda a$, $b''=\lambda b$ and $R''=\lambda R$.
}}
\end{figure}

\begin{figure}[H]
\centering
\vspace{-0.5cm}
\hspace{-0.5cm}(I)\includegraphics[scale=0.57,trim=0.4in 0.3in 0.7in 0.2in,angle=0]{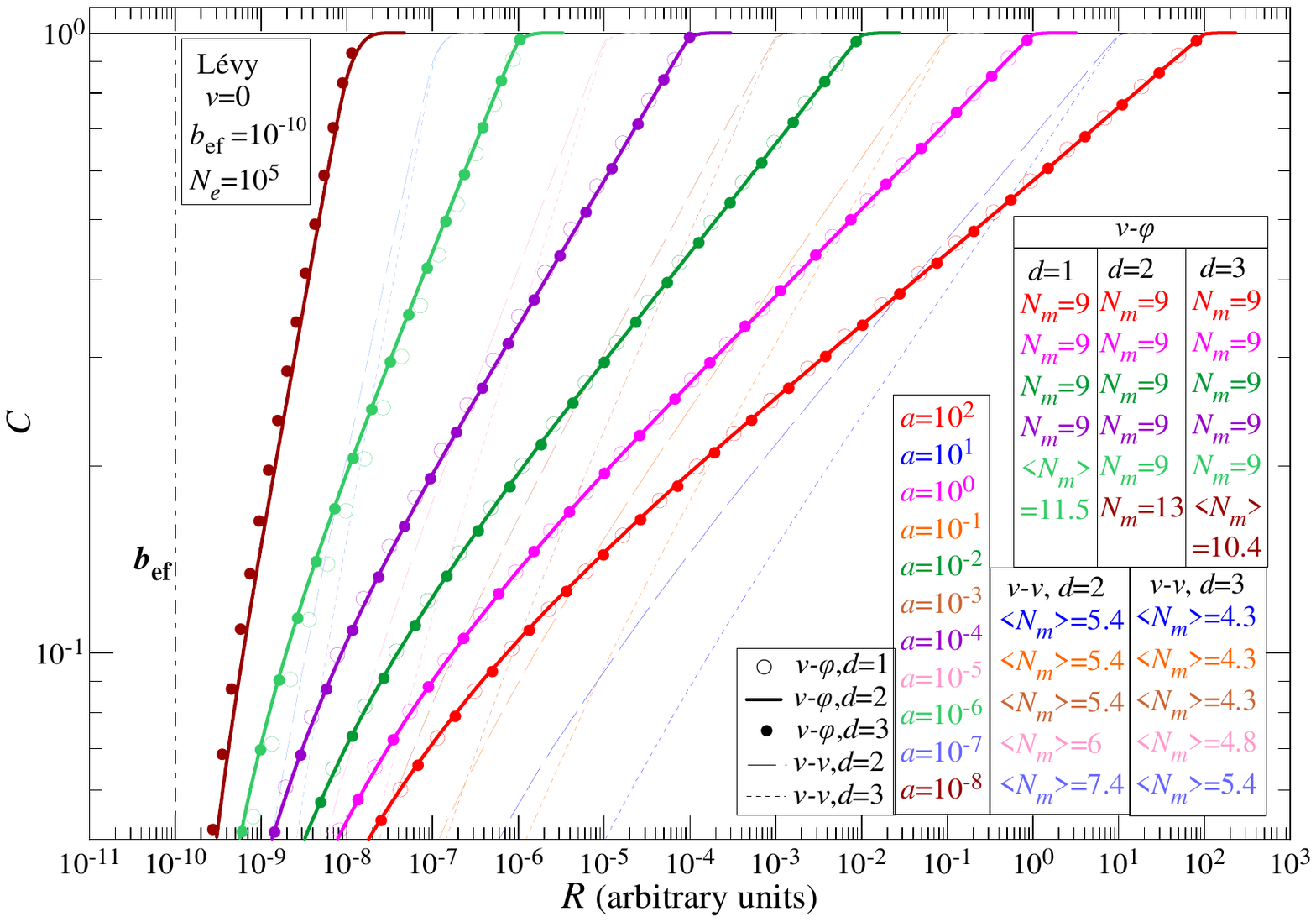}\\
\hspace{-0.7cm}(II)\includegraphics[scale=0.57,trim=0.4in 0.3in 0.7in 0.2in,angle=0]{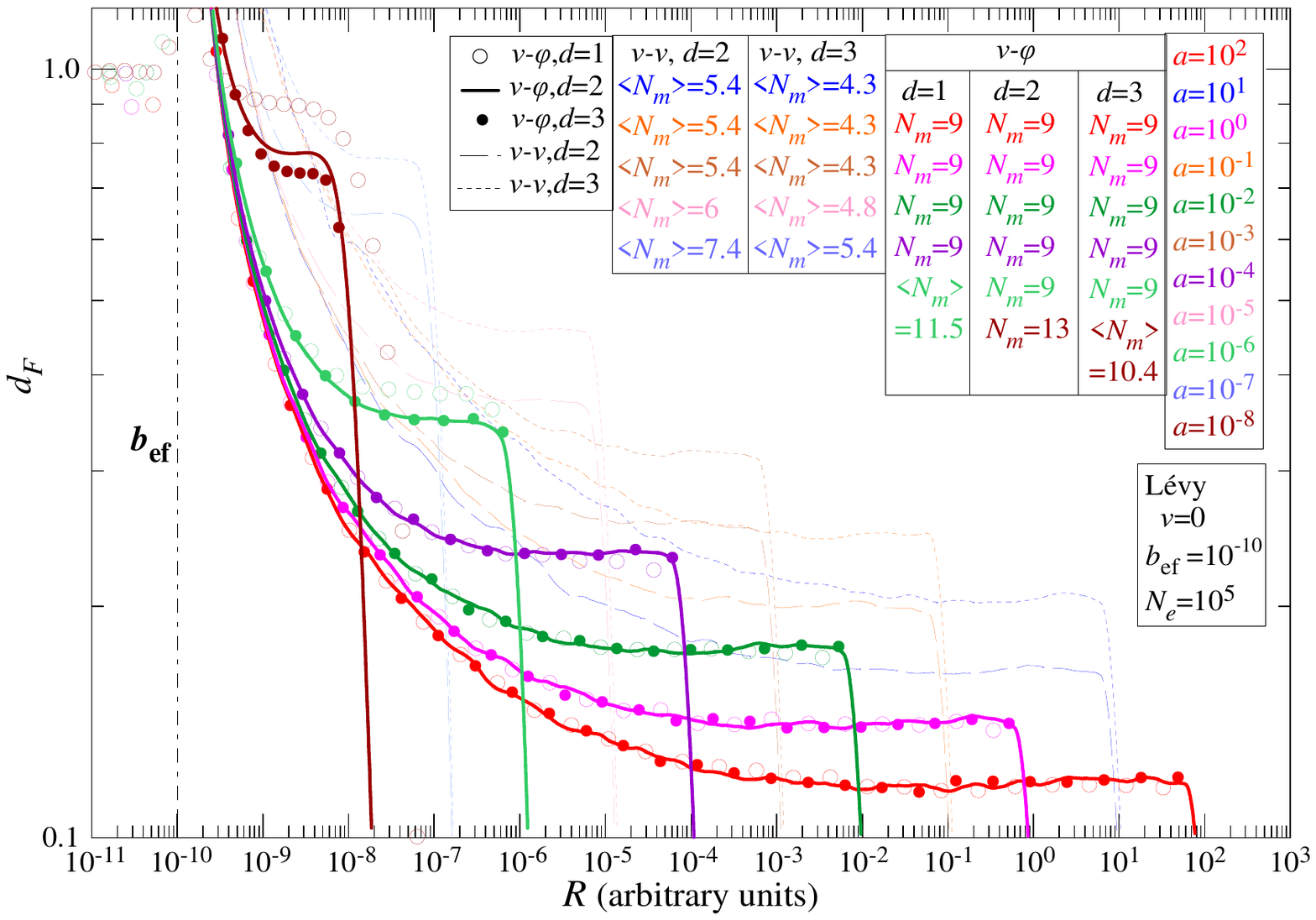}
\caption{\label{fig:nu=0_Nm_dF1} {\small L\'{e}vy walks with 
step probability exponent $\nu=0$ for $\nu-\phi$ and $\nu-\nu$ cases and $d=$1,2,3 dimensions, 
fixed lower step limit $b$ and varying upper step limit $a$ which produce for low number of
steps a slope of the type $d_{F1}$ very close to scale $a$. 
For a wide range of values of $a$ the needed number of steps remains unchanged, while this number is
increased as $a$ approaches $b$.
The $C$ distributions are depicted in (i), while the local log-log slopes, $d_F$, are depicted in (ii).
The graphs remain unchanged in the rescaling $a'=\lambda a$, $b'=\lambda b$ and $R'=\lambda R$.}}
\end{figure}

\noindent
In the same graph we, also, depict, another case with a different set of limits $a'$ and $b'$ to show
that the same pattern of behaviour is retained.

We turn, then, to the production of L\'{e}vy walks for the $\nu-\nu$ case. The results are
summarised in Fig.~\ref{fig:nu=0_Nm_nu-phi} for $d=$2,3 (the $d=$1 case is identical to the
$\nu-\phi$ case). We observe a similar behaviour with the $\nu-\phi$ case. However, now the
minimum slope, of the type $d_{F2}$, which stabilises for high $N_m$ occurs at different
scales. We find that these scales can be approximated by
\begin{equation}\label{eq:R_md}
R_{m,d} \cong b_{ef} \left( \frac{a}{b_{ef}} \right)^{\frac{1}{2} + \left( d-1 \right)\frac{1}{15}}\;,
\end{equation}
where $b_{ef}=b\sqrt{d}$.
Last equation converges to eq.~(\ref{eq:R_m1}) for $d=$1, since the $\nu-\nu$ and $\nu-\phi$ cases
are identical for $d=$1.

Another interesting result which can be inferred from 
Figs.~\ref{fig:nu=0_Nm_nu-phi}(II),\ref{fig:nu=0_Nm_nu-nu}(II) is that at very low $N_m$ the slope 
of $C$ is always decreasing having a completely transient character. At high $N_m$, as seen, it 
stabilises forming a minimum slope at scale $R_{m,d}$ with slopes increasing for greater scales.
So for high scales, as the number of steps increases, the log-log slope of $C$ gradually increases.
There is a specific number of steps for which the slope remains almost the same, thus, forming
a slope of the form $d_{F1}$. This situation is particularly interesting since the desired linear
part is formed very close to the higher scales, thus, limiting considerably boundary phenomena.
Also, the almost linear distribution is formed with only a few steps.
For the specific choice of limits of Fig.~\ref{fig:nu=0_Nm_nu-phi} and $\nu-\phi$ case this occurs
for $N_m=$9 steps. In Fig.~\ref{fig:nu=0_Nm_nu-nu} which corresponds to $\nu-\nu$ cases we see that
the $d_{F1}$ slope occurs for $d=$2 at $<N_m>=$5.4 and for $d=$3 at $<N_m>=$4.3.
In order to achieve the constant slope in these situations
we have formed events with mixed number of steps. Specifically each event has $N_{m1}$ steps with probability $r_1$ or 
$N_{m2}$ steps with probability $1-r_1$. Thus the average number of steps is
\begin{equation}\label{eq:<N_m>}
<N_m>=r_1 N_{m1} +(1-r_1)N_{m2}
\end{equation}
For $d=2$ we form events with multiplicity $N_{m1}=$5 with probability 0.6 and multiplicity $N_{m2}=$6
with probability 0.4. For $d=3$ we form events with multiplicity $N_{m1}=$4 with probability 0.7 and 
multiplicity $N_{m2}=$5 with probability 0.3.

We investigate the $d_{F1}$ slopes of $\nu=$0 L\'{e}vy walks further. The results are depicted in
Fig.~\ref{fig:nu=0_Nm_dF1}. We change the ratio of limits $a/b$ and observe the number of steps
which forms the constant slope. For the $\nu-\phi$ cases we observe that the $d_{F1}$ slope is formed 
for the same number of steps $N_m=$9 until a certain low value of $a/b$ (higher for $d=$1 and
lower for $d=$3). Below that value of $a/b$ a $d_{F1}$ slope can still be formed, but for a
higher number of steps which depends on the specific value of $a/b$. Similar behaviour we
observe for the $\nu-\nu$ cases. The $d_{F1}$ slopes are formed for a wide range of $a/b_{ef}$ values
for $<N_m>=$5.4 and for $<N_m>=$4.3 for $d=$2, 3, respectively.

The achieved values of $d_{F1,2}$ depend on the chosen value of $a/b_{ef}$. Normally, when we want to 
perform a simulation we want to have a data exhibiting a prescribed dimension at a scale interval.
So it is useful to know, if possible, a relation between $d_{F1,2}$ and $a/b_{ef}$.
For the $d_{F2}$ slopes the approximate scales at which they appear are known through
eqs.~(\ref{eq:R_m1}),(\ref{eq:R_md}).
So we produce events with a high number of L\'{e}vy steps (in order to have convergence of the slope to 
its final value) for each case with $\nu=$0
and varying ratio $a/b$. We record the slope $d_{F2}$ at the scales $R_{m,1}$ (for $\nu-\phi$ case
and $d$=1,2,3), $R_{m,2}$ (for $\nu-\nu$ case and $d$=2) and $R_{m,3}$ (for $\nu-\nu$ case and $d$=3).
For the $d_{F1}$ slopes we expect them to appear at high scales near upper limit $a$.
So we chose a scale interval $\Delta R= \left( c R_{m,d}, 0.5a \right)$, where $c=$10 for $\nu-\nu$ case and $d$=1,2,3 and
$c=$5 for $\nu-\phi$ case and $d$=2,3. 
We produce events with the specific number of L\'{e}vy steps for each case and calculate the
average slope in the interval $\Delta R$.
Our results are recorded in Fig.~\ref{fig:nu=0_dF1,2}.
The interesting situation that arises is that, if we plot the values of $d_{F1,2}$ as function of 
the variable $\eta=1/ \ln \left( \frac{a}{b_{ef}}\right)$, we see that until an upper value 
of $\eta$ for each case
 there is a 
simple proportionality relation (linear through the origin)

\begin{table}[h]
\centering
\begin{tabular}{|c|c|c|c|c|c|c|c|} \hline
$i$ & $j$ & case       & $d$ & $N_m$  &  $A_{ijd}$        & $\eta_{max}$ & $(a/b_{ef})_{min}$  
\\ \hline
1   & 2   & $\nu-\phi$ & 1   & 9      & 3.2356$\pm$0.0009 & 0.062       & 1.0$\cdot$10$^7$    \\
1   & 2   & $\nu-\phi$ & 2   & 9      & 3.2356$\pm$0.0009 & 0.086       & 1.1$\cdot$10$^5$    \\
1   & 2   & $\nu-\phi$ & 3   & 9      & 3.2356$\pm$0.0009 & 0.110       & 8.9$\cdot$10$^3$    \\
1   & 1   & $\nu-\nu$  & 2   & 5.4    &  4.143$\pm$0.004  & 0.072       & 1.1$\cdot$10$^6$    \\
1   & 1   & $\nu-\nu$  & 3   & 4.3    &  5.111$\pm$0.007  & 0.072       & 1.1$\cdot$10$^6$    \\
2   & 2   & $\nu-\phi$ & 1   & 10$^3$ & 4.0012$\pm$0.0021 & 0.040       & 7.2$\cdot$10$^{10}$ \\
2   & 2   & $\nu-\phi$ & 2   & 10$^3$ & 4.0012$\pm$0.0021 & 0.070       & 1.6$\cdot$10$^6$    \\
2   & 2   & $\nu-\phi$ & 3   & 10$^3$ & 4.0012$\pm$0.0021 & 0.110       & 8.9$\cdot$10$^3$    \\
2   & 1   & $\nu-\nu$  & 2   & 10$^3$ &  5.156$\pm$0.005  & 0.070       & 1.6$\cdot$10$^6$    \\
2   & 1   & $\nu-\nu$  & 3   & 10$^3$ &  6.204$\pm$0.003  & 0.110       & 8.9$\cdot$10$^3$    \\ \hline  
\end{tabular}
\vspace{-0.0cm}
\caption{\label{tab:d_F} {\small The constants $A_{ijd}$ in eq.~(\ref{eq:d_F-eta}) 
extracted from fits in the data accumulated from simulations of L\'{e}vy walks. 
$\eta_{max}$ is the upper bound of the $\eta$ interval where the fit in each case is performed
and eq.~(\ref{eq:d_F-eta}) appears to hold. The lower bound of the fit interval is
$\eta=$0.035 (or $(a/b_{ef})=$2.6$\cdot$10$^{12}$).
However eq.~(\ref{eq:d_F-eta}) can be extrapolated until $\eta \simeq$0.
For $d_{F1}$ the necessary number of steps remains constant in the value interval of $\eta$
where the fits are performed.
For $d_{F2}$ the necessary number of steps remains at a high value and higher values
should give the same outcome.}}
\end{table}


\begin{figure}[H]
\centering
\vspace{-0.2cm}
\hspace{-0.5cm}\includegraphics[scale=0.6,trim=0.4in 0.3in 0.7in 0.2in,angle=0]{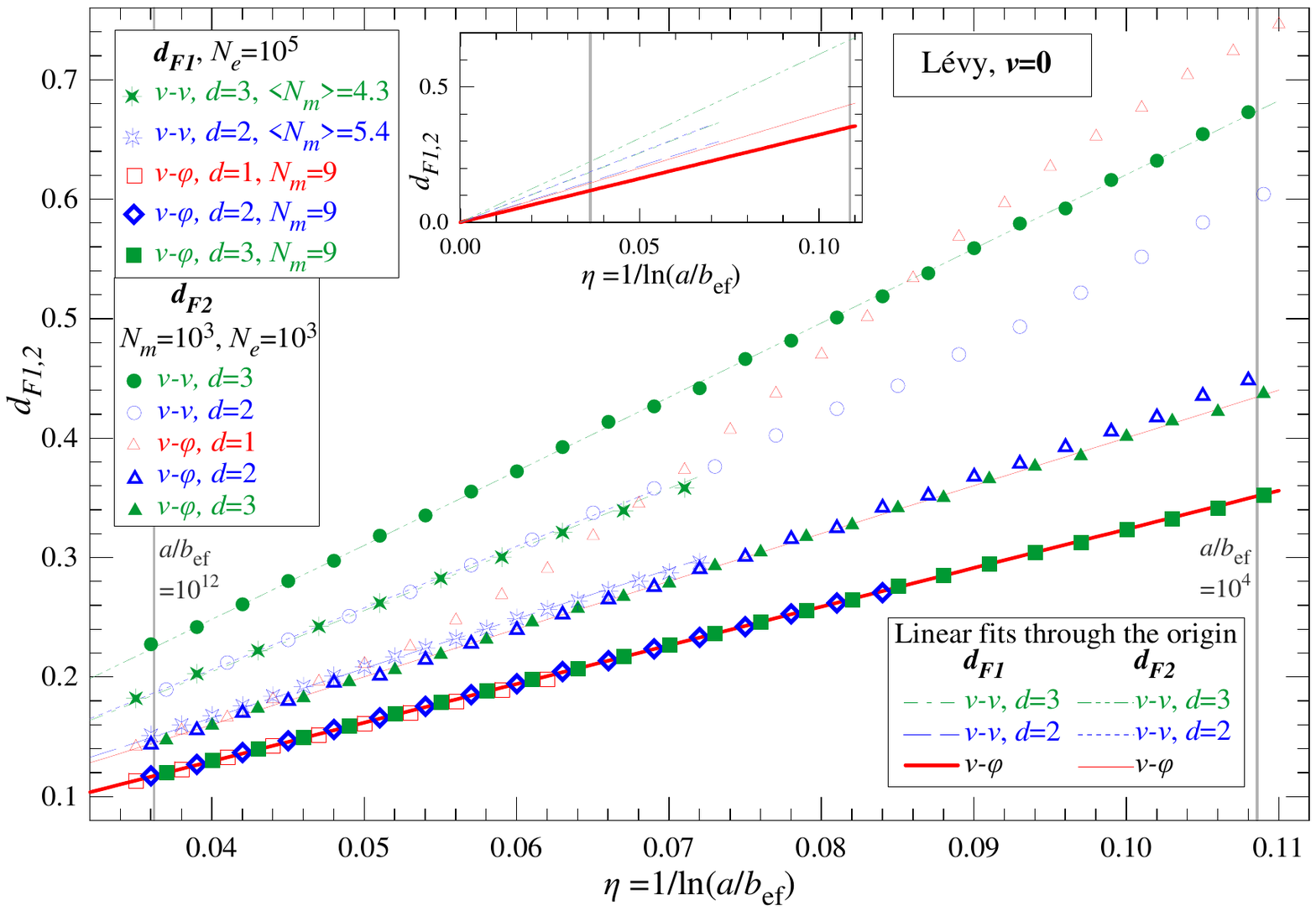}
\caption{\label{fig:nu=0_dF1,2} {\small 
The values of the slopes of the type $d_{F1}$ and $d_{F2}$ for L\'{e}vy walks with 
step probability exponent $\nu=0$ for $\nu-\phi$ and $\nu-\nu$ cases and $d=$1,2,3 dimensions
as function of the parameter $\eta$.
Slopes $d_{F2}$ are 
calculated at scales $R_{m,1}$ for all $\nu-\phi$ cases and $R_{m,d}$ ($d=$2,3) for $\nu-\nu$ cases.
Slopes $d_{F1}$ are calculated at chosen scale intervals greater than $R_{m,d}$ ($d=$1,2,3).
All slopes as function of $\eta$ show a proportional function relation up to a higher value to
$\eta$ before they begin to rise more rapidly. The slopes for $d=$1 start to deviate
at lower $\eta$ compared to the slopes for $d=$3.
Lines represent fits of the form $d_F= A \eta$ and are depicted up to the scale where these fits hold.
In the embedding graph only the lines of the fits are depicted, extrapolated up to $\eta \simeq$0.}}
\end{figure}

\noindent
of the form
\begin{equation}\label{eq:d_F-eta}
d_{Fi} \approx \frac{A_{ijd}}{\ln \left( {\frac{a}{b_{ef}}} \right)} 
\end{equation}

In last relation we encode the type of the slope with the index $i=$1,2, with $j$=1 the $\nu-\nu$ 
case, with $j$=2 the $\nu-\phi$ case and with $d$ the embedding space.
Performing a fit on the calculated values of $d_{F1,2}$ for the $\eta$ interval where the 
proportionality relation holds we are able to determine the constants $A_{ijd}$.
Their values are recorded in Table \ref{tab:d_F}, along with the intervals in each case where
eq.~(\ref{eq:d_F-eta}) holds. For $j=$2 it is evident from Fig.~\ref{fig:nu=0_dF1,2} that the 
slopes $d_{Fi}$ follow the same relation for all $d$. So we perform a common fit to all $d$
in these cases.
As the variable $\eta$ tends to zero the ratio $a/b_{ef}$ tends to infinity and at this point all
slopes $d_{Fi}$ will tend to zero. Thus, the L\'{e}vy walks with $\nu=$0 will achieve a minimum 
zero slope. However, long before that occurs, we will exhaust machine accuracy of our computer.
As the variable $\eta$ increases, the relation (\ref{eq:d_F-eta}) gradually stops to hold, the lower
and upper limits approach each other and the slopes $d_{Fi}$ start to rise towards the 
embedding space dimension $d$. We observe that for $d$=1 the slopes $d_{Fi}$ depart from
relation (\ref{eq:d_F-eta}) at lower $\eta$, for $d=$2 this occurs at a higher $\eta$, while for 
$d=$3 this occurs at even higher value of $\eta$. Even if the relation does not hold for
higher $\eta$, we can still locate slopes $d_{Fi}$ manually.

\begin{figure}[H]
\centering
\vspace{-0.0cm}
\hspace{-0.cm}\includegraphics[scale=0.6,trim=0.4in 0.3in 0.7in 0.2in,angle=0]{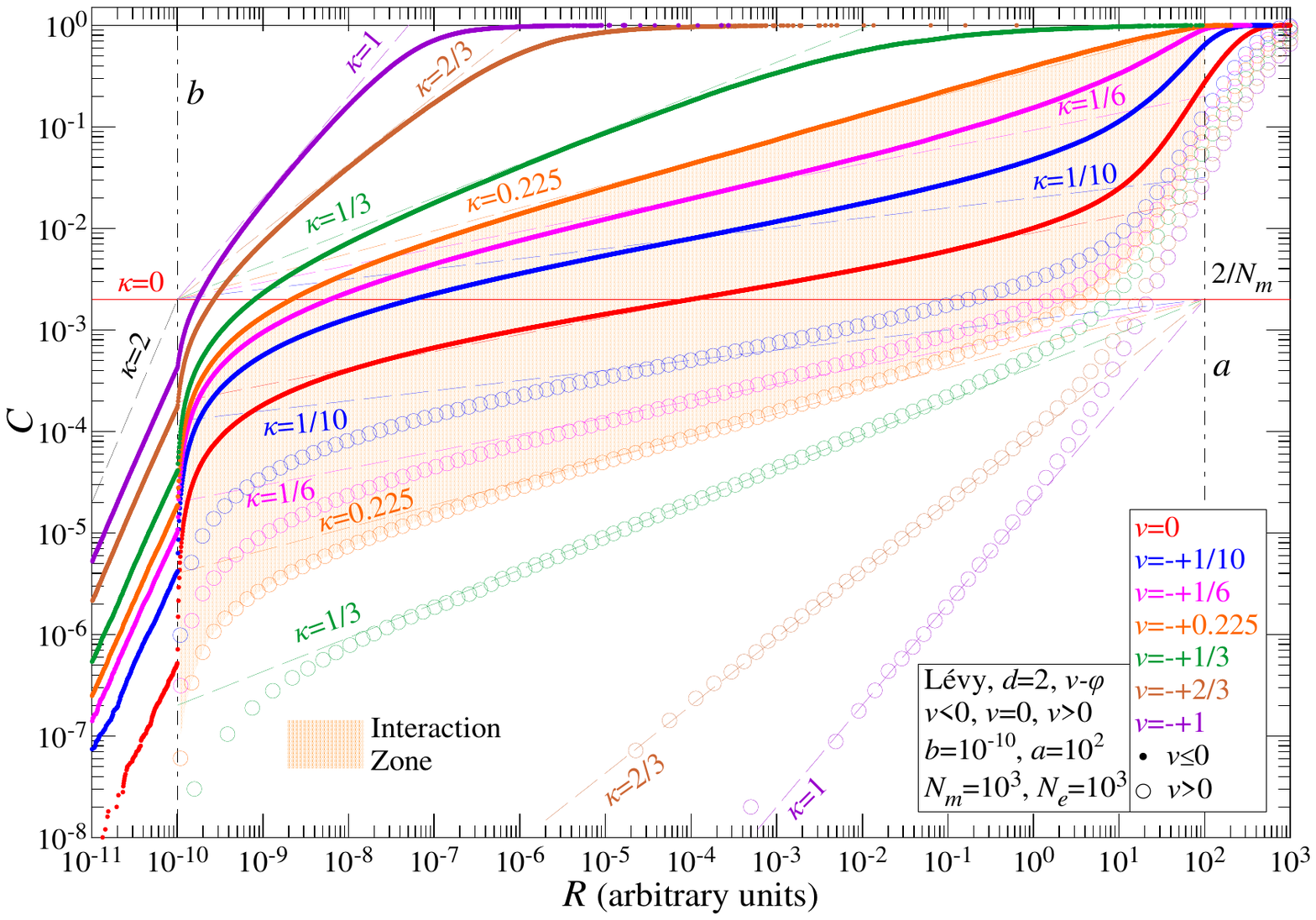}
\caption{\label{fig:int_zone} {\small
L\'{e}vy walks for the $\nu-\phi$ case, embedding spade dimension $d=$2, fixed number of steps
$N_m=10^3$, fixed step limits $a$ and $b$ for varying step exponent $\nu$ which ranges from
negative to positive values.
At higher values of $|\nu|$ the expected slope is acquired. For lower values of $|\nu|$ the slopes
stay higher than the expected value. The values of $\nu$ for the certain step limits $a$ and $b$
at which the acquired slopes start to deviate from the expected values mark the limits of an
``interaction zone''. Within this zone the boundaries ``interact'' considerably to change
the slope. The minimum slope acquired is the one for $\nu=$0 and it tends to zero only for
$a/b \rightarrow \infty$.
The graph remains unchanged in the rescaling $a'=\lambda a$, $b'=\lambda b$ and $R'=\lambda R$.
}}
\end{figure}

Let us consider now the situation where we have L\'{e}vy walks with negative $\nu$ and
high number of steps and have 
activated both limits, $a$ and $b$. We gradually diminish the absolute value of $|\nu|$.
This should give us the behaviour of the zero $\nu$ case which we saw that it cannot
acquire the zero slope anywhere between the limits.
After reaching $\nu$=0, if we keep increasing $\nu$ we can pass to the positive values.
In Fig.~\ref{fig:int_zone} we have plotted the results for this situation for the $\nu-\phi$ case 
and for $d=$2. 
Starting with negative $\nu$ away from the zero value and sufficiently
high number of steps we can achieve the predicted $|\nu|$ slope. However, at some limiting value
$-|\nu_c|$ the produced slope starts to deviate from the predicted value $|\nu|$, retaining
higher values. The situation is the same for positive values of $\nu$.
For the $\nu-\phi$ case we find that the value of positive $\nu$ until which there is deviation from 
the predicted slope is again $|\nu_c|$.
The territory in the $C(R)$ diagram where the slopes deviate from the predicted $\nu$ slope define a 
space which we shall call ``interaction zone''.
This is the outcome of the presence of both limits which 
for sufficiently low $\nu$ begin to ``interact'' and change the shape of the distribution. 
We call this phenomenon ``boundary interaction''. 
We can calculate approximately the value of $\nu_c$ which defines the onset of the interaction zone
in the $\nu-\phi$ case.
The approximated form of $C(R)$ is given for negative $\nu$ by eq.~(\ref{eq:nn_ap_nuph}).
If this function reaches unit at a scale lower than the upper limit $a$, no distortion is
observed. The distortion starts when this function meets the unit value at a scale which equals
$a$, so we may set: 
\begin{equation}\label{eq:nu_c}
\frac{2}{N_m} \left( {\frac{a}{b}} \right)^{\nu _c} = 1 \Rightarrow 
\left( \frac{a}{b} \right)^{\nu _c} = \frac{N_m}{2} \Rightarrow 
{\nu _c}\ln \left( \frac{a}{b} \right) = \ln \left( \frac{N_m}{2} \right) \Rightarrow 
\nu _c = \frac{\ln \left( \frac{N_m}{2} \right)}{\ln \left( {\frac{a}{b}} \right)}
\end{equation}
For the situation of Fig.~\ref{fig:int_zone}, where $N_m=10^3$ and $a/b=10^{12}$, eq.~(\ref{eq:nu_c})
gives us $\nu_c=$0.2249 which is about the value we find from our simulations.
Thus, it is inferred that the minimum slope we can achieve for a specific choice of limits $a$ and $b$ 
is the one given by the $\nu=0$ case.

If we do not keep both limits present, but we keep only the necessary one, we may still not be able
to attain a slope as close to zero as we want. The very low slope means that our distribution
passes through scales more ``quickly'', so the scale interval where our distribution lives becomes
vast. Around the necessary limit, $a$ or $b$, a scale interval is spend before reaching the predicted
slope. This interval may again become vast as $|\nu|$ diminishes.
So we may exhaust our machine accuracy before acquiring the desired slope.

\section{The limits revisited} \label{sec:lim_pm}

In section \ref{sec:dif_mul} we saw that the $C$ distributions of L\'{e}vy walks in the
$\nu-\phi$ case with only
the necessary limit present are approximated by a linear part given by formula (\ref{eq:nn_ap_nuph})
for negative $\nu$ and (\ref{eq:pn_ap}) for positive $\nu$.
So we may ask under what conditions we can produce $C$ distributions with different sign of $\nu$ 
which have the same approximate linear part.
Indeed, such a task is possible only for the same absolute value of $\nu$, but we have to use different number of multiplicities for each
sign of $\nu$. Denoting as $N_{m-}$ and $N_{m+}$ the multiplicities for negative and positive
$\nu$, respectively and equating the right hand side of the equations (\ref{eq:nn_ap_nuph}) and
(\ref{eq:pn_ap}), we get:

\begin{equation}\label{eq:npn_ap}
\frac{2}{N_{m-}} \left( \frac{R}{b} \right)^{|\nu|} =
\frac{2}{N_{m+}} \left( \frac{R}{a} \right)^{\nu}
\Rightarrow
N_{m-}=N_{m+} \left( \frac{a}{b} \right) ^{|\nu|}\;.
\end{equation}

If we now impose the additional limits for each case, we may set the additional limit for negative
$\nu$ to be equal to the necessary limit for positive $\nu$ and vice versa. We can, also, 
set the multiplicities to follow equation (\ref{eq:npn_ap}). If we do that we find that
the $C$ distributions for positive and negative $\nu$ almost coincide for the $\nu-\phi$ case.
It is remarkable that, although the slope changes due to the presence of an additional limit,
the distributions with opposite sign of $\nu$ ``react'' in the same way, thus, producing almost
identical curves. This finding holds for all embedding space dimensions $d=$

\begin{figure}[H]
\centering
\vspace{-0.5cm}
\hspace{-0.5cm}(i)\includegraphics[scale=0.57,trim=0.4in 0.3in 0.7in 0.2in,angle=0]{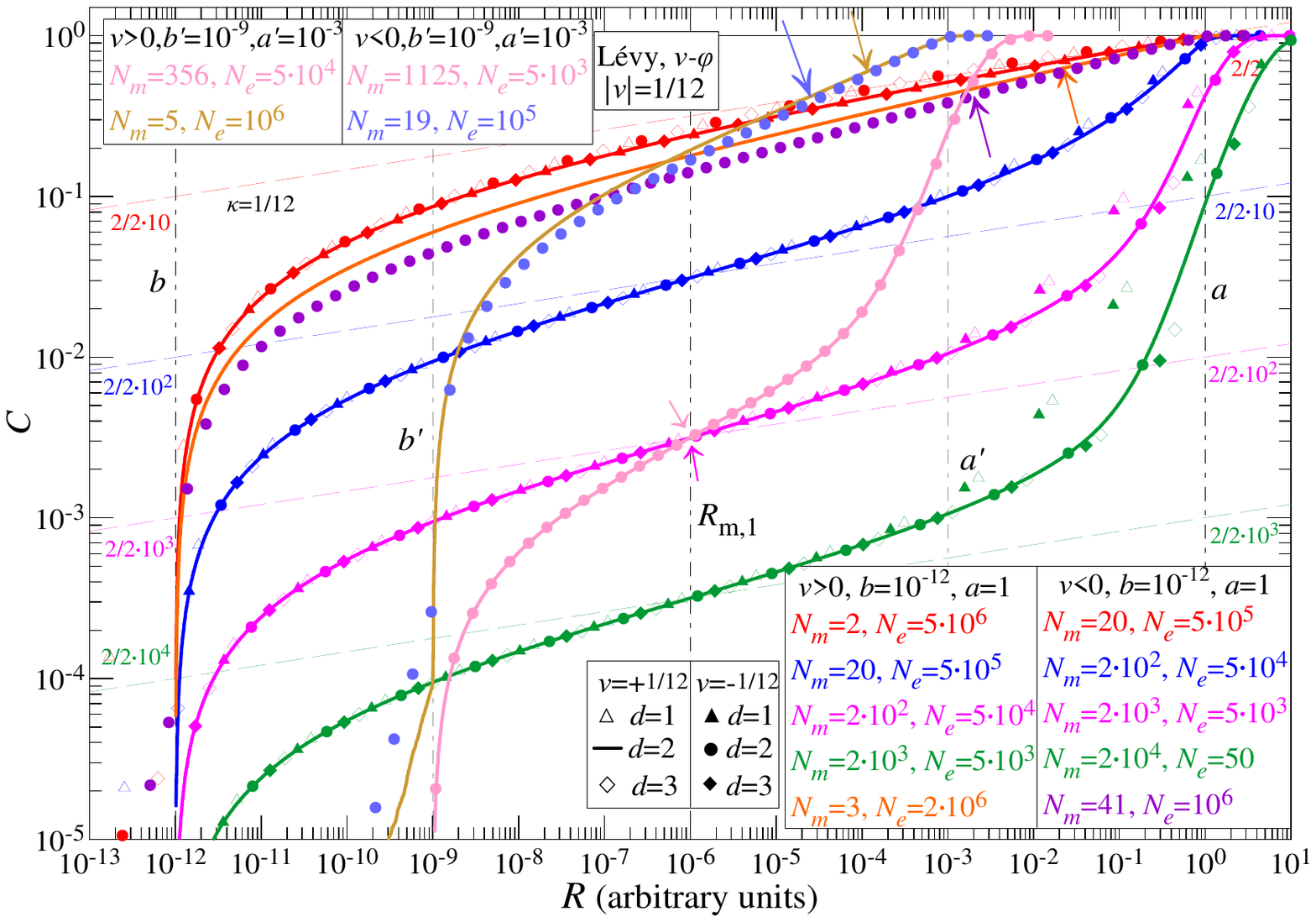}\\
\hspace{-0.7cm}(ii)\includegraphics[scale=0.57,trim=0.4in 0.3in 0.7in 0.2in,angle=0]{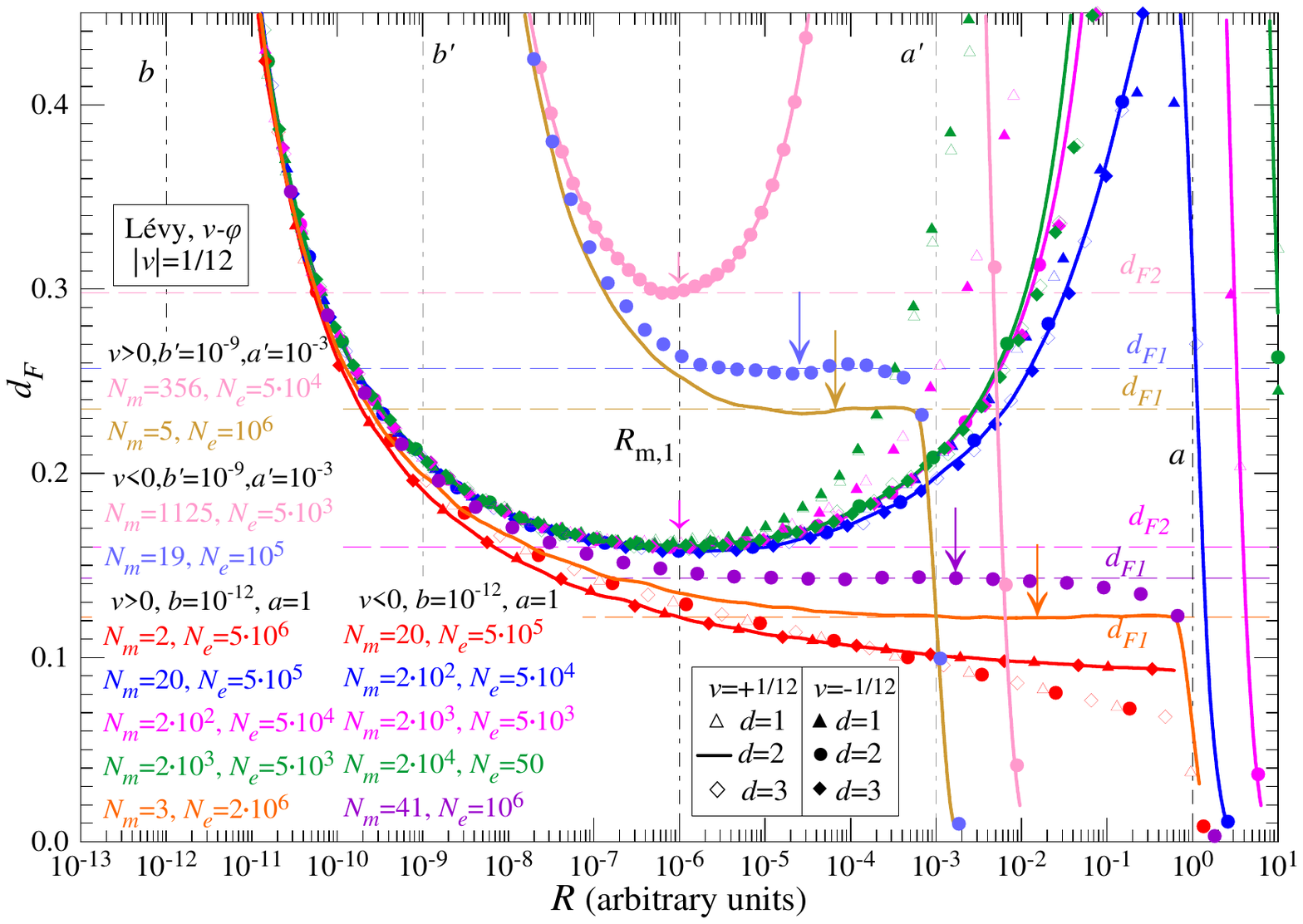}
\caption{\label{fig:pmnu_lim} {\small L\'{e}vy walks with 
step probability exponent $|\nu|=1/12$ and two signs of $\nu$ for the $\nu-\phi$ case, $d=$1,2,3 dimensions, fixed step 
limits $a$ and $b$ and with 
varying number of steps. 
Another case for $d=$2 and different step limits $a'$ and $b'$ is depicted.
At high number of steps the slope at $R_{m,1}$ stabilises at $d_{F2}$ 
(short arrows).
At a certain low number of steps a slope of type $d_{F1}$ is formed (long arrows). 
The $C$ distributions are depicted in (i), while the local log-log slopes, $d_F$, are depicted in (ii).
The graphs remain unchanged in the rescaling $a''=\lambda a$, $b''=\lambda b$ and $R''=\lambda R$.}}
\end{figure}

\begin{figure}[H]
\centering
\vspace{-0.5cm}
\hspace{-0.5cm}(i)\includegraphics[scale=0.57,trim=0.4in 0.3in 0.7in 0.2in,angle=0]{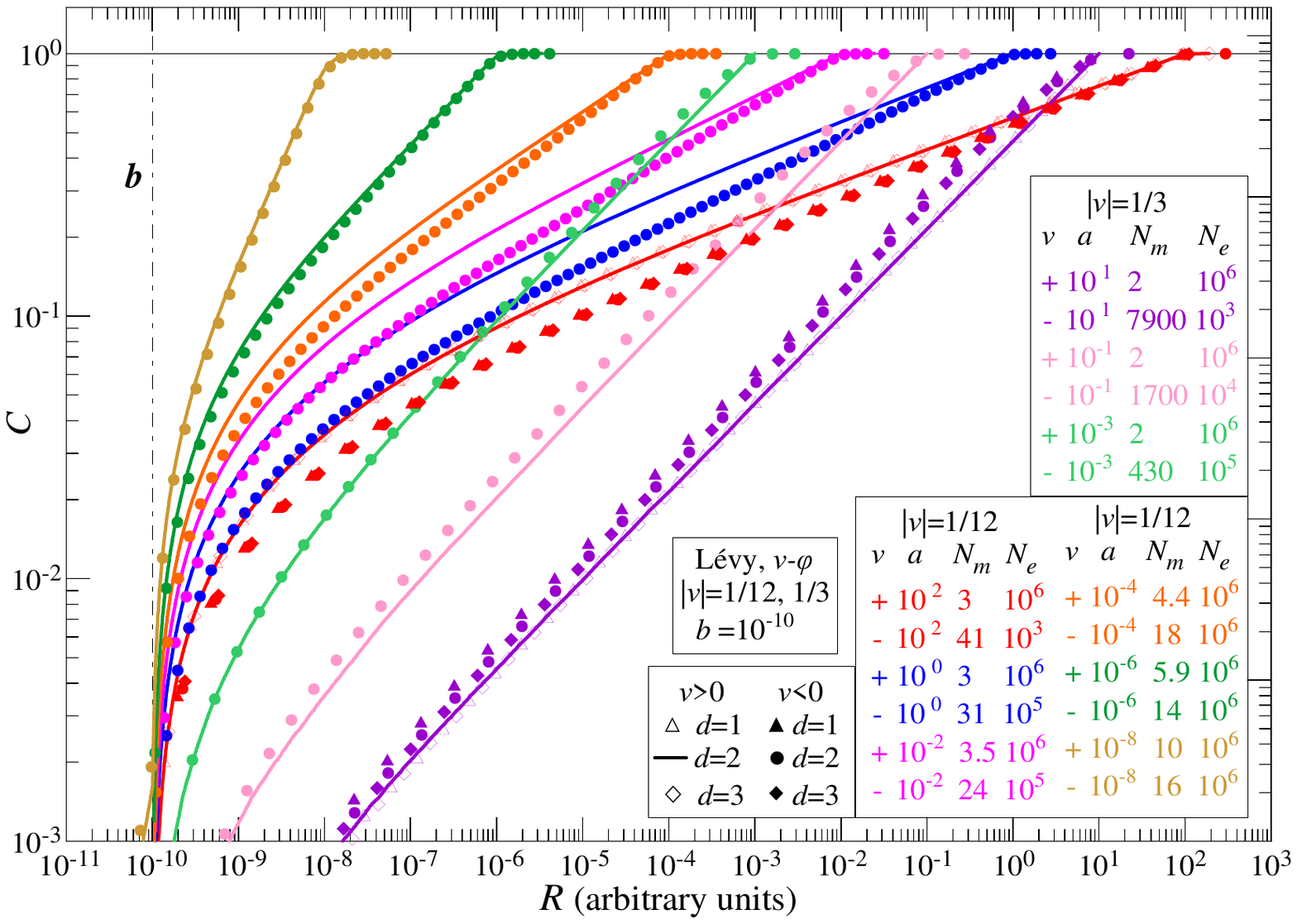}\\
\hspace{-0.7cm}(ii)\includegraphics[scale=0.57,trim=0.4in 0.3in 0.7in 0.2in,angle=0]{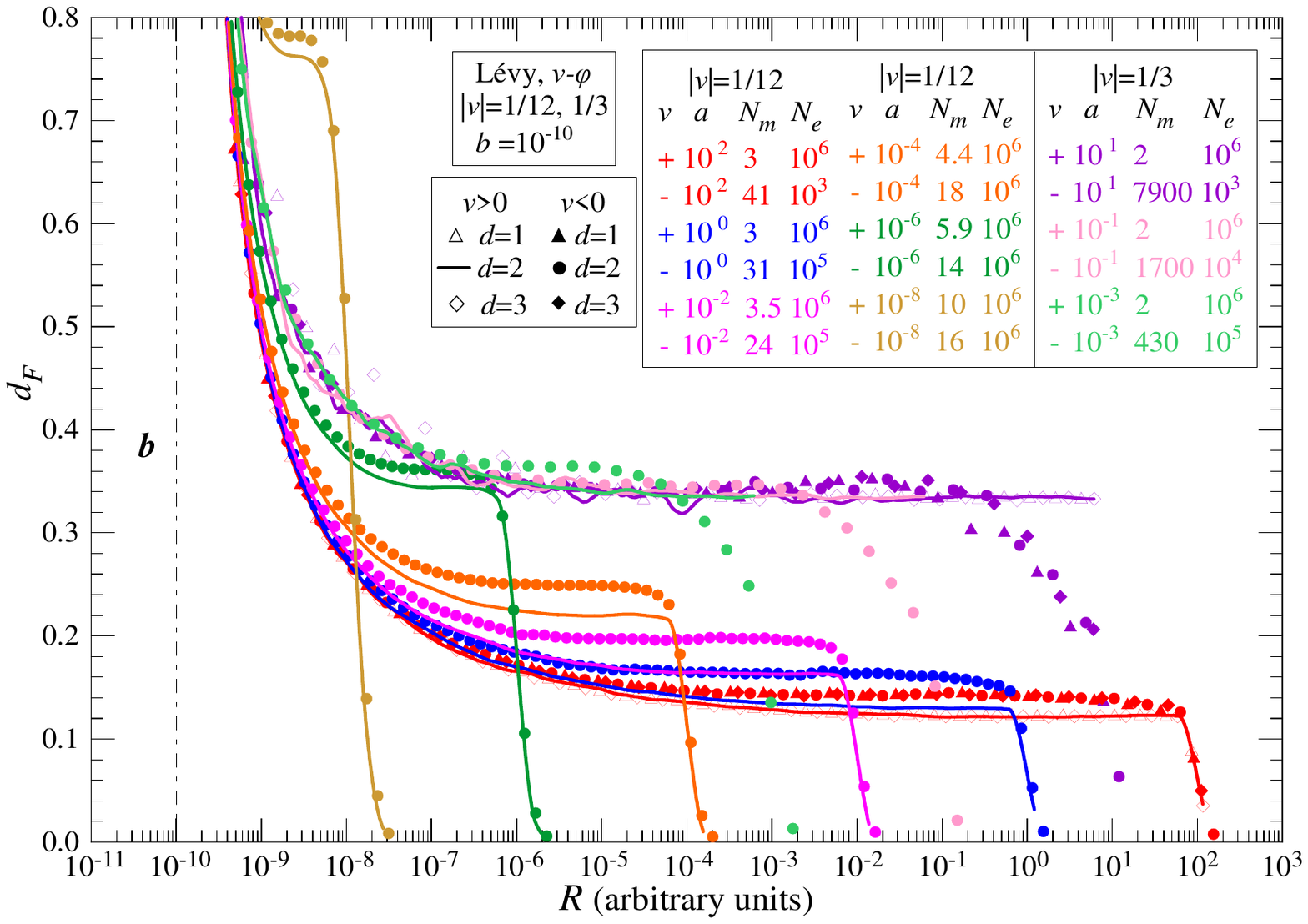}
\caption{\label{fig:pmnu_lim_df1} {\small L\'{e}vy walks with 
step probability exponent $|\nu|=$1/12 and 1/3 and two signs of $\nu$ for $\nu-\phi$ cases and $d=$1,2,3 dimensions, 
fixed lower step limit $b$ and varying upper step limit $a$ which produce for certain number of
steps a slope of the type $d_{F1}$ very close to scale $a$. 
For high values of $a$ the needed number of steps for $\nu>0$ is $N_{m+}$=3(2) for $\nu=$1/12(1/3).
The $C$ distributions are depicted in (i), while the local log-log slopes, $d_F$, are depicted in (ii).
The graphs remain unchanged in the rescaling $a'=\lambda a$, $b'=\lambda b$ and $R'=\lambda R$.}}
\end{figure}

\noindent
1, 2, 3.
 
 In this way we have a connection between L\'{e}vy walks with different sign of
step exponent and we can use them both to produce similar distributions.
Of course, since in the presence of both upper and lower limits $a>b$, $N_{m-}>N_{m+}$, 
greater multiplicity for negative $\nu$ is needed to produce distribution with the same 
characteristics with the distribution for positive $\nu$.

In Fig.~\ref{fig:pmnu_lim} we present the results for L\'{e}vy walks with the same $|\nu|$, but
 different signs of $\nu$ and the 
same upper $a$ and lower $b$ limits. We have chosen $a/b=10^{12}$ and $|\nu|=1/12$, so, according to
eq.~(\ref{eq:npn_ap}), $N_{m-}=10N_{m+}$.
We observe that, as the multiplicity increases, the distributions in the log-log plot start to move
downwards in a parallel way, retaining the same slope for a specific scale.
This figure can be treated as the $\nu\neq 0$ generalisation of Fig.~\ref{fig:nu=0_Nm_nu-phi}  
which stands for $\nu=0$. The slashed lines in Fig.~\ref{fig:pmnu_lim}(i) represent the common
approximate linear part with absent the unnecessary limit for both signs of $\nu$ with the 
multiplicities given by
 eq.~(\ref{eq:npn_ap}). These lines pass through the points  $\left ( a, \frac{2}{N_{m+}} \right )$ at the right end
and $\left ( b,\frac{2}{N_{m-}} \right )$ at the left end, while their slope is $|\nu|$.
The corresponding slashed lines in Fig.~\ref{fig:nu=0_Nm_nu-phi} are horizontal, since $\nu=0$ there.
The lower slope between scales $a$ and $b$ is found again at about $R_{m,1}$ given by
 eq.~(\ref{eq:R_m1}).
In Sec.~\ref{sec:nu=0} for $\nu=0$ we formed linear $C$ distributions at large scales for certain
multiplicities. The same is true for $\nu\neq 0$. But we can form such a linear part for positive
$\nu$ or for negative $\nu$. 
The behaviour of $C$ distributions for positive and negative $\nu$ can be
slightly different at maximum scales, although the relevant multiplicities satisfy
eq.~(\ref{eq:npn_ap}). So this equation can serve as a crude estimation of $N_{m-}$ for given $N_{m+}$
before the final adjustment which will give the final value of $N_{m-}$.
In Fig.~\ref{fig:pmnu_lim} we depict two pairs of cases of linear $C$ distributions near the maximum
scales for two values of $a/b$.

Since the linearity is interesting subject in our analysis, in Fig.~\ref{fig:pmnu_lim_df1} 
we depict results for 
two values of $|\nu|=$1/12 and 1/3 and each for different values of $a/b$.
For the low value of $\nu=$+1/12, the necessary multiplicity is $N_{m+}=3$
\footnote{One can compare this value with $N_m=9$ for $\nu=0$ in Sec.~\ref{sec:nu=0}.} 
for large value of the
ratio $a/b$ and increasing tendency as this ratio decreases. 
For the higher value of $\nu=$+1/3, the necessary multiplicity has reached the lowest value 
$N_{m+}=2$ (it is known that for $N_m=2$ the $C$-distribution is a perfect straight line in the
log-log plot right from the higher scales).
The multiplicities $N_{m-}$, in any case, can be crudely calculated by eq.~(\ref{eq:npn_ap}).

For the $\nu-\nu$ case ($d$=2,3) the $N_{m+}$ and $N_{m-}$ multiplicities  can be calculated as in the 
$\nu-\phi$ case, but using eq.~(\ref{eq:nn_ap_nunu}). However, the $C$ distributions for 
the two signs of $\nu$ in the $\nu-\nu$ case
differ and are not almost alike as in the $\nu-\phi$ case. Differences arise which can be attributed to
the different behaviour for positive and negative $\nu$.
At this point we would like to remind the reader that for positive $\nu$ we have to perform walks 
in each dimension with exponent
$\nu'=\nu/d$ to achieve a slope of $\nu$, whereas for negative $\nu$ the situation is that $\nu'=\nu$.

\section{Applications for $\bf d_F$=1/3}\label{sec:d_F=1/3}

We shall use the accumulated information of the previous sections to perform an application
where the desired slope of a data set has to be $1/3$ in an embedding space of dimension $d=$2.
This represents a fractal dimension which
is a signature of the QCD critical point 
\cite{{intermittency},{Antoniou2006}}.
Also, in this application we must have low
multiplicity events. This situation is the one which appears in heavy-ion collisions where experiments
search for critical correlations \cite{NA49_protons}.

In such a case we want to simulate an experimental data set where in different events there is
a mixture of measurements of a physical quantity (e.g. momentum), called tracks.
Some of the tracks are considered to represent ``noise'', i.e. they smear the signal which is
to be extracted and other are considered to be ``critical'' and carry the relevant signal.
A typical probability of the presence of critical tracks is of the order of $p \sim$0.01, while the 
average number of tracks per event is of the order of $<N_m> \sim$3.

\begin{figure}[H]
\centering
\vspace{-0.0cm}
(i)\includegraphics[scale=0.6,trim=0.4in 0.3in 0.7in 0.2in,angle=0]{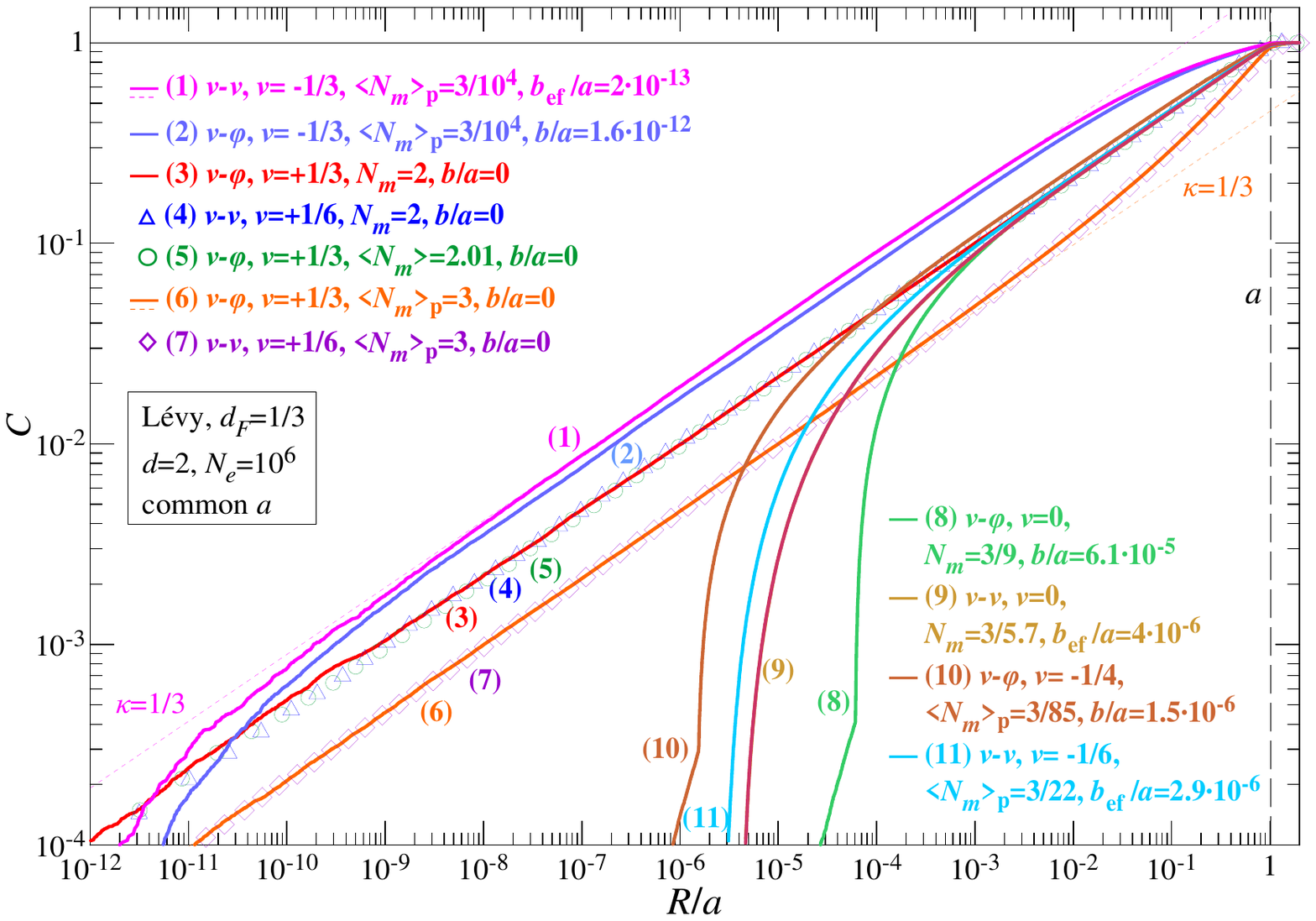}\\
(ii)\includegraphics[scale=0.6,trim=0.4in 0.3in 0.7in 0.2in,angle=0]{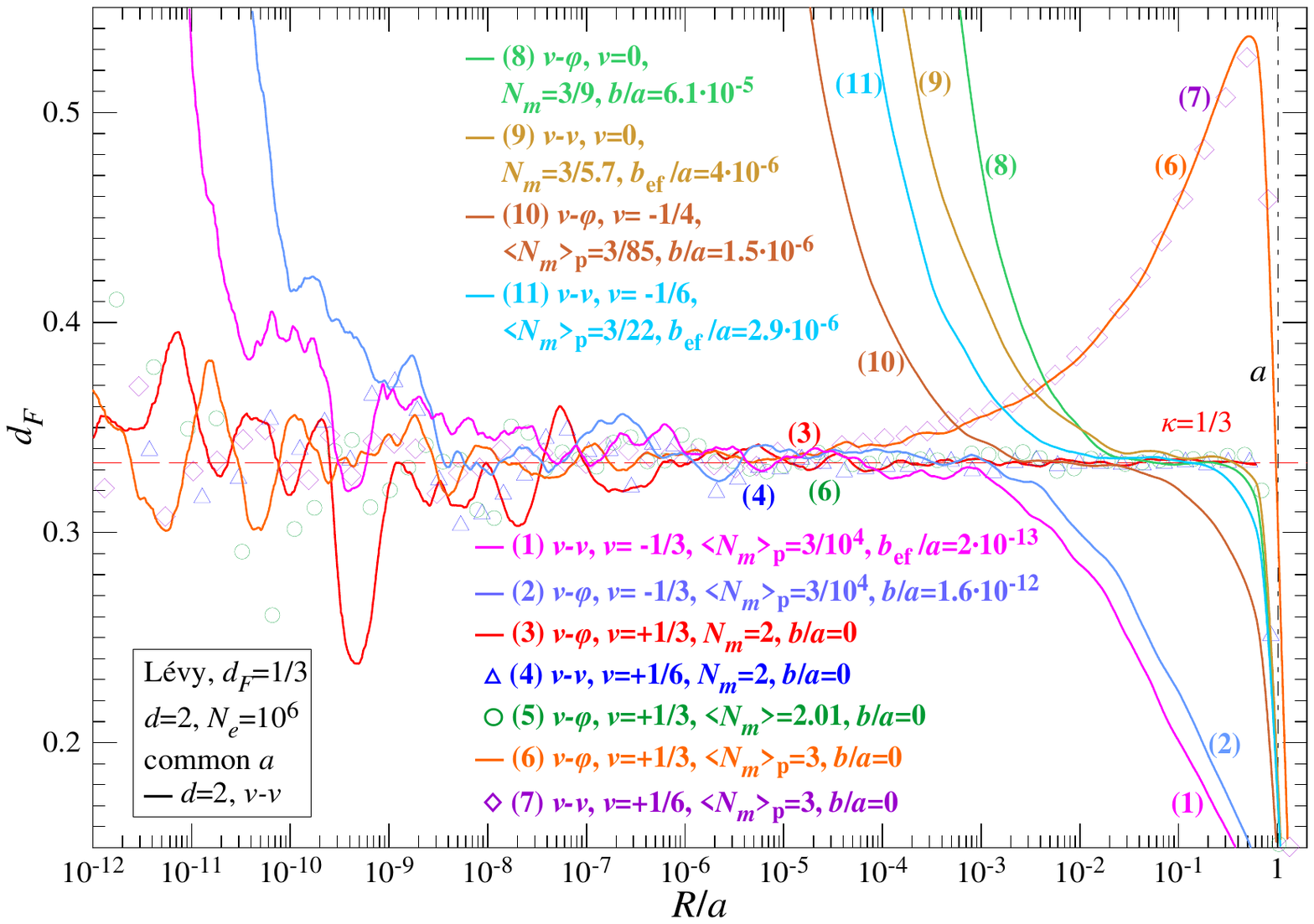}\\
\end{figure}

\begin{figure}[H]
\centering
\vspace{-0.0cm}
(iii)\includegraphics[scale=0.6,trim=0.4in 0.3in 0.7in 0.2in,angle=0]{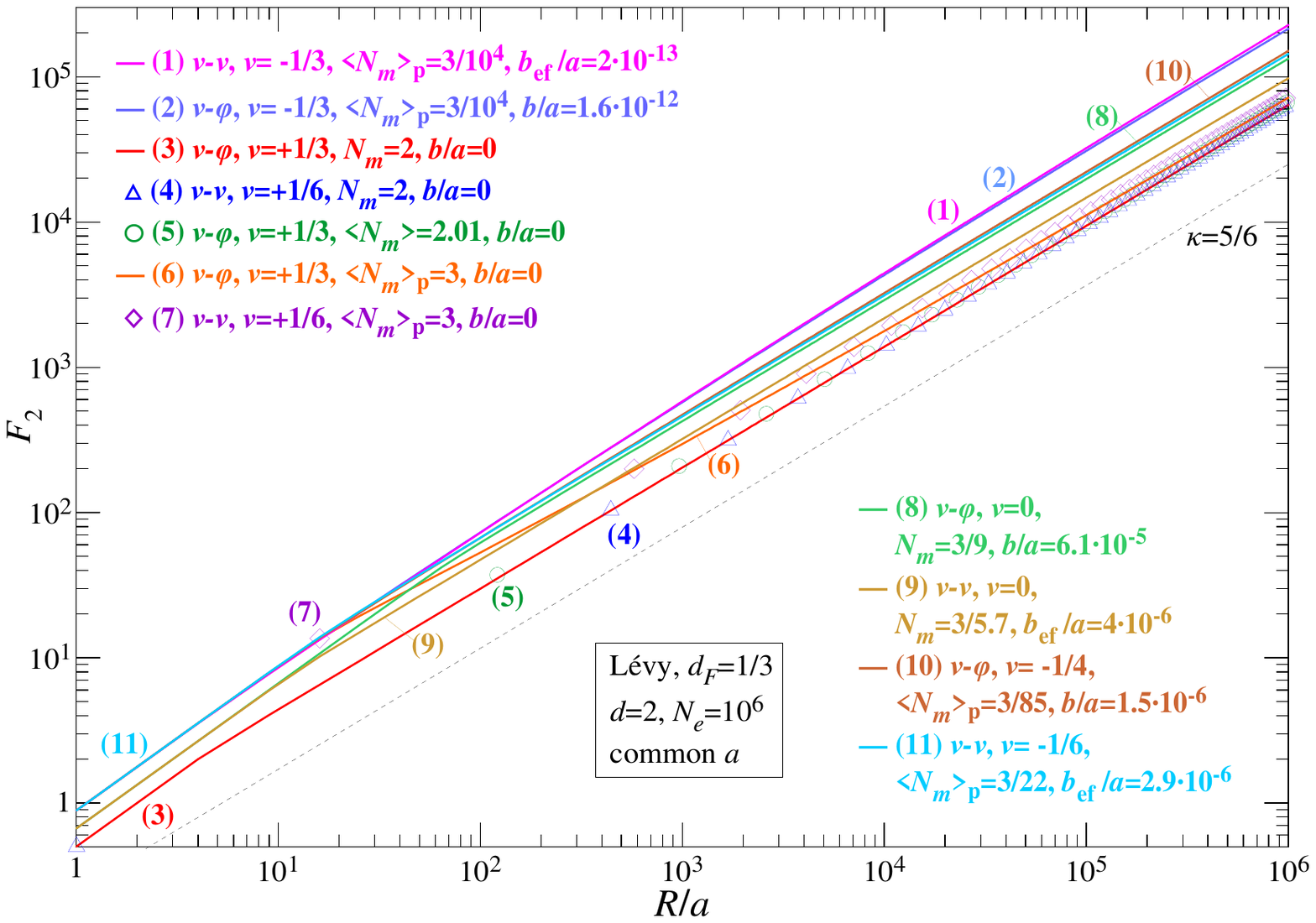}
\caption{\label{fig:d_F=1/3} {\small 
Various methods to produce a log-log slope in the $C(R)$ distribution equal to 1/3 
with L\'{e}vy walks at an embedding space dimension $d=$2. 
Curves (1)-(2) and (10)-(11) correspond to $\nu<0$, curves (3)-(7) to $\nu>0$ and
curves (8)-(9) to $\nu=0$. All walks have the same upper step limit $a$.
The walks (3)-(7) have no lower step limit, while the rest do.
(i) The correlation integral $C$ of each set as function of the scale $R/a$
in log-log plot. The linear asymptotic lines for the $C$ distributions of curves
(1) and (6) are shown.
(ii) The local slopes $d_F$ of the walks presented in (ii).
(iii) The second order factorial moments $F_2$ of the curves presented in (i), where
the expected slope should be $1-(d_F/d)=$5/6, calculated through the correlation integral with 
the ring technique developed in \cite{Kapoyannis2022}.}}
\end{figure}

To achieve this we must have data sets which show the ``critical'' behaviour and combine them with
data sets which behave like noise. We will focus here on various methods to produce critical
data sets which for a scale interval their correlation integral shows in the log-log plot
a slope of specific value, in this case $1/3$. Our results are shown in Fig.~\ref{fig:d_F=1/3}.

In Section \ref{sec:dif_mul} we saw that for negative $\nu$ the $C$ distribution acquires the 
slope $|\nu|$ for high number of steps. In Section \ref{sec:alter_Nm} we show that if we pick
randomly a smaller number of points from a set of
number of points produced from a 
walk, then the two sets follow the same distribution. Combining these two facts, we can
produce L\'{e}vy walks with high number of steps, $N_m=10^4$ and pick some steps from it.
Thus, we can produce events with multiplicity that follows e.g. a Poisson distribution with 
$<N_m>_p=$3. All the events, even if they have different number of tracks, since they are all drawn 
from the same walk, follow the same distribution, that of the walk of the $10^4$ steps.
Also, we set an upper limit $a$, sufficiently higher than $b$, which will exclude wildly high values
and improve the high boundary phenomena, bringing the $C$ distribution closer to its asymptotic 
form. We use the $\nu-\phi$ (line (1), along with its asymptotic line), 
as well as the $\nu-\nu$ (line (2)) case.
We see that in this way we can have fractal dimension $d_F$ exhibited for a vast interval of
scales. Moreover, since the sets of each event are drawn from a high step walk, we do not have to 
worry whether the upper number of steps needed to be drawn reaches the number of total number
of the complete walk. 

In Section \ref{sec:dif_mul} we saw that for positive $\nu$ the $C$ distribution acquires the  
prescribed slope, $\nu$ or $d\nu$ for cases $\nu-\phi$ and $\nu-\nu$ respectively, which, once reached, 
extends to infinitely low scales. For two steps a perfect slope in immediately reached below the
upper limit $a$. In the Appendix we show that for fixed number of total tracks per event, $N_m$,
the mean value of critical tracks is $pN_m$. For $p=$0.01 and $N_m=$3 this 
gives an average number of critical tracks per event $pN_m$=0.03, which is indeed much lower than 2. 
Also, the probability for needing to draw 3 critical tracks in total of 3 tracks, is $p^3=10^{-6}$.
Excluding this possibility is equivalent in neglecting 1 event in total of $10^6$ events.
So we can form events containing critical and noisy tracks
by drawing the critical tracks from a distribution of positive $\nu$ and $N_m=$2, thus resulting to
a perfect slope at all scales which represents the critical part of our events.
The relevant distribution of critical tracks alone is curve (3) for $\nu-\phi$ case and curve (4)
for $\nu-\nu$ case.

In the Appendix we, also, show that for a number of total tracks per event which follow a Poisson
distribution (common phenomenon in heavy-ion collisions) with a mean value $\tau$,
the mean value of critical tracks is again $p\tau$. In this case, the ratio of the events
with 3 critical tracks to the ratio of events with 2 critical ones (see Appendix) is 
$\frac{p\tau}{2+1}=\frac{0.03}{3}=0.01$. For 3 fixed tracks this ratio is even lower 
(also, see Appendix): $\frac{1}{\tau}\frac{p}{1-p}=\frac{0.01}{3 \cdot 0.99}=0.0034$.
We can then draw critical tracks from a distribution with $N_{m,c}=$2 if we need 2 critical ones and 
a distribution with $N_{m,c}=$3 if we need 3 critical ones. The critical part of the distribution
will be one with average at most $<N_{m,c}>=$2.01. This is very close to the previous cases and it
is represented by line (5) in Fig.~\ref{fig:d_F=1/3}.

Next we turn to a case where we have fully critical events with multiplicities that follow a
Poisson distribution with average $<N_m>_p=$3. These events can be used with a certain percentage 
along with other events that represent the noise to simulate the experimental data. 
For this reason when a certain multiplicity
is chosen we produce a L\'{e}vy walk with $\nu>0$ and exactly the same number of steps. 
Our distribution is then an one of events with different multiplicities. The set of events with
specific number of tracks follows its own distribution, different from the rest. However, all the
sets have asymptotic curves with the same slope, so the mixture of all the sets retains the same slope.
The exact asymptotic curve is calculated in the Appendix. We present in Fig.~\ref{fig:d_F=1/3} two
curves of this kind, line (6) for a $\nu-\phi$ case which is plotted along with its asymptotic line 
and line (7) for a $\nu-\nu$ case.

In the following we utilise the findings of Section \ref{sec:nu=0} to produce distributions with slope 
of the $d_{F1}$ type for $d=$2. These distributions have the ability to acquire the needed slope
at scales very close to the upper limit $a$.
So for the $\nu-\phi$ case, we use the L\'{e}vy walk with $\nu=$0 and 
9 steps. Through eq.~(\ref{eq:d_F-eta}) we evaluate the necessary ratio $a/b_{ef}$ to produce
$d_{F1}=1/3$ for this case. Then we draw randomly 3 steps from the 9 step walk. The result is
shown as curve (8) in Fig.~\ref{fig:d_F=1/3}.
For the $\nu-\nu$ case the distribution which has the needed form is one which has events with
5 step L\'{e}vy walks with probability $r_1\simeq$0.39 and 6 step L\'{e}vy walks with probability 
$1-r_1\simeq$0.61. This gives and average multiplicity of about 5.61 steps. 
We are slightly above the average of 5.4 steps which produce a linear part in
$C$ for $d=2$ in the $\nu-\nu$ case.
These two walks, with their relative ratio, combine to give our final $C$ distribution.
We want each of our events to have 3 tracks. So we randomly draw 3 from the 5 step walk and the
corresponding distribution follows that of the 5 steps. The same holds for the distribution of the 
3 tracks taken out of the 6 ones. However, when the two distributions combine the outcome depends,
also, on the number of pairs in each event. In the Appendix we show how we can combine the two 
different 3 step walks in such a way as to arrive at the same distribution with the case of the
5 and 6 step walks. So in this case where we pick 3 steps, the walks of 5 and 6 steps have to
be used with probabilities 0.3 and 0.7, instead of 0.39 and 0.61, as if the original average was 
$<N_m>=$5.7, instead of 5.61. The results are represented by curve (9).

Lastly we use the information of Section \ref{sec:ad_lim}. We want to produce slopes of the
$d_{F1}$ type using the negative $\nu$ walks. Since we want to produce a slope equal to 1/3, we 
have to begin with $|\nu|$ less than 1/3. Then we regulate the ratio $a/b_{ef}$ and the 
multiplicity $N_m$ to achieve the slope we want. Two solutions are presented in Fig.~\ref{fig:d_F=1/3}.
One corresponds to the $\nu-\phi$ case (line (10)) and one to the $\nu-\nu$ case (line (11)).
Since the two cases correspond to 85 and 22 steps, respectively, which are considerably higher than
3, we can host events with multiplicities that follow a Poisson distribution with average 3.

\section{Conclusions}\label{sec:con}

Fractals observed in nature, unlike their idealized mathematical counterparts, exhibit self-similar structure only over a finite range of scales. Truncated L\'{e}vy flights provide a convenient framework for modeling such systems, as they can generate point sets with a wide range of fractal dimensions that can be controlled in a systematic manner. In the present work, we investigate generalized L\'{e}vy-type flights with step-length distributions of the form $P(r) \sim r^{-1+\nu}$, allowing $\nu$ to take arbitrary real values. We show that the scaling properties of these L\'{e}vy-like flights (truncated or not) span a broad spectrum, enabling their application to the description of geometrical features in a wide variety of  systems.

Our analysis is not restricted to flights derived from ``stable'' distributions ($-2 \le \nu < 0$), since many systems are inherently bounded at large scales, rendering the asymptotic behavior of the distribution less relevant. We therefore consider the full range of $\nu$—negative, zero, and positive—and extend our study to embedding spaces of up to three dimensions.

We employ two distinct mechanisms for generating the walks. In the first, only the step length is drawn from the L\'{e}vy-like distribution, while the direction is determined by uniformly distributed angles appropriate to the embedding dimension. In the second, independent L\'{e}vy walks are generated along each coordinate axis, producing the corresponding multi-dimensional trajectory.

Our results reveal two regimes capable of producing fractal point sets over extended scale ranges. For negative $\nu$ and a sufficiently large number of steps, fractal scaling emerges within a finite interval defined by lower and upper bounds. From such long walks, shorter subsets can be randomly sampled while preserving the same statistical and geometric properties. For positive $\nu$, the fractal structure extends from an upper bound down to arbitrarily small scales, without a lower cutoff. Remarkably, even walks consisting of as few as two steps can generate a perfect fractal below the upper scale, making this regime particularly suitable when only a limited number of steps is available.

By introducing additional cutoffs and adjusting the number of steps, the effective scaling interval can be tuned, allowing the construction of point sets with prescribed fractal dimensions over selected ranges. The case $\nu = 0$ also plays a significant role in these constructions, exhibiting nontrivial behavior, particularly for low-multiplicity datasets.

Overall, our findings provide a flexible framework for generating simulated point sets with fractal dimensions spanning a continuous range, bounded above by the dimension of the embedding space. When both upper and lower cutoffs are imposed on the step-length distribution, the minimal attainable fractal dimension is governed by the properties of the $\nu = 0$ walks.

\section*{Acknowledgements}
KC acknowledge support by the ERC-AdG grant MOSE No. 101199196.

\vspace{0.5cm}
\appendix

\section*{Appendix}

\renewcommand{\theequation}{A.\arabic{equation}}
\setcounter{equation}{0}

We have a set containing a fixed number of $N_m$ tracks. The probability of the presence of a
critical track is $p$. The probability of having exactly $k$ critical tracks is:
\begin{equation}\label{eq:k,Nm_fixed}
P\left( {k;N_m} \right) = \left( \begin{array}{*{20}{c}}
{N_m}\\
k
\end{array} \right){p^k}{\left( {1 - p} \right)^{N_m - k}}
\end{equation}
This a binomial distribution, so the average number of of $k$ is $<k>=pN_m$.
The ratio of the probability of having $k+1$ tracks to the probability of having $k$
tracks is
\[
\frac{P\left( k+1;N_m \right)}{P\left( k;N_m \right)}
=\frac{{\left( {\begin{array}{*{20}{c}}
{{N_m}}\\
{k + 1}
\end{array}} \right){p^{k + 1}}{{\left( {1 - p} \right)}^{{N_m} - k - 1}}}}
{{\left( {\begin{array}{*{20}{c}}
{{N_m}}\\
k
\end{array}} \right){p^k}{{\left( {1 - p} \right)}^{{N_m} - k}}}} 
= \frac{\frac{N_m!}{{\left( k+1 \right)!\left( N_m-k-1 \right)!}}}{\frac{{{N_m}!}}{{k!\left( {{N_m} - k} \right)!}}} \frac{p}{1 - p} =
\]
\begin{equation}\label{eq:k+1/k,Nm_fixed}
= \frac{\frac{1}{\left( k+1 \right)}}{\frac{1}{\left( N_m-k \right)}}
\frac{p}{1 - p}
=\frac{\left( N_m - k \right)}{\left( k + 1 \right)}\frac{p}{1 - p}
\end{equation}

We have a set containing a number of tracks which follow a Poisson distribution with average $\tau$.
The probability of the presence of a
critical track is $p$. The probability of having exactly $k$ critical tracks is:
\[
P\left( {k;\tau } \right) = \sum\limits_{n = k}^\infty  {P\left( {k;n} \right)\left( {\begin{array}{*{20}{c}}
n\\
k
\end{array}} \right){p^k}{{\left( {1 - p} \right)}^{n - k}}}  = \sum\limits_{n = k}^\infty  {\frac{{{\tau ^n}{e^{ - \tau }}}}{{n!}}\frac{{n!}}{{k!\left( {n - k} \right)!}}{p^k}{{\left( {1 - p} \right)}^{n - k}}}  = 
\]
\[
= \frac{{{\tau ^k}{e^{ - \tau }}}}{{k!}}{p^k}\sum\limits_{n = k}^\infty  {\frac{{{\tau ^{n - k}}{{\left( {1 - p} \right)}^{n - k}}}}{{\left( {n - k} \right)!}}}  = \frac{{{{\left( {p\tau } \right)}^k}{e^{ - \tau }}}}{{k!}}\sum\limits_{n' = 0}^\infty  {\frac{{{{\left[ {\tau \left( {1 - p} \right)} \right]}^{n'}}}}{{n'!}}}  = \frac{{{{\left( {p\tau } \right)}^k}{e^{ - \tau }}}}{{k!}}{e^{\tau \left( {1 - p} \right)}} = 
\]
\begin{equation}\label{eq:k,Nm_Poisson}
= \frac{{{{\left( {p\tau } \right)}^k}{e^{ - p\tau }}}}{{k!}}
\end{equation}
Thus, the critical tracks follow a Poisson distribution with average $p\tau$.
The ratio of the probability of having $k+1$ tracks to the probability of having $k$
tracks in this situation is
\begin{equation}\label{eq:k+1/k,Nm_Poisson}
\frac{{P\left( {k + 1;\tau } \right)}}{{P\left( {k;\tau } \right)}} = \frac{{\frac{{{{\left( {p\tau } \right)}^{k + 1}}{e^{ - p\tau }}}}{{\left( {k + 1} \right)!}}}}{{\frac{{{{\left( {p\tau } \right)}^k}{e^{ - p\tau }}}}{{k!}}}} = \frac{{p\tau }}{{k + 1}}
\end{equation}

We need, now, to find the asymptotic form of $C$ when the multiplicities $n$ follow a Poisson distribution
with average $\tau$ and all have the same upper limit $a$.
According to the Appendix of \cite{Kapoyannis2022} in such case the correlation
integral is
\[
C(R) = \sum\limits_{n=2}^{\infty} \frac{N_{e,n}}{N_e} 
\frac{n(n - 1)}{\left\langle n (n - 1) \right\rangle_e} C_n (R)\;.
\]
In last equation, $n$ acquires values greater or equal to 2, since for $n$=0,1 there are zero pairs
in an event, so these events do not contribute to the correlation integral. We will denote
with an asterisk the Poisson distribution where the $n$=0,1 numbers are excluded.
Further $N_{e,n}$ is the number of events with multiplicity $n$, $N_e$ is the total number of 
events which do not have 0 or 1 tracks and $C_n (R)$ is the correlation integral of the $n$ tracks. 
Proceeding we have 
$\frac{N_{e,n}}{N_e}=P^*(n;\tau)$ and the average over the events becomes average of the Poisson
distribution. Then, using eq.~\ref{eq:pn_ap}, the asymptotic form of $C(R)$ becomes:
\[
C(R) \simeq \sum\limits_{n=2}^{\infty} P^*(n;\tau) 
\frac{2}{n} \frac{n(n - 1)}{\left\langle n (n - 1) \right\rangle_{p*}} 
\left( \frac{R}{a} \right)^{\nu}
= \frac{\left\langle n - 1 \right\rangle_{p*} }{\left\langle n (n - 1) \right\rangle_{p*}} 
2 \left( \frac{R}{a} \right)^{\nu}
=\frac{\left\langle n  \right\rangle_{p*} - 1 }
{\left\langle n^2 \right\rangle_{p*}-\left\langle n \right\rangle_{p*}} 
2 \left( \frac{R}{a} \right)^{\nu}
\]
We calculate the necessary average values:
\[
\left\langle n  \right\rangle_{p*}
=\frac{\sum\limits_{n=2}^{\infty} n P(n;\tau)}{1-P(0;\tau)-P(1;\tau)}
=\frac{\left\langle n  \right\rangle_p - 1 \cdot P(1;\tau)}{1-e^{-\tau}-\tau e^{-\tau}}
=\frac{\tau -\tau e^{-\tau}}{1-e^{-\tau}-\tau e^{-\tau}}
\]
\[
\left\langle n^2  \right\rangle_{p*}
=\frac{\sum\limits_{n=2}^{\infty} n^2 P(n;\tau)}{1-P(0;\tau)-P(1;\tau)}
=\frac{\left\langle n^2  \right\rangle_p - 1^2 \cdot P(1;\tau)}{1-e^{-\tau}-\tau e^{-\tau}}
=\frac{\tau^2 +\tau -\tau e^{-\tau}}{1-e^{-\tau}-\tau e^{-\tau}}
\]
So the asymptotic form becomes
\begin{equation}\label{eq:C_Poisson}
C(R) \simeq \frac{\tau -\tau e^{-\tau} - 1+e^{-\tau}+\tau e^{-\tau} }
{\tau^2 +\tau -\tau e^{-\tau}-\tau +\tau e^{-\tau}} 
2 \left( \frac{R}{a} \right)^{\nu}
\Rightarrow 
C(R) \simeq \frac{\tau - 1+e^{-\tau}}{\tau^2} 
2 \left( \frac{R}{a} \right)^{\nu}
\end{equation}

Next, we have some events with multiplicity $N_{m1}$ and probability $r_1$ and some events with
multiplicity $N_{m2}$ and probability $1-r_1$. The events of multiplicity $N_{m1}$ follow the 
correlation integral $C_1$ and the rest the correlation integral $C_2$.
The average multiplicity is
\[
\left\langle {{N_m}} \right\rangle  = {r_1}{N_{m1}} + \left( {1 - {r_1}} \right){N_{m2}}
\]
Defining
\[
{N_{pi}} = {N_{mi}}\left( {{N_{mi}} - 1} \right)\;,
\]
the correlation integral of all the events is
\[
C\left( R \right) = {g_1}{C_1}\left( R \right) + \left( {1 - {g_1}} \right){C_2}\left( R \right)\;,
\]
where
\[
{g_1} = \frac{{{r_1}{N_{p1}}}}{{{r_1}{N_{p1}} + \left( {1 - {r_1}} \right){N_{p2}}}}
\]
Then, we draw randomly $N_{m3}$ points from the set of $N_{m1}$ points. These will follow the distribution
$C_1$. We, also, draw $N_{m3}$ points from the set of $N_{m2}$ and these will follow the 
distribution $C_2$. We combine the events with probability $r'_1$ for the ones with points drawn
from the $N_{m1}$ points. Obviously the average multiplicity is $N_{m3}$. The correlation
function of all the events will be:
\[
C'\left( R \right) = {g'_1}{C_1}\left( R \right) + \left( {1 - {g'_1}} \right){C_2}\left( R \right)\;,
\]
where
\[
g'_1 = \frac{{{{r'}_1}{N_{p3}}}}{{{{r'}_1}{N_{p3}} + \left( {1 - {r'_1}} \right){N_{p3}}}} 
= r'_1
\]
We want the $C'$ to be the same as $C$, so
\begin{equation}\label{eq:r1_r'1}
{g_1} = {g'_1} \Rightarrow {g_1} = {r'_1} \Rightarrow 
{r'_1} = \frac{{{r_1}{N_{p1}}}}{{{r_1}{N_{p1}} + \left( {1 - {r_1}} \right){N_{p2}}}}
\end{equation}
This means that if we had used all the points the average multiplicity would change to
\[
\left\langle {{N_m}} \right\rangle  = {r'_1}{N_{m1}} + \left( {1 - {r'_1}} \right){N_{m2}}
\]
For initial probability $r_1\simeq$0.39 and $N_{m1}=$5, $N_{m2}=$6 and $N_{m3}=$3 we find
\[
{r'_1} = \frac{{0.39 \cdot 5 \cdot 4}}{{0.39 \cdot 5 \cdot 4 + 0.61 \cdot 6 \cdot 5}} = \frac{{0.39 \cdot 4}}{{0.39 \cdot 4 + 0.61 \cdot 6}} = \frac{{1.56}}{{5.22}} \cong 0.3
\]
\end{document}